\documentclass[fleqn,usenatbib]{mnras}

\usepackage{newtxtext,newtxmath}
\usepackage[T1]{fontenc}

\DeclareRobustCommand{\VAN}[3]{#2}
\let\VANthebibliography\thebibliography
\def\thebibliography{\DeclareRobustCommand{\VAN}[3]{##3}\VANthebibliography}

\usepackage{graphicx}	% Including figure files
\usepackage{amsmath}	% Advanced maths commands
\usepackage{float}
\newcommand{\teff}{$T_{\rm eff}$}

\newcommand{\logg}{$\log g$}
\newcommand{\vsini}{$v\sin i$}

\newcommand{\cd}{d$^{-1}$}
\newcommand{\eleanor}{{\sf Eleanor}}
\newcommand{\lightkurve}{{\sf Lightkurve}}

\title[Investigating Pulsating Variables and Eclipsing Binaries in NGC\,2126]{Investigating Pulsating Variables and Eclipsing Binaries in NGC\,2126 using Ground and Space-based Photometry, Astrometry, Spectroscopy and Modeling}

\author[Dileep et al.]{
Athul Dileep$^{1,2},$\thanks{E-mail: dileep@aries.res.in}
Santosh Joshi$^{1}$,
Sneh Lata$^{1}$,
Patricia Lampens$^{3}$,
Peter De Cat$^{3}$,
\newauthor
Sebastian Zúñiga-Fernández$^{4}$,
David Mkrtichian$^{5}$,
Pramod Kumar S$^{6}$,
Mrinmoy Sarkar$^{1,2}$,
\newauthor
Alaxender Panchal$^{1,7}$,
Yogesh C. Joshi$^{1}$,
Rishi C.$^{1}$,
Neelam Panwar$^{1}$,
Arjav Jain$^{1}$
and Neeraj Rathore$^{8}$
\\
$^{1}$Aryabhatta Research Institute of Observational Sciences (ARIES), Manora Peak, Nainital 263002, India  \\
$^{2}$Department of Applied Physics, M.J.P. Rohilkhand University, Bareilly 243006, Uttar Pradesh, India \\
$^{3}$Royal Observatory of Belgium (ROB), Av. Circulaire 3, 1180 Uccle, Belgium \\
$^{4}$Astrobiology Research Unit, Universit\'e de Li\`ege, All\'ee du 6 Ao\^ut 19C, B-4000 Li\`ege, Belgium  \\
$^{5}$National Astronomical Research Institute of Thailand (NARIT), Chiang Mai 50180, Thailand \\
$^{6}$Indian Institute of Astrophysics (IIA), Bangalore 560034, Karnataka, India \\
$^{7}$Physical Research Laboratory (PRL), Ahmedabad 380009, Gujarat, India \\
$^{8}$Bareilly College, Bareilly 243005, Uttar Pradesh, India
}

\date{Accepted XXX. Received YYY; in original form ZZZ}

\pubyear{\the\year{}}

\begin{document}
\label{firstpage}
\pagerange{\pageref{firstpage}--\pageref{lastpage}}
\maketitle

% Abstract of the paper
\begin{abstract}
Pulsating variables are prevalent in the classical $\delta$ Scuti instability strip of intermediate-age open star clusters. The cluster membership of these stars facilitates a comparative analysis of their evolution in analogous environments. In this study, we integrate ground-based observations, TESS Full Frame Images (FFIs), and Gaia DR3 data to investigate variable stars in the intermediate-age open star cluster NGC\,2126. We performed ground-based time-series observations of NGC 2126 to identify variable stars within its vicinity. Next, we determined the membership of these stars using parallax and the proper motions from Gaia DR3 archive. Then, we searched the TESS Full Frame Images (FFIs) for counterparts to the variables identified above and performed their frequency analysis and classification. Finally, we modeled the light curves (LCs) of detected eclipsing binaries (EBs), including V551\,Aur, which has a pulsating component. We found 25 members and 85 field variable stars. In TESS FFIs, we found LCs for 11 known variables and a new rotational variable. We determined that the pulsating EB V551\,Aur is a member of the cluster. The low- and medium-resolution spectra revealed the line profile variation and the basic parameters for the star, respectively. Simultaneous modeling of the eclipses and the embedded pulsations resulted in improved orbital parameters for the binary system. We also report the determination of orbital parameters for the previously uncharacterized EB system UCAC4 700-043174.

% The research demonstrates that the inadequate spatial resolution of TESS can be mitigated by cross-referencing the frequencies identified from ground observations with those of TESS. Additional ground-based observations of NGC\,2126 are required to perform a comprehensive frequency analysis of the variable stars. The present spectroscopic analysis indicates the need for high-resolution observations to further delineate the binary system V551\,Aur.
\end{abstract}

% Select between one and six entries from the list of approved keywords.
% Don't make up new ones.

%%%%%%%%%%%%%%%%%%%%%%%%%%%%%%%%%%%%%%%%%%%%%%%%%%

%%%%%%%%%%%%%%%%% BODY OF PAPER %%%%%%%%%%%%%%%%%%

\section{INTRODUCTION}

Open star clusters comprise stars that have emerged from a common molecular cloud and are gravitationally bound to one another. Consequently, they possess analogous ages, distances, and initial chemical compositions. The membership of stars within an open cluster imposes constraints, especially regarding their age in comparison to the ambiguously determined ages of field stars. Thus, examining stars that are constituents of an open cluster is more advantageous than analyzing field stars. Diverse independent methodologies for stellar characterization, such as asteroseismology \citep{2021Aerts}, can gain from open clusters by imposing supplementary constraints on parameters like age and mass \citep{2024A&A...686A.142L, 2023ApJ...946L..10B, 2023A&A...675A.167P}. The intermediate-age open star clusters possess their turn-off point adjacent to the classical instability strip, thereby hosting a variety of pulsating variable stars of A- and F-type, including $\delta$ Scuti stars and $\gamma$ Doradus stars \citep{2018chehlah}.

Ground-based studies of open clusters yield ensemble photometry for a collection of stars within a defined area. The Aryabhatta Research Institute of Observational Sciences (ARIES) in Nainital, India, possesses photometric observational facilities in the northern hemisphere, that are very suitable for studying open clusters. Details regarding the telescopes at the site will be presented in Section~\ref{observations} later in this paper. To capitalize on the advantages of studying open clusters, ARIES, Nainital is conducting long-term photometric monitoring using 1-meter class telescopes to examine pulsating variable stars in both young open clusters \citep{2020YCyoung, 2023lata} and intermediate-age open clusters \citep{2020YC, 2023maurya}. Additionally, we also performs long-term surveys for pulsating A- and F-type field stars \citep{2000ashoka, 2001martinez, joshi2003, 2006joshi, joshi2009, joshi2010, joshi2012, joshi2016, joshi2017}. We have recently augmented this exploration with space-based observations from K2 and TESS concerning field stars \citep{joshi2022, 2021otto, 2023otto, 2024mrinmoy, 2024dileep}. In this study, we integrate the space-based observations for open star clusters with the ground-based observations.

Similar to open star clusters, eclipsing binaries provide an independent assessment of fundamental stellar parameters. They provide accurate measurements of stellar masses and radii through orbital dynamics and eclipse depths \citep{2023panchal}. If one of the binary components is pulsating, the target becomes increasingly unusual. The examination of pulsations in eclipsing binary (EB) components provides insights into the internal structure of the pulsating star, system evolution, and the influence of tidal forces inducing oscillations \citep{pigulski2006, murphy_2019_2533474, 2021Galax...9...28L}. The existence of these unusual variable systems in open clusters serves as valuable test beds for validating various theoretical models for pulsation and stellar evolution.

We conducted a comprehensive ground- and space-based investigation of the intermediate-age open star cluster NGC 2126 \citep{gasper2003}. We utilized ground-based photometric observations from telescopes of varying diameters at ARIES, Nainital, low- to medium-resolution spectra, TESS Full Frame Images (FFIs), and Gaia DR3 data.

The manuscript is structured as follows. Sec.~\ref{sample} delineates the sample selection, while Sec.~\ref{observations} examines ground-based observations for the detection of variable stars. In Sec.~\ref{membership}, we ascertain the membership of the detected variables within the cluster, and in Sec.~\ref{sec:CLparam}, we evaluate the cluster parameters. Subsequently, in Sec.~\ref{tess}, we examine the TESS light curves for all stars and categorize the identified variables in Sec.~\ref{classification}. Ultimately, we present the modeling of the binary LCs in Sec.~\ref{binary} and provide a summary of our findings in Sec.~\ref{results}. We present our conclusions in Sec.~\ref{conclusion}.

\section{SAMPLE SELECTION}\label{sample}

We selected NGC\,2126 (RA = $06^{h} 02^{m} 37.9^{s}$, Dec = $+49^{o} 52' 59''$), which is an intermediate-age open star cluster hosting various pulsating stars located in the $\delta$ Scuti instability strip. The first CCD photometric study of the cluster was performed by \citet{gasper2003}. Later, it was studied by \citet{liu2012} and more recently \citet{2018chehlah} presented new ground-based photometric observations for NGC\,2126 and reported 11 variables, including two new $\delta$ Scuti variables. This makes this cluster as a potential target for asteroseismic investigation of pulsating variables of A- and F- type. Another reason that makes this cluster interesting is the presence of the EB V551\,Aur, which shows pulsational variability on top of the eclipses. The membership of this star to the cluster will place strong constraints on its evolution and pulsation mechanism. All of the previous ground-based studies had limitations due to their short and discrete observations. Thus, aliasing and frequency resolution limit the accuracy of detected frequencies. Furthermore, NGC\,2126 has one EB, UCAC4 700-043174, but there was insufficient data to model the system or report an orbital period \citep{2018chehlah}. Thanks to the nearly continuous TESS observations spanning approximately a month in each sectors, we can address the limitations due to lack of continuous data in ground-based observations. The long and short cadence of TESS enables us to study both faint and bright targets of the cluster. 

\section{GROUND-BASED PHOTOMETRIC OBSERVATIONS}\label{observations}

Time-series photometric observations for NGC\,2126 were acquired in the $V$ and $R$ bands with the 1.3-m Devasthal Fast Optical Telescope (DFOT), Nainital, India \citep{2022JAI....1140004J}. This telescope is equipped with a 2k $\times$ 2k CCD, gives a field of view of about 18 x 18 arcmin$^2$ with a plate scale of 0.54 arcsec/pixel. For standardization of our targets, the standard field SA\,98 of \citet{1992landolt} was also observed in \texttt{UBVRI} filters. For the pre-processing of the images, several bias frames and twilight flats were taken during the observing nights. We furthermore observed the cluster in the \texttt{V} and \texttt{R} bands during one more night using the 1.04-m Sampurnanand Telescope (ST), Nainital, India. This telescope is equipped with a 4k $\times$ 4k CCD, which gives a field of view of about 15 x 15 arcmin$^2$ \citep{2022Sampu}. The CCD was used in 4 $\times$ 4 binning mode to increase the signal-to-noise ratio, this gives a plate scale of 0.92 arcsec/pixel after binning. Additionally, we also observed the pulsating EB V551\,Aur in NGC\,2126 for a couple of nights with the 3.6-m Devasthal Optical Telescope (DOT) \citep{2018DOT2}. This telescope is equipped with a 4k $\times$ 4k CCD Imager that covers a field of view of 6.5 x 6.5 arcmin$^2$ \citep{2018SBP} and has a plate scale of 0.2 arcsec/pixel after 2 $\times$ 2 binning. An overview with the details of the collected ground-based observations is given in Table~\ref{tab:obser}.

The images were pre-processed with Image Reduction and Analysis Facility (\texttt{IRAF}) tasks: \texttt{zerocombine, flatcombine, and CCDPROC}. We applied the method of point-spread function (PSF) photometry to obtain the instrumental magnitudes for the stars in the target field, for this, we used the code \texttt{DAOPHOTII} by \citet{Stetson1992}. 

The standard field SA\,98 was used to convert the instrumental magnitudes to standard magnitudes. The following equations define the resulting transformations \citep{1987PASP...99..191S}:
\begin{equation}
v = V + (2.064 \pm 0.005) + (0.071 \pm 0.005)(V-I) + 0.21 Q
\end{equation}
\vspace{-5mm}
\begin{equation}
b = B + (2.652 \pm  0.007) - (0.143 \pm  0.006)(B-V) + 0.32 Q
\end{equation}
\vspace{-5mm}
\begin{equation}
i = I + (2.446 \pm  0.008) - (0.036 \pm 0.007)(V-I) + 0.08 Q
\end{equation}
\vspace{-5mm}
\begin{equation}
r = R + (1.910 \pm  0.010) + (0.088 \pm  0.016)(V-R) + 0.13 Q
\end{equation}
\vspace{-5mm}
\begin{equation}
u = U + (4.641 \pm  0.010) - (0.090  \pm 0.011)(U-B) + 0.49 Q
\end{equation}
where v, b, i, r, u are the instrumental magnitudes, and $V, B, I, R, U$ refer to the standard magnitudes and $Q$ is the airmass.

\begin{table*}
\centering
\caption{Observational log for NGC\,2126. The telescopes used for multi-band photometric observations and the date of observations are listed. Each of the columns named $V, R, U, B$ and $I$ have entries of the form `number of frames $\times$ exposure time'.} 
\begin{tabular}{c c c c c c c}
\hline
\hline
\\
Telescope& Date of & $V$ & $R$ & $U$ & $B$ & $I$  \\
&Observation &  &  &  &  &   \\
\hline
\\

1.04-m ST  & 2023 Nov 22 & $119\times50$s  & $119\times50$s &     -           &     -           &     -          \\
           &             &                 &                &                 &                 &                \\
1.3-m DFOT & 2022 Oct 29 & $ 71\times40$s  & $ 70\times30$s &     -           &     -           &     -          \\
           & 2022 Oct 30 & $ 12\times40$s  & $ 12\times30$s &     -           &     -           &     -          \\
           & 2022 Nov 11 & $200\times50$s  &      -         &     -           &     -           &     -          \\
           & 2023 Feb 11 & $ 59\times50$s  & $ 59\times30$s &     -           &     -           &     -          \\
           & 2023 Feb 14 & $ 44\times50$s  & $ 44\times30$s &     -           &     -           &     -          \\
           & 2023 Feb 18 & $103\times50$s  & $103\times30$s &     -           &     -           &     -          \\
           & 2023 Feb 19 & $ 17\times50$s  & $ 17\times30$s &     -           &     -           &     -          \\
           & 2023 Oct 15 & $ 50\times50$s  & $  3\times50$s &     -           &     -           &     -          \\
           & 2023 Nov 14 & $100\times50$s  & $100\times30$s &     -           &     -           &     -          \\
           & 2023 Nov 18 & $111\times50$s  & $111\times50$s &     -           &     -           &     -          \\
           & 2023 Dec 02 & $102\times50$s  & $103\times30$s & $  3\times300$s & $  3\times12$0s & $  3\times20$s \\
           &             &                 &                &                 &                 &                \\
3.6-m DOT  & 2021 Dec 13 & $ 12\times300$s &     -          &     -           & $ 11\times100$s &     -          \\
            &              &     -           &     -          &     -           & $  1\times150$s &     -          \\
           & 2021 Dec 14 & $ 12\times100$s &     -          &     -          & $  9\times200$s &     -          \\
           &             & $ 10\times150$s &     -          &     -           & $ 12\times300$s &     -          \\
            &             & $  1\times300$s &     -          &     -           &     -           &     -          \\
           & 2021 Dec 15 & $ 16\times100$s &     -          &     -           & $ 11\times270$s &     -          \\
           &             & $ 11\times80$s  &     -          &     -           & $ 12\times160$s &     -          \\
           & 2023 Mar 28 & $ 50\times50$s  &     -          &     -           &     -           &     -          \\

\hline
\end{tabular}
\label{tab:obser}
\end{table*}

The search for variables in the cluster field was done based on the LCs obtained with 1.3-m DFOT. The $V$-band photometric LCs from each night were combined together and then phase-folded with the period that matches the first main peak in the Lomb-Scargle periodogram. The same period was used to phase-fold the combined $R$-band LCs to see any corresponding variations in the $R$-band. The stars that showed brightness variation greater than three standard deviations with respect to the mean brightness, at least in the $V$-band were considered to be variable stars and were chosen for further analysis. The LCs for newly detected member variables from the ground are given in Fig.~\ref{fig:kpvars} and the field stars in Appendix Fig.~\ref{fig:newvars1}-\ref{fig:newvars5}. We also list the detected dominant freqeuncies and the corresponding amplitudes of the field stars in Table~\ref{tab:newvars}.

\begin{figure}
    \includegraphics[width=\columnwidth]{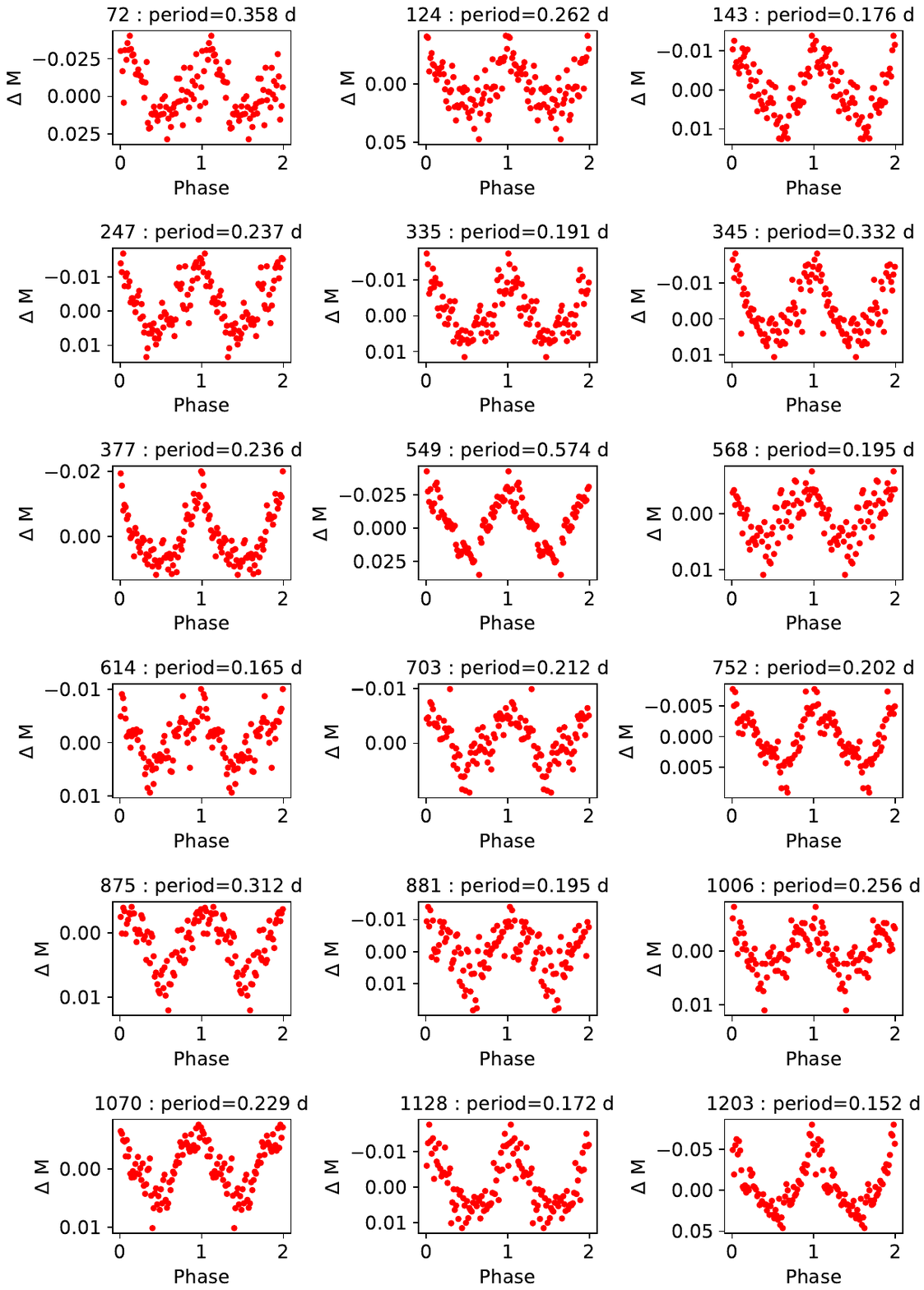}
    \caption{Phase-folded LCs of the newly detected variables, which are the members of open cluster NGC\,2126. Their respective IDs and detected periods are given in the title of each plot.}
    \label{fig:kpvars}
\end{figure}

\section{CLUSTER MEMBERSHIP}\label{membership}

\begin{figure}
    \includegraphics[width=\linewidth]{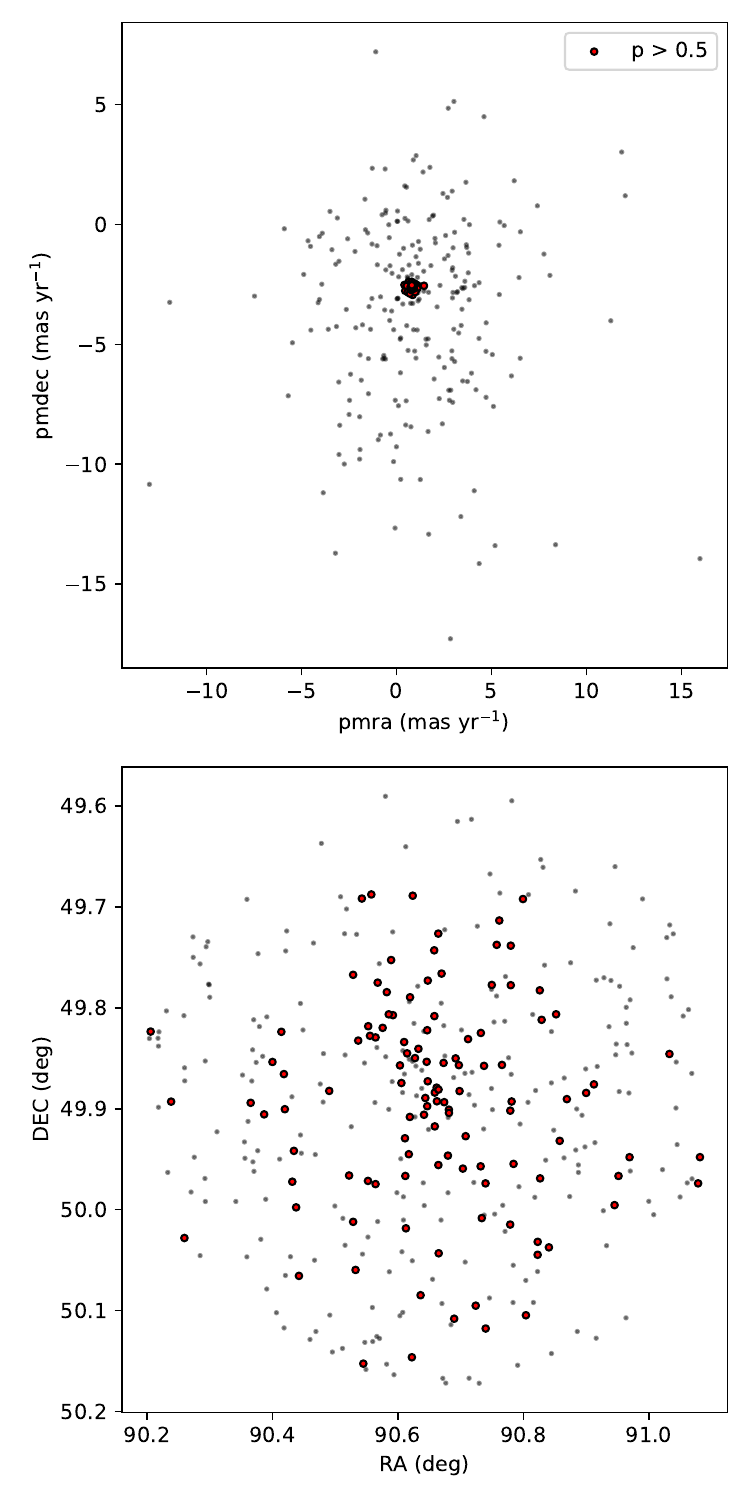}
    \caption{Top panel shows the vector point diagram of NGC 2126 using Gaia DR3 proper motion data. The Bottom panel shows the position of the cluster members and field stars in the sky as observed by Gaia DR3. Both panels have stars within an 18 arcmins radius from the center which was selected for membership determination. In both panels red dots represents the member stars for which the membership probability is greater than 50 \% and gray dots represents the field stars.}
    \label{fig:VPD}
\end{figure}

\begin{figure}
    \includegraphics[width=\columnwidth]{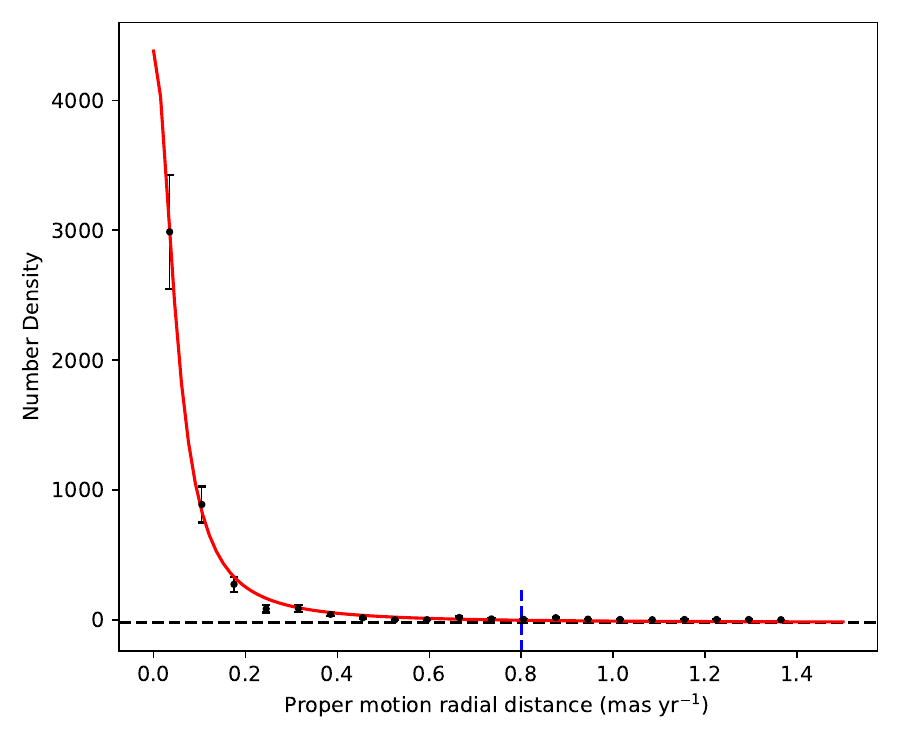}
    \caption{Number density plot for the proper motion space from Gaia data. The red curve shows the best fitting RDP. The black horizontal dashed line shows the background density. The blue vertical dashed line marks the chosen limiting radius.}
    \label{fig:king}
\end{figure}

For the determination of membership, we used astrometric data from Gaia DR3. The Gaia mission \citep{gaia2016} is an all-sky survey that collects multi-band photometric, astrometric, and spectroscopic data. Gaia's astrometric data have proven to be very successful in identifying and determining the membership of open clusters and globular clusters. The proper motion and parallax of all stars within a radius of 18 arcminutes around the cluster center were obtained by querying the Gaia DR3 catalogue \citep{gaiadr3}. Stars with large error bars were excluded from our study. We constructed the vector point diagram (VPD) of the cluster region to identify stars with similar proper motion (Fig.~\ref{fig:VPD}). Furthermore, we used the parallaxes to select samples with similar distances. For the membership probability, we used the method described in \citet{2002Wu} by identifying a cluster population and field population from the VPD. For this purpose, we plotted the radial density profile (RDP) of the VPD by calculating the number density in concentric annular regions from the center of the VPD. For the center of VPD, we smoothed the proper motion space with a 2D Gaussian kernel and chose the point with the maximum kernal density. The resultant RDP is shown in Fig.~\ref{fig:king} along with a fitted King's profile \citep{king1962structure}. The number density was seen to approach the background density at a radial distance $\sim$ 0.8 mas/yr. We chose this as the limiting radius for probable cluster members and the stars outside this radius as field stars. 

\begin{figure}
    \includegraphics[width=\columnwidth]{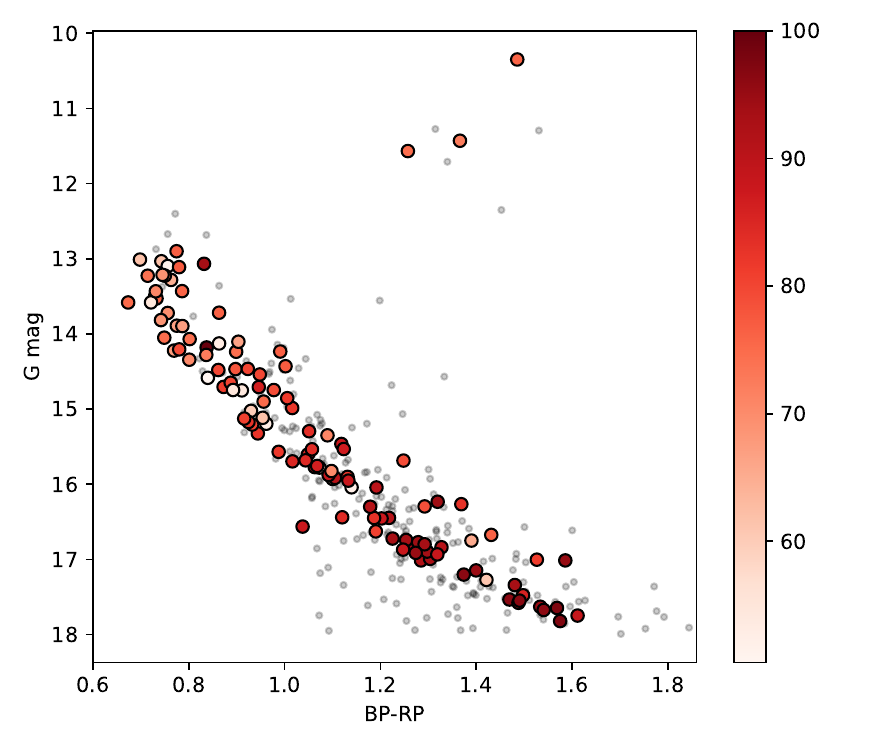}
    \caption{The CMD of NGC\,2126 with the field stars (grey dots) and the member stars (color-coded dots according to their calculated membership probability). The probability scale is represented by the color bar on the right side (50 - 100 \%).}
    \label{fig:CMD}
\end{figure}

The Gaia Color-Magnitude Diagram (CMD) for stars having membership probability greater than 50 \% is shown in Fig.~\ref{fig:CMD}. The members that are highly probable are situated along the main sequence, the evolved red giant branch, and the turn-off point. These members will introduce substantial constraints on the system's age by means of isochrone fitting (see Sec.~\ref{sec:CLparam}).

\section{CLUSTER PARAMETERS}\label{sec:CLparam}
\begin{figure}
    \includegraphics[width=\columnwidth]{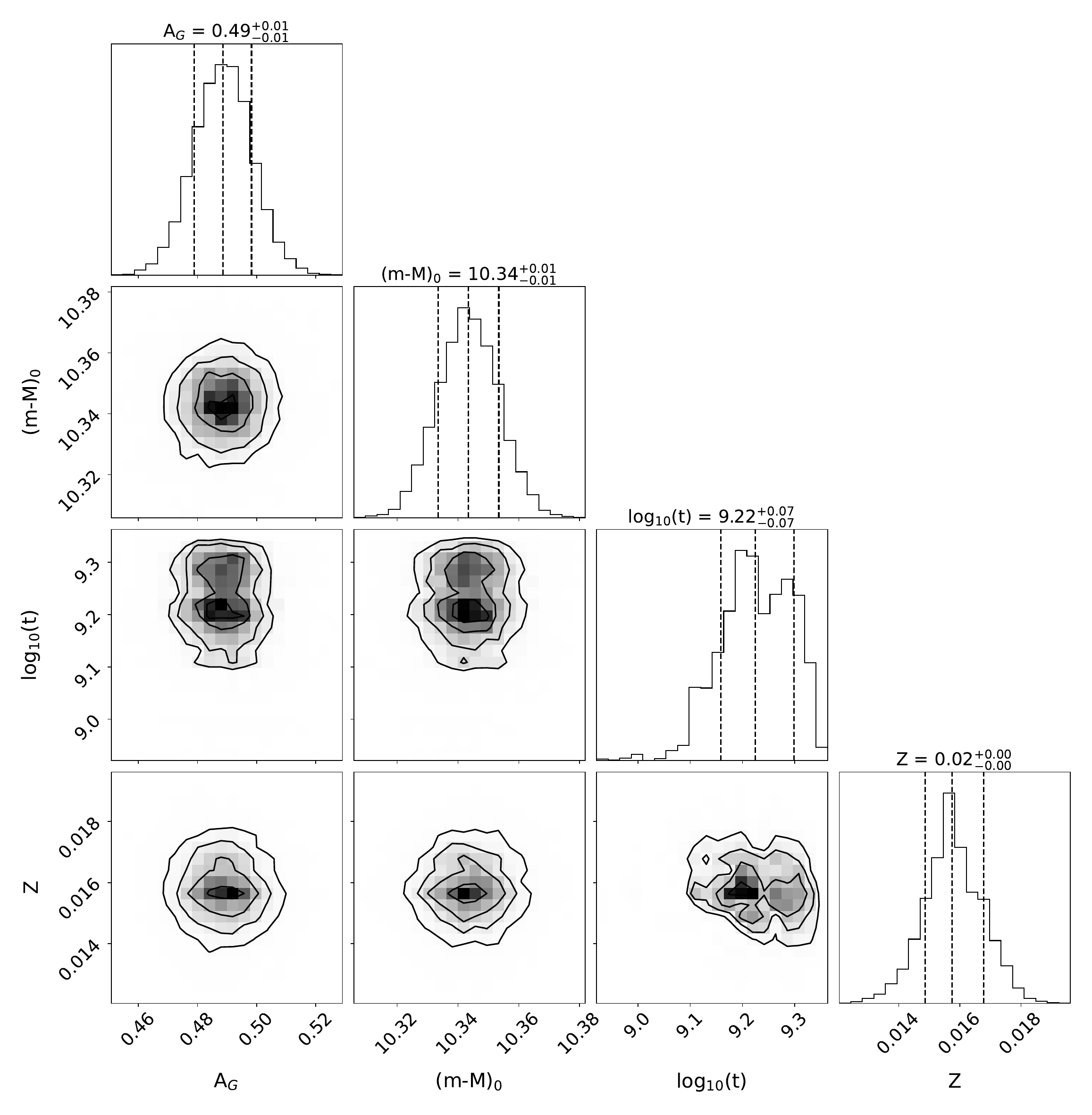}
    \caption{The resultant posterior distribution for the best fit isochrone.}
    \label{fig:asteca_iso_corner}
\end{figure}

\begin{figure}
    \includegraphics[width=\columnwidth]{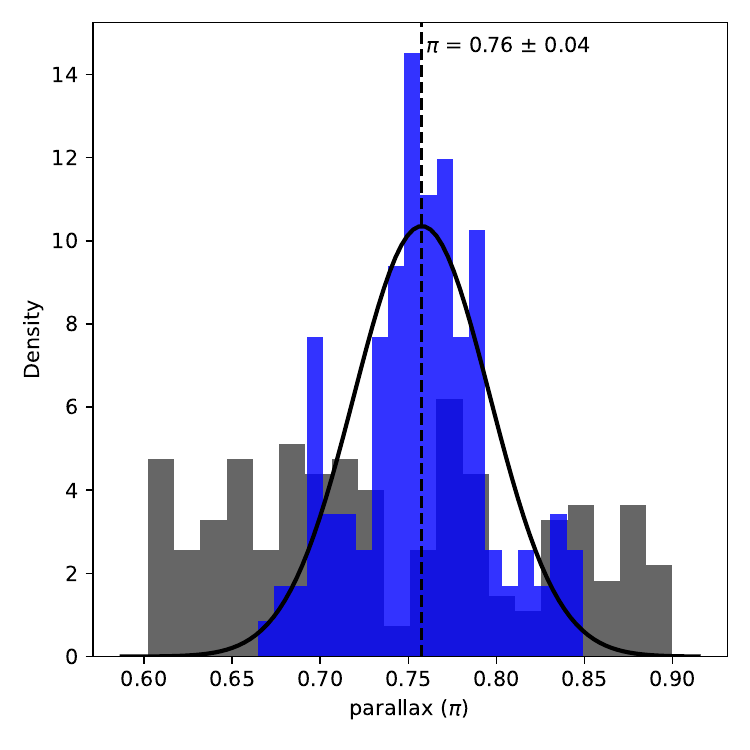}
    \caption{The histogram in blue displays the parallax distribution of the cluster members and the histogram in grey shows the parallax distribution of the field stars. By fitting a Gaussian to the cluster member parallaxes, we obtained the mean cluster parallax, which was then used to determine the distance to the cluster.}
    \label{fig:parallax}
\end{figure}

\begin{figure}
    \includegraphics[width=\columnwidth]{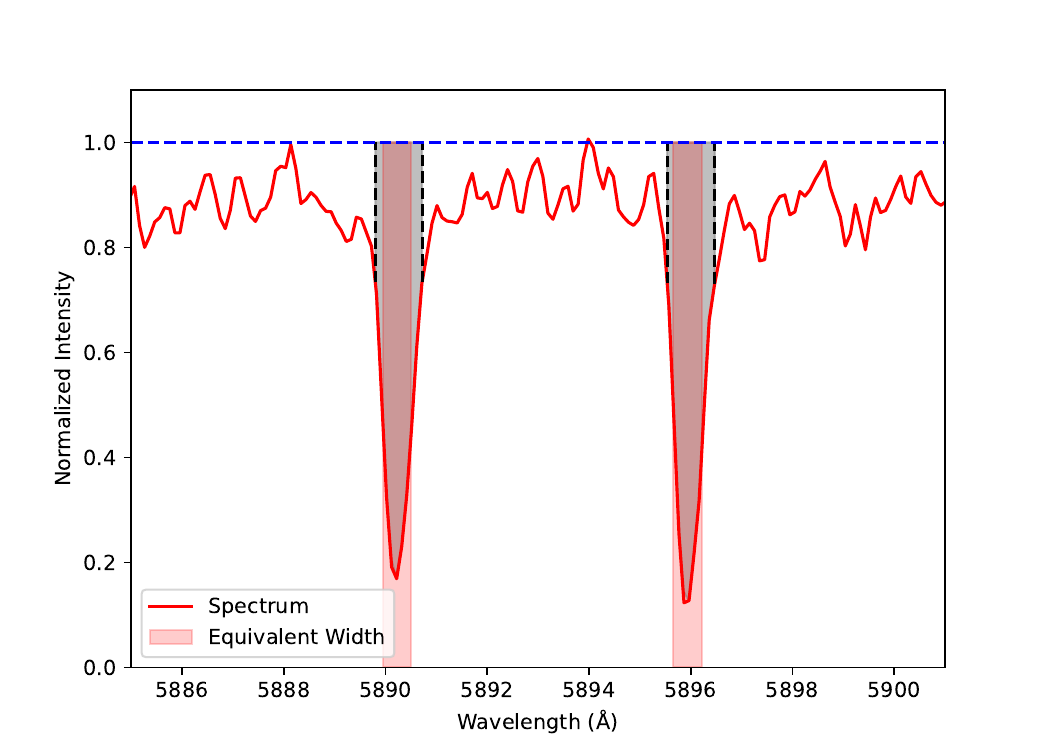}
    \caption{The interstellar absorption lines of Sodium in the medium-resolution spectrum of V551\,Aur. The blue horizontal dashed line is the continuum. We selected the regions between the black dashed lines to calculate the equivalent width of Na D1 \& D2 lines, filled with gray color in the figure. The resultant equivalent width is the horizontal width of the rectangles drawn between 0 and 1 filled with red color.}
    \label{fig:medres}
\end{figure}

\begin{figure}
    \includegraphics[width=\columnwidth]{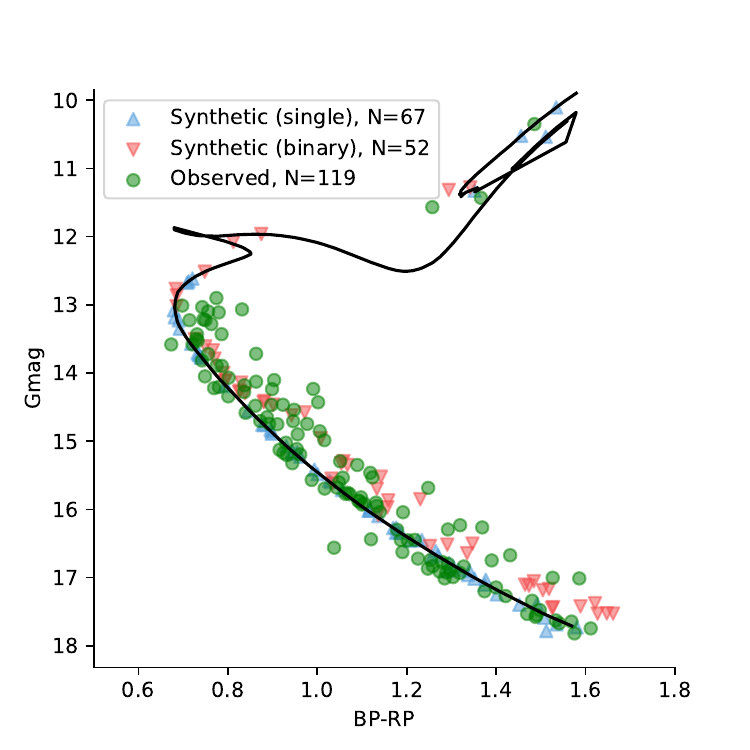}
    \caption{The figure shows the best-fit isochrone using the code \texttt{ASteCA} represented by the black curve. The synthetic members calculated by \texttt{ASteCA} are represented with blue straight and red inverted triangles, respectively. And our input observed members from Gaia DR3 are represented as green circles.}
    \label{fig:asteca}
\end{figure}

The CMD can be used to fit theoretical isochrones and determine cluster parameters such as age, metallicity, and reddening. The age and metallicity of the member stars should be comparable as a result of their shared origin. Thus, by comparing theoretical isochrones of varying ages and metallicities, we can estimate the cluster parameters. The Gaia CMD consists of the BP-RP color and the G-band mean magnitude. We used the isochrones downloaded from the CMD service \href{http://stev.oapd.inaf.it/cgi-bin/cmd}{CMD 3.7} in the Gaia EDR3 photometric system, and the evolutionary tracks from the PARSEC release v1.2S \citep{2012Bressan, 2014chen, 2015chen, Tang2014}. We utilized the Gaia DR3 parallaxes for the cluster members to determine an initial distance modulus and used prior literature values for an initial value for reddening. Utilizing these corrections, we drew a series of isochrones next to the observed CMD. Subsequently, the fit was optimized using the \texttt{ASteCA} code \citep{asteca2015} by selecting the above guessed range of values (metallicity = [0.010   ,  0.030], log age =  [9.0   ,  9.3], distance modulus = [9, 12], extinction = [0.2, 1] ) as uniform priors. \texttt{ASteCA} computes synthetic isochrones between the given grid of isochrones and uses a genetic algorithm \citep{1995charbo} to minimize the negative log likelihood, -log[$L_{i}$(z,a,d,e)], where z is the metallicity, a is the age, d is the distance, and e is the extinction. The genetic algorithm (GA) is an optimization technique derived from natural selection, wherein a population of candidate solutions evolves over generations through selection (optimal fit), crossover (various combinations of parent sets), and mutation (random alterations introduced at each stage) to identify the most effective solution for a specific problem. It is particularly effective in exploring complex, multi-dimensional spaces and avoiding local minima. For more thorough details regarding the algorithm please refer to \citep{1995charbo}. Additionally to quantify parameter uncertainties we employed the Markov Chain Monte Carlo (MCMC) method within \texttt{ASteCA}, the resultant posterior distribution is shown in Fig.~\ref{fig:asteca_iso_corner}. By fitting a Gaussian to the parallax histogram (Fig.~\ref{fig:parallax}), we calculated the mean parallax, which allowed us to calculate the distance to the cluster, which came out to be 1.3 $\pm$ 0.1 kpc. Also, we calculated distance to the cluster from the distance modulus of the isochrone fit using standard relations. We used the best fit coefficients published by Gaia team based on the empirical relation used by \citet{2018gaiaav} to convert extinction in gaia G-band ($A_{G}$) to extinction in V-band ($A_{V}$), and used $R_{v}$=3.1 to get the reddening $E(B-V)$ for the cluster. We also verified this estimate using color-color diagrams from ground-based observations (Fig.~\ref{fig:color_color}). Apart from that a medium-resolution spectrum (see Sec.~\ref{spectro}) of a member variable, V551 Aur, was used to estimate the reddening towards the cluster. We used the calibration published by \citet{2012ebv}, between the equivalent width of Na doublets and the reddening (Fig.~\ref{fig:medres}). The final values for the HR diagram were found by taking the average of the distance estimates from the isochrone fit and the parallax fit, as well as the reddening estimates from the isochrone fit and the Na doublets. The best-fit parameters for the cluster are tabulated in Table~\ref{iso} and the best fit isochrone is plotted in Fig.~\ref{fig:asteca}.

\begin{table}
\centering
\begin{minipage}{\linewidth}
%\begin{minipage}{150mm}
\caption{The table gives the best-fit parameters for isochrone fitting with \texttt{ASteCA}. Here, Z is the metal fraction, t is the age in years, A$_{G}$ is the extinction in Gaia G band, $(m-M)_{0}$ is the distance modulus of the cluster, D gives the distance to the cluster and E(B-V) is the reddening. The uncertainty in the last digit is given in brackets for each values.}
\label{iso}
\end{minipage}
\bigskip
\begin{tabular}{cccc}
\hline
\hline
\\
\textbf{Parameter}  &\textbf{Isochrone fit} & \textbf{Parallax Fit} & \textbf{Na D lines} \\

\hline
\\
$log_{10}(t)$  & 9.22(7)  & -& - \\
$Z$   & 0.016(1) & - & -\\
$A_{G}$  &    0.49(1) mag & - & -\\
$(m-M)_{0}$  & 10.34(1) mag & - & -\\
$D$  &    1.169(5) kpc & 1.3(1) kpc & -\\
$E(B-V)$ & 0.20(4)& -& 0.29(8) \\

\hline
\end{tabular}
\end{table}

\section{TESS PHOTOMETRY}\label{tess}

The Transiting Exoplanet Survey Satellite (TESS) is an all-sky survey telescope whose prime objective is to detect exoplanets orbiting nearby bright stars using the transit method \citep{2014JAVSO..42..234R}. Apart from its primary mission, it also provides time-series photometry for asteroseismic studies of stars with magnitudes up to 12. TESS Target Pixel Files (TPF) and Light Curve Files are available with short cadences (20\,s and 120\,s) while Full Frame Images (FFIs) are taken with a longer cadence (30 min). Recently, the FFIs were taken at a shorter cadence of 200 seconds, which we used in our study. These data products can be downloaded from Mikulski Archive for Space Telescopes (\href{https://mast.stsci.edu/portal/Mashup/Clients/Mast/Portal.html}{MAST}).

Given the large pixel size of the TESS CCD, which is 21$''$ per pixel, resolving crowded fields like we see in open star clusters will be difficult, so our search for variability was primarily based on ground-based observations. The higher spatial resolution of ground-based observations enables the resolution of individual stars in the crowded field. Consequently, we concentrated our investigation of variability in the TESS FFIs on sources that had confirmed variable signals from the ground.

TESS observed the field of NGC\,2126 during Sector 19 with an 1800-s cadence and during Sectors 59, 60, and 73 with a 200-s cadence. The observations from sector 60 were excluded from this study because the targets were too close to the edge of the TESS CCD, preventing useful photometry. The log of observations taken from the TESS archive is given in Table.~\ref{Tab:TESS}. We performed the FFI photometry of our targets in the field using the python packages \eleanor\ \citep{2019adina} and \lightkurve\ \citep{2018Lk}. To perform the photometry, we relied on the cut-outs taken from the FFI using the package {\sf tesscut} \citep{2019brasseur}. 

\begin{table}
\centering
\begin{minipage}{\linewidth}
%\begin{minipage}{150mm}
\caption{NGC\,2126 was observed in the sectors listed below. The observation period, cadence (in seconds), and duration (in days) are also given.} 

\label{Tab:TESS}
\end{minipage}
\bigskip
\begin{tabular}{ccccc}
\hline
\hline
\\
 \textbf{TESS }  &\textbf{Observation Period} & \textbf{Cadence} & \textbf{Duration} \\
 \textbf{Sector}          &              &    (s)          & (days)            \\
\hline
\\
19  & 28 Nov - 23 Dec, 2019 & 1800 & 24.5\\
59  & 26 Nov - 23 Dec, 2022 &  200 & 26.4 \\
73  & 07 Dec - 03 Jan, 2024 &  200 & 26.9\\

\hline
\end{tabular}

\end{table}

\eleanor\ is a Python package that automates the extraction of LCs from FFIs. 
For sector 19, the \eleanor\ package was used and the resulting sky-subtracted LCs were free of telescope systematic and suitable for analysis. 
To identify our targets in the FFI, we overplotted the Gaia sources in the field and the apertures were carefully selected to reduce the contamination from nearby targets. 
%Whereas, in 
\eleanor\ did not perform well in sectors 59 and 73, necessitating a manual approach utilizing the \lightkurve\ package. Here, we also conducted an investigation into the FFI by overplotting the Gaia sources to minimize contamination from nearby sources and used custom apertures to extract the sky-subtracted LCs. As a final step, we used the \texttt{RegressionCorrector} in \lightkurve\ package. The task creates a design matrix, which is a matrix of flux time series of the background pixels. For which a principal component analysis (PCA) \citep{2009A&A...507.1729D} is performed to reduce its dimensionality and select the first few principal components, which contain most of the instrumental and systematic effects in that particular part of the sector. Then a linear combination of these components is used to remove the long term trends in the light curve.

Furthermore, we attempted a global photometric search for variables in TESS FFIs independent of the ground-based observations with a PSF-based technique utilizing the flux distribution from Gaia observations, implemented in the Python package \texttt{tglc} \citep{Han_2023}. We used \texttt{tglc} for only long cadence data (sector 19), since the computational expense for short cadence was high and not possible for us at the moment. This resulted in the discovery of one new variable in addition to the known variables. Due to contamination-related uncertainty, we had to discard a lot of other targets with variable signals. We discuss how we dealt with the effect of contamination in our data in Sec.~\ref{contamin}.

The frequency analysis for the variable stars in TESS data was done using the program \texttt{Period04} \citep{2004perio04}. We adopted the criteria SNR $>=5.6$ given by \citet{2016Zong} for space-based observations. The SNR was calculated after prewhitening all of the dominant frequencies to obtain the residual spectra, followed by calculating the noise using a box size of one cycle per day along both sides of the target frequency. The errors in frequency and amplitude were calculated using the analytical relations provided by \citet{1999mont}. The Appendix Table.~\ref{tab:pulsa1} \& \ref{tab:pulsa3} list the frequencies for the known variables detected using TESS and from the ground and the light curves in the respective sectors are plotted in Fig.~\ref{fig:lcs19}, \ref{fig:lcs59} and \ref{fig:lcs73}.

\subsection{Contamination}\label{contamin}

The large pixel size of TESS causes the blending of signals from nearby sources. The resulting LCs for a target can only be trusted when (i) the source is much brighter than the nearby contaminating sources so that the background source only adds to the noise and (ii) the contaminating sources are non-variables and, hence, only increase the noise level rather than introduce signals in the periodogram. 

\begin{figure}
    \includegraphics[width=\columnwidth]{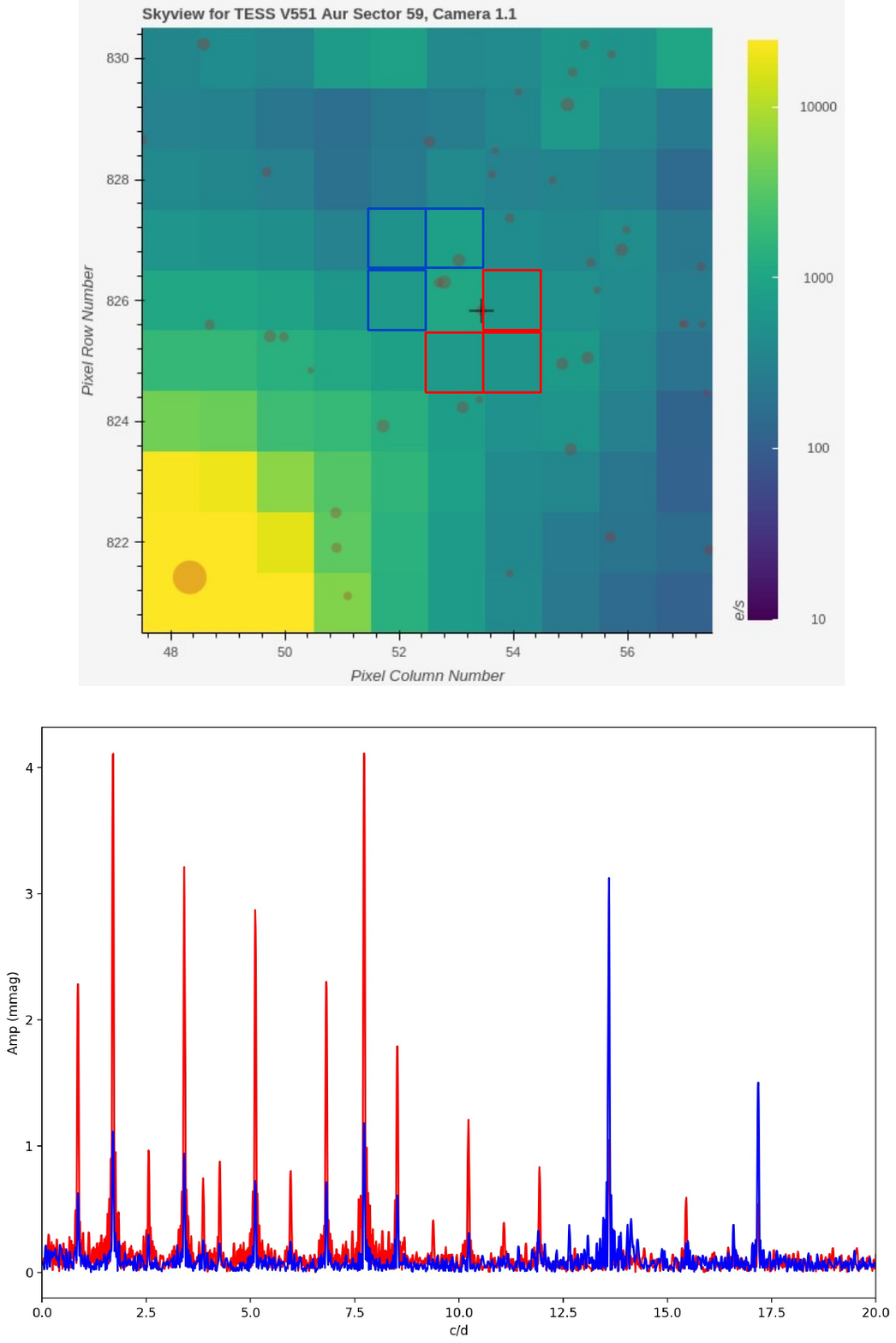}
\caption{The top panel depicts the TPF of star V6 overlaid with nearby Gaia DR3 targets. V551 Aur (V6) is identified with a black `+' symbol. Custom apertures were chosen near V6 (marked with red boxes) and N1 (marked with blue boxes). The bottom panel depicts the periodogram generated by the two apertures, V6 in red and N1 in blue, which are superimposed on each other. The harmonic signals and pulsation are stronger in the red aperture and weaker in the blue aperture. The two dominant frequency signals in blue behave in opposite ways, indicating they are from the nearby star N1.}
    \label{fig:Pixv6}
\end{figure}

In order to investigate the contamination in TESS LCs, we first overplotted the Gaia sources on the target pixel file (TPF) in the vicinity of each target to see if there are any bright nearby objects that might potentially contribute to contamination. After that, we selected a variety of custom apertures to determine the origin of a specific set of frequencies within the frequency spectrum. An example of this investigation is shown in Figure~\ref{fig:Pixv6}, where V551 Aur (V6) is a pulsating EB that is contaminated with a $\delta$ Scuti star (N1). Next, we cross-checked whether any of these frequencies are predefined in the ground-based data obtained from \cite{2018chehlah}. For, example in case of V6 in Fig.~\ref{fig:Pixv6}, the two strong blue peaks towards the right were detected by \citet{2018chehlah} for N1 and few of the dominant red peaks to the left were reported by \citet{2018chehlah} for V6.

We found that two of the pulsating stars ZV2 (hybrid star) and N1 ($\delta$ Scuti star) were contaminated significantly by the nearby binary stars V4 and V6 respectively and found that the contamination can be addressed with the above technique. Furthermore, to quantify the extent of contamination endured by the variable stars detected in our study, we calculated a contamination factor for each of these stars using Gaia DR3 positions and magnitudes, where we took the ratio of the target flux to the total flux in a 21$''$ radius. 

\section{CLASSIFICATION OF VARIABLE STARS}\label{classification}

Variable stars are classified according to the shape of their LCs, amplitude of variability, frequency of variability, and location on the HR diagram. The pulsating stars can be identified from the HR diagram by plotting the instability strips for $\delta$ Scuti and $\gamma$ Doradus pulsator. In our study, we used the theoretical $\delta$ Scuti and $\gamma$ Doradus instability strips, computed by \citet{2005dupret}.

\begin{figure}
    \includegraphics[width=\columnwidth]{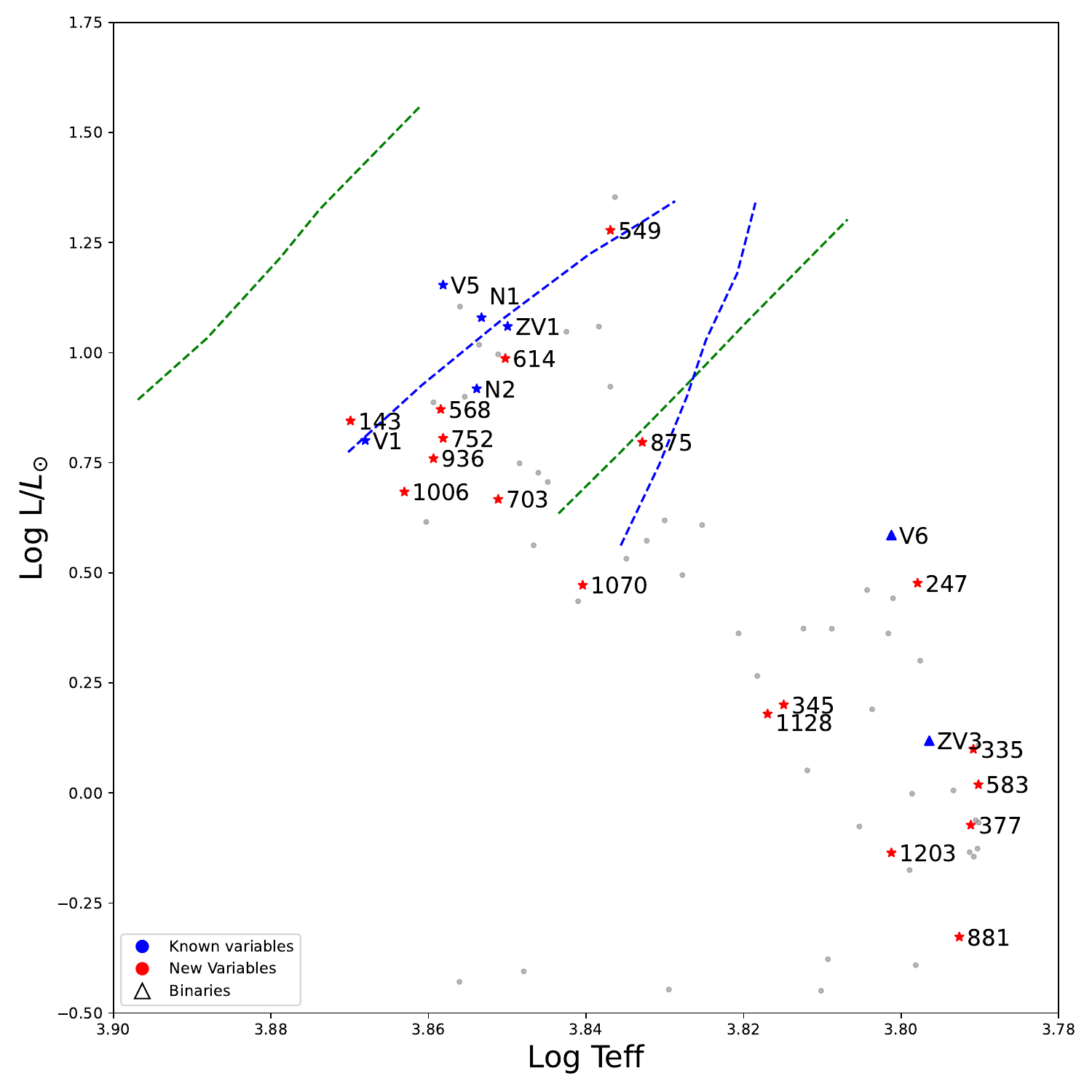}
    \caption{HR-diagram for the member stars based on the ground-based observations. Blue `star' symbols and the blue triangles indicate the known pulsating variables and the EBs respectively. The red `star' symbols represent the newly detected variable stars. The green and blue dashed lines represent the edges of the theoretical $\delta$ Scuti and $\gamma$ Doradus instability strips, respectively, computed by \citet{2005dupret}.}
    \label{fig:HRdiagram}
\end{figure}

The standard B- and V-band magnitudes from our ground-based observations with the 1.3-m DFOT were used to generate the HR diagram for the cluster members. After extinction correction, the color index B-V was used to find the effective temperature (\teff), which was based on an empirical relationship from \cite{2010torres}. Next, the empirical relation provided by \cite{2010torres} was employed to calculate the bolometric correction ($BC_{v}$) for the stars in the V-band. Then the luminosity can be estimated using the standard relations,
where we took bolometric magnitude of sun as $M_{bol_{\odot}}$ = 4.73 \citep{2010torres}. The HR diagram in Fig.~\ref{fig:HRdiagram} shows the known variables, the newly found variables, and the instability strips for the $\delta$ Scuti and $\gamma$ Doradus pulsations.

$\gamma$ Doradus variables are mainly non-radial $g$-mode pulsators showing periodic variations in their brightness with pulsation frequencies ranging from 0.5 to 3\,\cd\ \citep{2010astero} and amplitude of variation up to 0.1 mag in the V-band. The convective blocking drives the $g$-mode pulsations, with buoyancy as the restoring force   \citep{2005dupret}. In the HR diagram, the $\gamma$ Doradus instability strip is seen overlapped with the cooler edge of the $\delta$ Scuti instability strip in the main sequence with an early F spectral type. 

$\delta$ Scuti stars are primarily $p$-mode pulsators with brightness variations ranging from 0.001 to 0.8 mag in the V band and pulsation frequencies ranging from 5 to 50\,\cd\ \citep{2010astero}. The $\kappa$-mechanism drives the pulsations in these stars, which occur at the partial ionization zones in the outer radiative region \citep{2000pamya}. These stars are found at the lower part of the classical instability strip, which intersects the main sequence, called the $\delta$ Scuti instability strip. They are A- to early F-type stars with masses ranging from 1.5 to 2.5 $M_{\odot}$. They pulsate in both radial and non-radial $p$-modes. Some of the stars pulsate both in $g$-modes and $p$-modes, with characteristics similar to $\gamma$ Doradus and $\delta$ Scuti stars. These stars are known as hybrid pulsators \citep{2009handler}.

The G- and K-type stars are located below the $\delta$ Scuti and $\gamma$
Doradus instability strips. These stars, with a convective envelope, exhibit rotational variability due to spots on their surface \citep{2008mamajek}. The LCs of these rotational variables produces a frequency spectrum with the rotational frequency below 5\,\cd\ and at least one harmonic of the rotational frequency \citep{2013balona}. Thus we classified the main sequence variables with a frequency less than 5 \cd\ falling in the G to K spectral type as rotational variables. Also in this region one can find the solar type oscillations. The stars whose LCs show (partial or full) eclipses can be classified as EBs. 

All of the member stars, as well as those with LCs in both TESS and ground-based observations, were classified according to the criteria listed above. Table~\ref{tab:newmem} shows the results for newly discovered member stars, while Table~\ref{tab:known} shows the results for stars with variability in TESS and ground-based observations. We observe that ground-based classification is incomplete due to insufficient data.

\begin{table*}
\centering
\caption{
The newly detected variable cluster members are listed with their ID, RA, DEC, standard magnitude, first dominant frequency used for phase folding, corresponding semi-amplitude, assigned variability type, membership, and contamination in TESS.} 
\begin{tabular}{c c c c c c c c c}
\hline
\hline
\\
ID &  RA & DEC & Mag &Frequency & Amp  & Variability type  & Mem. & Cont. \\
 &(deg)&(deg)&  (mag) & (\cd) & (mag)  & & (\%) &(\%)\\
\hline
\\
  72 &90.5885   & 49.752833&17.423 & 2.796 & 0.022 & -        & 90.8 & 49.40\\ 
  124 &90.527917   &  49.767417&18.18 & 1.905 & 0.029&   -     & 85.8 & 0 \\

 143 &90.567028 & 49.775222 &13.817 & 5.690  & 0.009 & DSCT      & 75.2 & 0 \\ 
 247 & 90.657528  & 49.808472 &14.785 & 4.224 & 0.010  & -         & 77.3 &33.17 \\ 
  
 335 &90.711278   & 49.831306&15.739 & 5.234 & 0.009 & -         & 70.9 & 3.79\\ 
 345 & 90.536139  & 49.832639&15.455 & 3.016 & 0.009 & -     & 61.9 &3.44\\ 
 377 & 90.607278   &49.84275 &16.169 & 4.231 & 0.011 & ROT         & 57.3 &89.78\\ 

 549 &90.663472   & 49.881278&12.743 & 1.741 & 0.024 & -     & 89.8 &51.58\\ 
 568 & 90.606944  & 49.884222&13.752 & 5.119 & 0.005 & DSCT      & 54.9 &89.35\\ 

 614 &90.643167   &49.889639 &13.465 & 6.062 & 0.006 & DSCT      & 64.8  &46.79\\ 
 703 & 90.680778   & 49.90125&14.264 & 4.722 & 0.005 & - & 74.4  &37.33\\ 
 752 &90.6185 & 49.908417  &13.916 & 4.948 & 0.005 & - & 69.8 &20.28\\ 
 875 & 90.610778  & 49.929528&13.949 & 3.202 & 0.005 & GDOR & 76.3  &7.07\\ 
 881 &90.857528   &  49.932306&16.802 & 2.568 & 0.008&  ROT      & 83.1 & 38.97\\

1006 & 90.611389   & 49.966889 &14.22  & 3.906 & 0.005 & - & 74.8 & 1.89\\ 
1070 & 90.679444  & 49.983611&14.756 & 4.359 & 0.005 & - &  50.5  &3.21\\
1128 &90.437278   &  49.997917&15.505 & 2.907 & 0.012&     -   & 82.7 & 24.59\\

1203 &90.778667   &  50.015194&16.312 & 3.284 & 0.055&   ROT     & 90.2 & 3.19\\

\\

\hline
\\
\multicolumn{6}{l}{DSCT : $\delta$ Scuti, GDOR : $\gamma$ Doradus, ROT : Rotational variable } \\

\end{tabular}
    \label{tab:newmem}
\end{table*}

\begin{table*}
\centering
\begin{minipage}{\textwidth}

\caption{The variables detected in TESS with their ID's, names (if any), the orbital or rotational period ($P$(d)) from TESS, the pulsational period ($P_{pul}$(d)) from TESS, assigned variability type, membership from Gaia DR3 (Mem.) and TESS contamination (Cont.). The error values at the last digit are represented inside brackets for the periods.}
\label{tab:known}
\end{minipage}
\bigskip
\begin{tabular}{ccccccccc}
\hline
\hline
\\
ID & Name   & RA & DEC &$P$(d)& $P_{pul}$(d) & Variability & Mem. & Cont.\\
&&&&&&Type&(\%)&(\%)\\
\hline
\\
V1 & V546 Aur &90.433639  &  49.941861    & - &   0.8094(2) & GDOR & 68.7 &0.24\\
V2 & V547 Aur   &90.488972  &  49.982056   & - &   0.9394(2) &  GDOR &-&0.37\\
V3 & V548 Aur    &90.521556  &  49.819861   & - & 0.078040(6) &  DSCT &-&6.38\\
V4 & V549 Aur     &90.588583  &  49.877167 & 3.300284(1) & - & EB  &-& 88.50\\
V5 & V550 Aur     &90.609806  &  49.865889   & -& 0.082706(7)  &   DSCT &- &7.36\\
V6 & V551 Aur     &90.658250   &  49.884250  &  1.1731767(4) & 0.129453(8) & EB + DSCT & 76.5&88.52\\

ZV1 & UCAC4 700-043212  &90.637472  &   49.880111  & -& 0.08161(1) & DSCT  & 51.6 &15.05\\
ZV2 & UCAC4 700-043178    &90.590278  &  49.873361 & - &  0.067336(9) & Hybrid & - &38.08\\
ZV3 & UCAC4 700-043174    &90.584750  & 49.806611 &  3.224040(2) &-& EB  & 83.7&59.84\\
N1 & UCAC4 700-043245     &90.661139  & 49.879417 & - &  0.073534(4) & DSCT & 61.8&65.29\\
N2 & UCAC4 700-043196    &90.614056  & 49.845361  & -&  0.068719(9) & DSCT & 70.7&10.51\\
839 & - &  90.648306  & 49.920833  & 0.7390(3)& -  & ROT & - &  2.17\\
\hline
\\
\multicolumn{6}{l}{DSCT : $\delta$ Scuti, GDOR : $\gamma$ Doradus, EB : Eclipsing Binary, ROT : Rotational variable } \\

\end{tabular}
\end{table*}

\begin{figure*}
    \centering
    \includegraphics[width=0.95\linewidth]{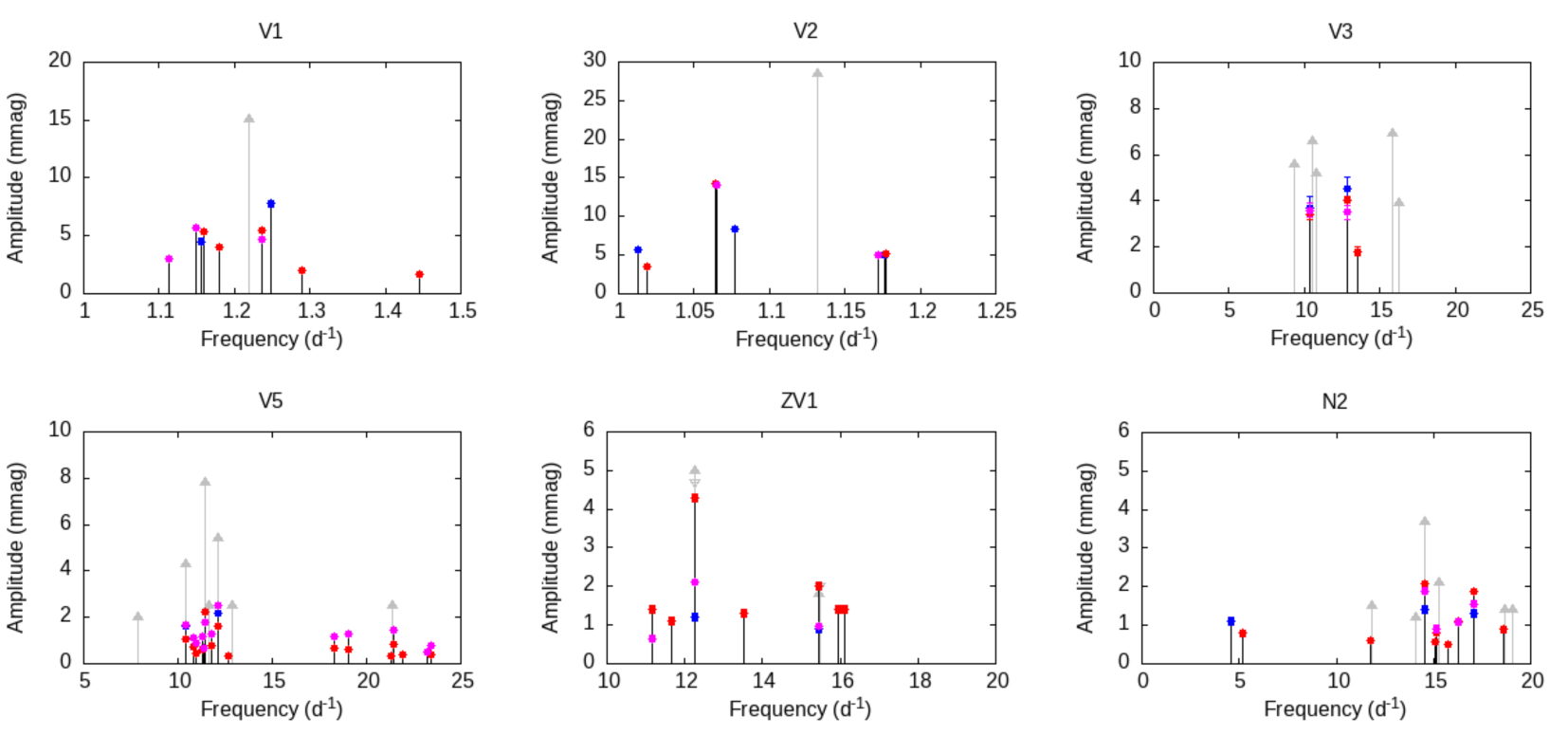}
    \caption{
    Representation of the results of the frequency analysis of the TESS data of the stars without contamination. The filled circles with error bars in blue, red, and magenta show the observed amplitudes in the TESS data of sectors 19, 59, and 73, respectively. They are compared to those published by \citet{2018chehlah} (grey upward triangles) and \citet{2012zhang} (grey downward triangles), if available. 
    }
    \label{fig:freq1}
\end{figure*}

\begin{figure*}
    \centering
    \includegraphics[width=0.95\linewidth]{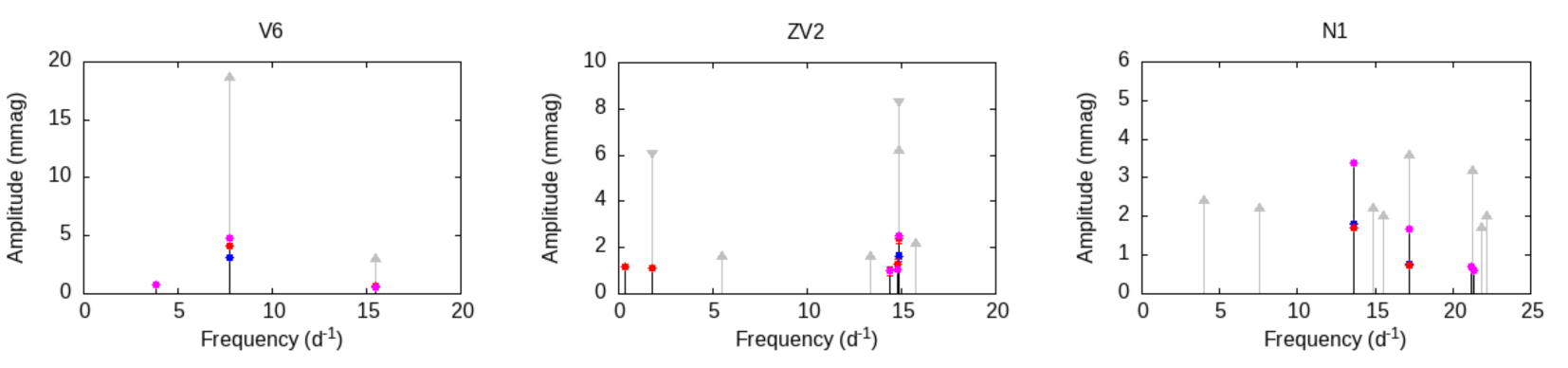}
    \caption{Same as Fig.\,\ref{fig:freq1}, but for stars affected by contamination. 
    }
    \label{fig:freq2}
\end{figure*}

\subsection{Intrinsic Variables}

\subsubsection{$\gamma$ Doradus Stars}

NGC 2126 contains two known $\gamma$ Doradus variables (V1 and V2). Our membership analysis revealed that V2 is not likely to be a cluster member. The two stars were detected by TESS. Both stars have a frequency lower than 3 c/d. Fig.~\ref{fig:freq1} shows a comparison of our results with previous frequency estimates from the literature. There were no harmonics of the fundamental frequency in the frequency spectrum, showing that they are not rotational variables. The position of V1 in the HR diagram is within the $\gamma$ Doradus instability strip. Since, V2 is not a member star, we used the Gaia temperature \teff = 7134 K, indicating it is an early F-type star. Thus, we classify V1 and V2 as $\gamma$ Doradus variables. From our ground observations, we classified the star 875, as $\gamma$ Doradus variable, based on the frequency and position in the HR diagram. The TESS data for this star produced null results.

\subsubsection{$\delta$ Scuti Stars and Hybrid Pulsators}

\citet{2018chehlah} reported six $\delta$ Scuti stars (V3, V5, ZV1, N1, N2) and a hybrid pulsator (ZV2). All of these stars were detected in TESS data. ZV1, N1, and N2 were determined to be probable cluster members. Fig.~\ref{fig:freq1} and Fig.~\ref{fig:freq2} show a comparison between the frequencies detected in literature and those we detected from TESS's three sectors. Pulsation frequencies in ZV1, N1 and N2 were similar to those of a $\delta$ Scuti star, ranging from 5 to 30 c/d. They belong to the $\delta$ Scuti instability strip on the HR diagram. Thus, we classify these stars as $\delta$ Scuti stars. However, N1 is contaminated by a nearby binary star V6. To address this we used the method described in the previous section. Since, V3 and V5 are not cluster members we used the Gaia temperatures, V3 has \teff = 7454 K, while V5 has \teff = 6166 K, this corresponds to a spectral type F. The spectral type along with the pulsation frequencies from TESS indicates these are $\delta$ Scuti stars. According to \citet{2012zhang}, ZV2 is a hybrid pulsator because it has a low frequency of 1.812\,\cd in addition to the $\delta$ Scuti type frequencies. Our analysis of TESS data recovered the dominant frequency 14.851\,\cd\ , as well as the low frequency reported by \citet{2012zhang}. As ZV2 is not a member, we used the Gaia temperature, \teff = 7081 K, indicating an F-type star. This means the star is in the $\gamma$ Doradus instability strips where $\delta$ Scuti instability strip also overlaps. Thus, we classify it as a hybrid pulsator. This star was also contaminated by a nearby EB, V4. We used the method described in previous section in this case also to identify the frequencies. Using ground-based data, we identified three new $\delta$ Scuti variables, 143, 568, and 614. The classification was made based on their positions in the HR diagram and oscillation frequencies. TESS detected no signals from these stars due to high contamination.

\subsection{Extrinsic Variables}
\subsubsection{Rotational Variable}

\begin{figure}
    \centering
    \includegraphics[width=0.95\linewidth]{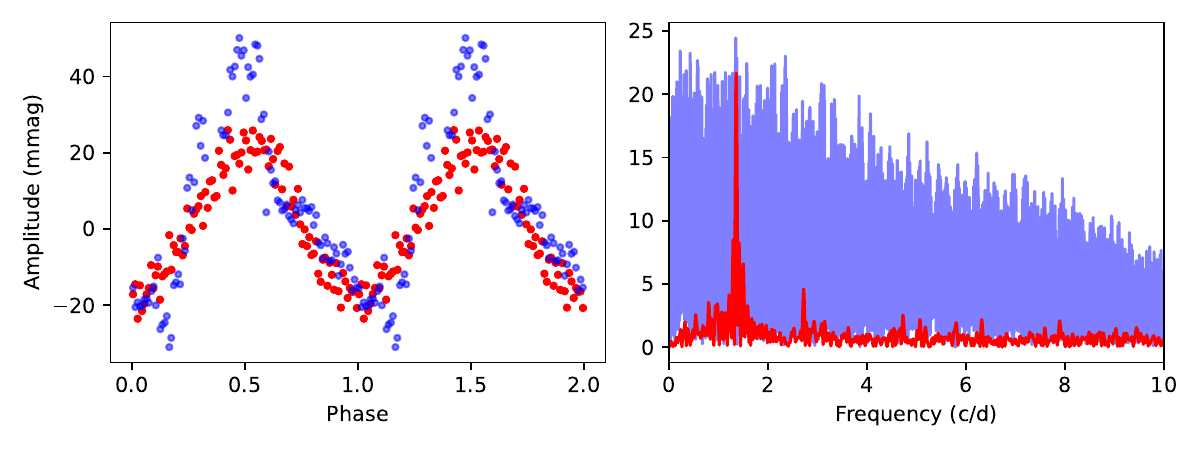}
    \caption{Left panel shows the LCs for the newly detected rotational variable 839 phase folded to the dominant peak in its frequency spectrum and binned with a bin size of 0.02 in phase. The red points are from TESS Sector 19 and the blue points are from the ground-based observations. In the right panel, the corresponding frequency spectra for the respective LCs are given. We see the advantage of frequency resolution from TESS compared to the ground-based.}
    \label{fig:new839}
\end{figure}

We discovered a new variable, 839, in the field of NGC\,2126. The star shows variable signals on the ground, as well as TESS data. Fig.~\ref{fig:new839} compares the TESS LC and frequency spectra of this star. This star demonstrates TESS's improved frequency resolution. The dominant frequency is the same in both spectra, however, the ground-based spectrum has many aliases. The frequency is less than 5 c/d and appears as harmonics in the TESS frequency spectra. The Gaia temperature, \teff = 5725, indicates a G-type star. Thus, we classify this star as a rotational variable. From the ground-based observations, we also identified three additional rotational variables, 377, 881, and 1203. We were unable to detect any of these stars using TESS due to contamination.

\subsubsection{Eclipsing Binaries}

\citet{2018chehlah} identified three EBs in NGC 2126, V4, V6, and ZV3. They were able to cover a large enough phase range for two stars, V4 and V6, allowing them to model the LCs and derive basic parameters such as orbital periods, inclination, mass ratio, and temperature ratio. For ZV3, however, modeling and period determination were not possible due to a lack of data covering the full eclipse. We used continuous time series TESS observations to fully cover the three EBs' phase range in our study. ZV3 and V6 are members of the cluster, while V4 is not a member. Given the cluster membership, the case of V6 is more interesting because it is a pulsating EB.

According to available data, \citet{2018chehlah} estimated an orbital period of 1 to 1.5 days for ZV3. However, with the help of TESS data that included multiple eclipses, we were able to calculate the orbital period for this star. We used all of the primary minima found in sectors 19, 59, and 73 by fitting a parabola to their dips. To determine the system's linear ephemeris, the orbital cycle vs. time of minima (TOM) curve was fitted with a linear relation.

\begin{figure}
    \includegraphics[width=\columnwidth]{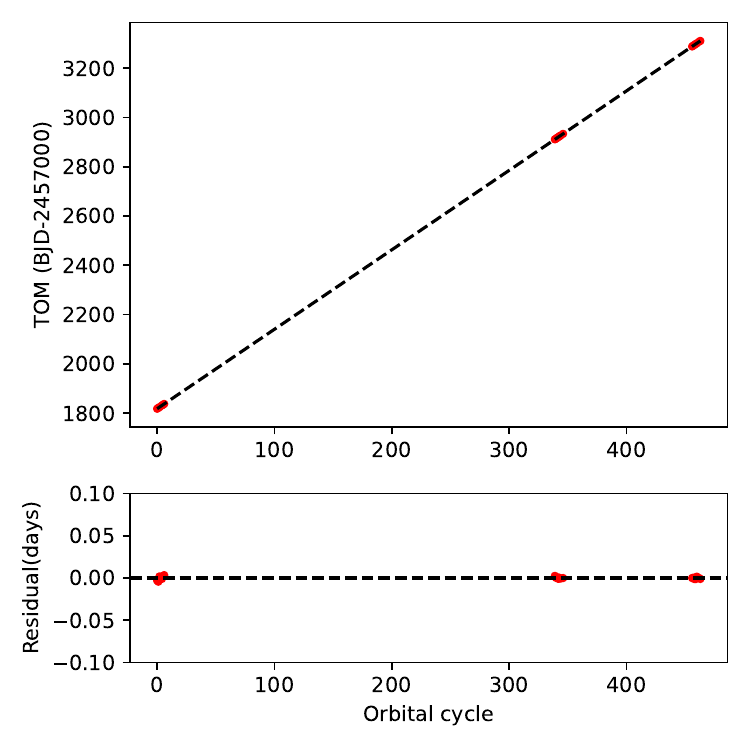}
    \caption{The top panel shows the orbital cycle vs. times of primary minima (TOM) detected from all the three TESS Sectors. The dashed line is the straight line fit to this data. The bottom panel shows the residuals of the fit.}
    \label{fig:Eph3}
\end{figure}

The resulting linear ephemeris for ZV3 is :
\begin{equation}
\begin{split}
BJD_{min} &=   2458818.0594 \pm  0.0008 \\
        &\quad +  3.224040 \pm  0.000002 \times E
\end{split}
\end{equation}
where $BJD_{min}$ is the time of primary minima at epoch E. Fig.~\ref{fig:Eph3} shows the epheremis fit.

For V4, \citet{2018chehlah} discovered an orbital period of 3.300281 days. We collected LCs for this star from TESS sectors 19, 59, and 73. In addition to the primary minima detected from TESS sectors, two minima reported in \citet{2018chehlah} were included in the ephemeris fit.

The subsequent linear ephemeris for V4 is :
\begin{equation}
\begin{split}
BJD_{min} &= 2452308.388   \pm  0.002 \\
        &\quad + 3.300284  \pm  0.000001 \times E
\end{split}
\end{equation}

\citet{gasper2003} described V6 as an EB with a pulsating component. \citet{2018chehlah} revisited the nature of this variable star, as well as the parameters of the open cluster that hosts it. They concluded that the pulsation could be caused by tidally excited $g$-modes, which exhibit tidal resonance with the orbital period. Although previous research concluded that V551 Aur is unlikely to belong to the open cluster, our analysis of Gaia DR3 data revealed that it most likely does. As in the previous case, we determined the linear ephemeris from all the primary minima detected in TESS sectors, as well as the ground-based minima listed in Table 9 of \citet{2018chehlah}.

The final linear ephemeris for V6 is:
\begin{equation}
\begin{split}
BJD_{min} &=  2452307.410 \pm  0.003 \\
        &\quad + 1.1731767  \pm  0.0000004 \times E
\end{split}
\end{equation}

\begin{figure*}
    \includegraphics[width=\linewidth]{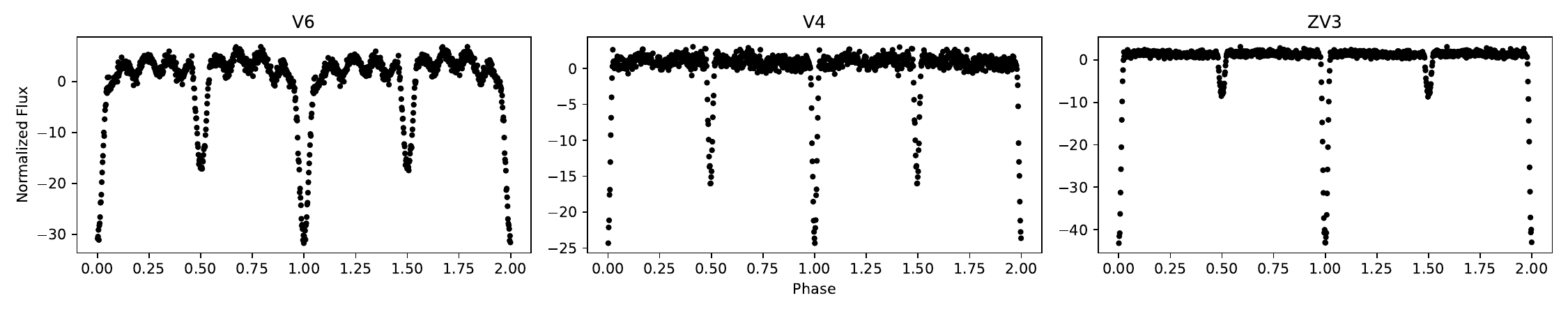}
    \caption{From left to right each panel shows the phased LCs of V6, V4 and ZV3, using all the available TESS sectors (19, 59 and 73) and binned using a bin size 0.002 in phase. For V6 we can clearly see the pulsations on top of the eclipses.}
    \label{fig:phas3}
\end{figure*}

Fig.~\ref{fig:phas3} depicts the phased and binned TESS LCs for three EBs. Appendix Fig.~\ref{fig:Eph1} \& \ref{fig:Ephv549} shows the ephemeris fit for V6 and V4, respectively.

\section{Eclipsing Binary Modeling}\label{binary}

EBs provide an independent method for calculating stellar masses and radii. By modeling the eclipsing LCs and the radial velocity curve, we can determine the best fitting parameters for masses, radii, inclination, and so on. However, we only have TESS LCs for our EBs. Due to the faintness of these stars, no high-resolution spectra could be obtained using the facilities we have direct access to. Thus we used the eclipsing LCs alone to model our EBs. While modelling the EB LC, one should take care of the intrinsic variability that it may possess either due to some intrinsic process in the system such as pulsations, mass transfer, star spots etc. \citep{pigulski2006} or external reasons like telescope systematics or contamination in case of TESS. Therefore, it was imperative to incorporate the intrinsic variability or long-term systematics into the LC modeling process.

The primary and secondary minima of all three of our EBs (V4, V6, and ZV3) are clearly distinct, with the out-of-eclipse region flat and the secondary minima precisely located at the 0.5 phase of the orbital period. Consequently, it is reasonable to assume that they are detached binaries in a circular orbit. We modeled the intrinsic variability in EB LCs using the celerite package \citep{celerite1, celerite2}. This package is part of the \texttt{exoplanet}\footnote{https://github.com/exoplanet-dev/exoplanet} package \citep{exoplanet:joss} and features a scalable Gaussian process model with a \texttt{SHOTerm} (a term representing a stochastically-driven, damped harmonic oscillator). This package has the capability to simultaneously model the intrinsic variability with a scalable Gaussian process and the eclipses for a detached binary system. The exoplanet code employs \texttt{PyMC} \citep{2015Pymc} for its probabilistic modeling. It utilizes the \texttt{starry} package \citep{2019Starry} to generate limb-darkened LCs and employs a fast and efficient solver for Kepler's equations.

\begin{figure}
    \includegraphics[width=\columnwidth]{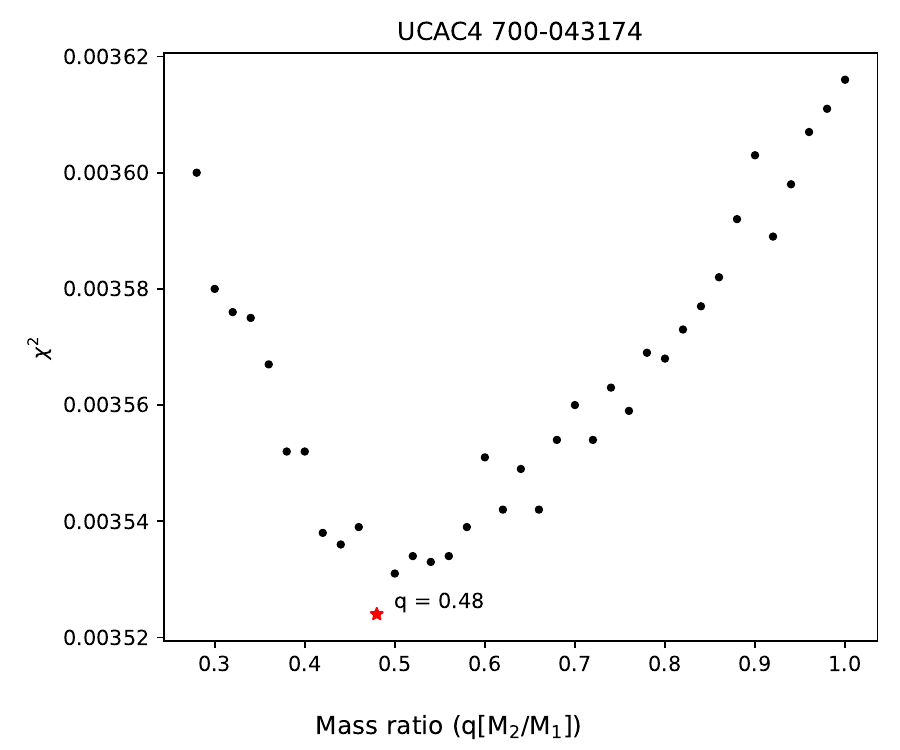}
    \caption{ The figure shows the $q$ (mass ratio) search for UCAC4 700-043174 (ZV3) using \texttt{PHOEBE 1.0} legacy version. The $q$ value with minimum $\chi^{2}$ is marked with a red asterisk and was chosen for further modeling.}
    \label{fig:qval}
\end{figure}

The q-search method \citep{1994PASP..106..921W, 2016joshi} can be employed to estimate the mass ratio (q-value) of the system in the absence of spectroscopic data. Previous studies by \citet{liu2012} and \citet{2018chehlah} calculated the mass ratio for V6 and V4, respectively, using the photometric q-search method using the \texttt{PHOEBE 1.0} legacy version based on Wilson-Devinney code \citep{2005phoebe}. We used the q-values calculated for V4 and V6 from previous studies. For ZV3, we used a similar approach and calculated the q-value with \texttt{PHOEBE 1.0}. The best fitting q-value was $q$ = 0.48 based on the q-value vs. chi square plot, as shown in Fig.~\ref{fig:qval}.

These photometrically derived q-values served as priors for modeling the EB LCs. The primary star's mass was determined by comparing evolutionary tracks on the HR diagram, and the radius was calculated using the Stefan-Boltzmann's relation, with the primary star's temperature being the B-V color temperature in the case of V4 and ZV3. For V6, we used the spectroscopic temperature obtained through medium-resolution spectroscopy.

\begin{figure}
    \includegraphics[width=\columnwidth]{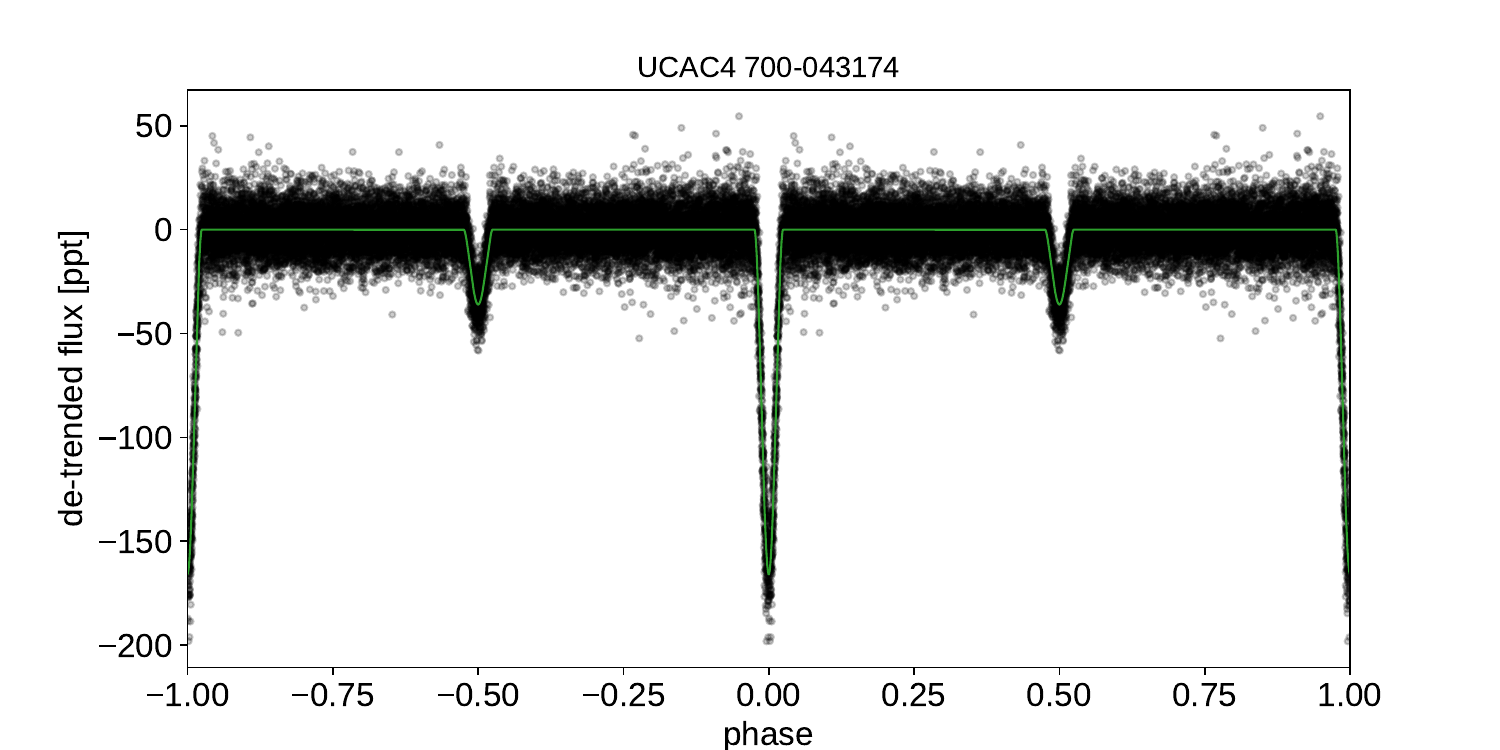}
    \caption{The phased folded LC of UCAC4 700-043174 (ZV3) with the best fitting model.}
    \label{fig:model3}
\end{figure}

\begin{figure}
    \includegraphics[width=\columnwidth]{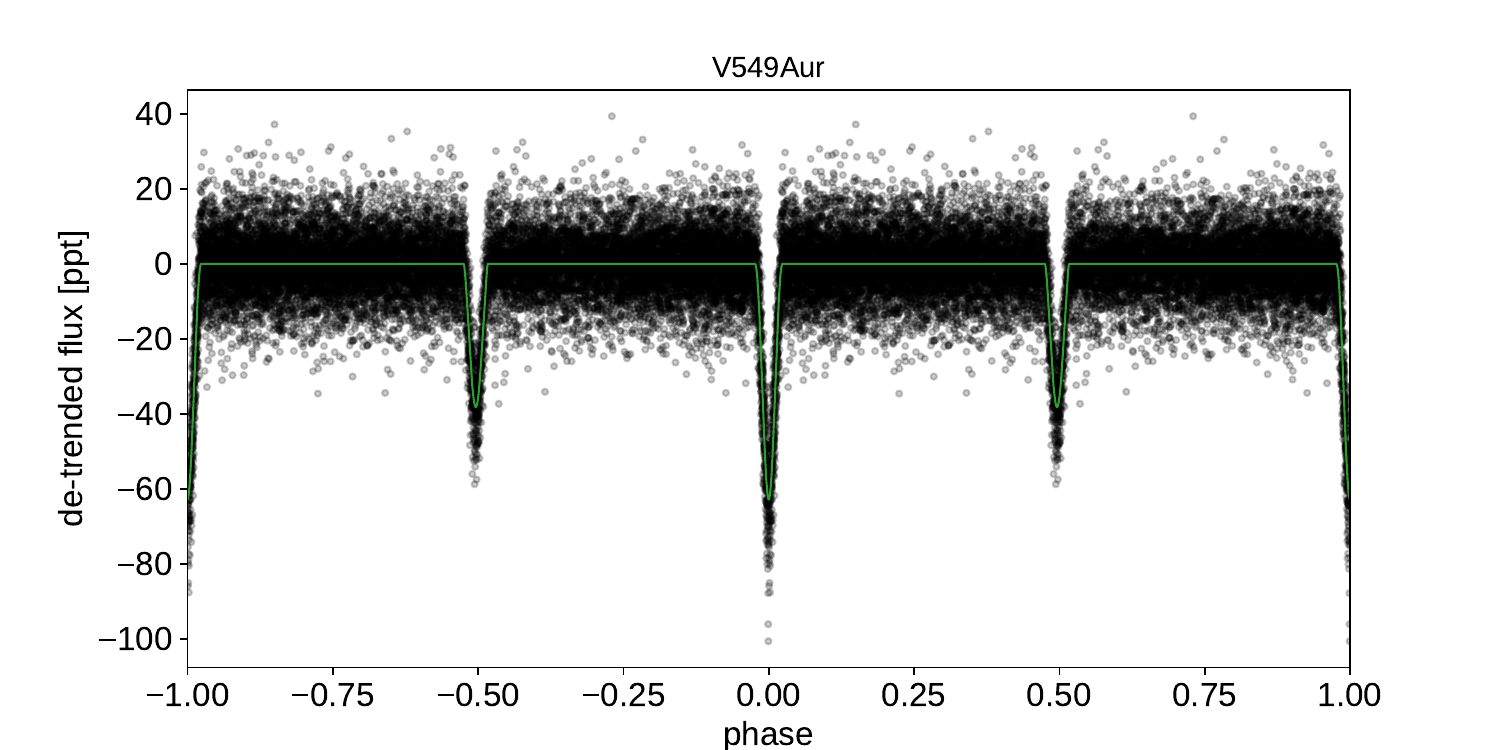}
    \caption{The phased folded LC of V549\,Aur (V4) with the best fitting model.}
    \label{fig:model1}
\end{figure}

\begin{figure}
    \includegraphics[width=\columnwidth]{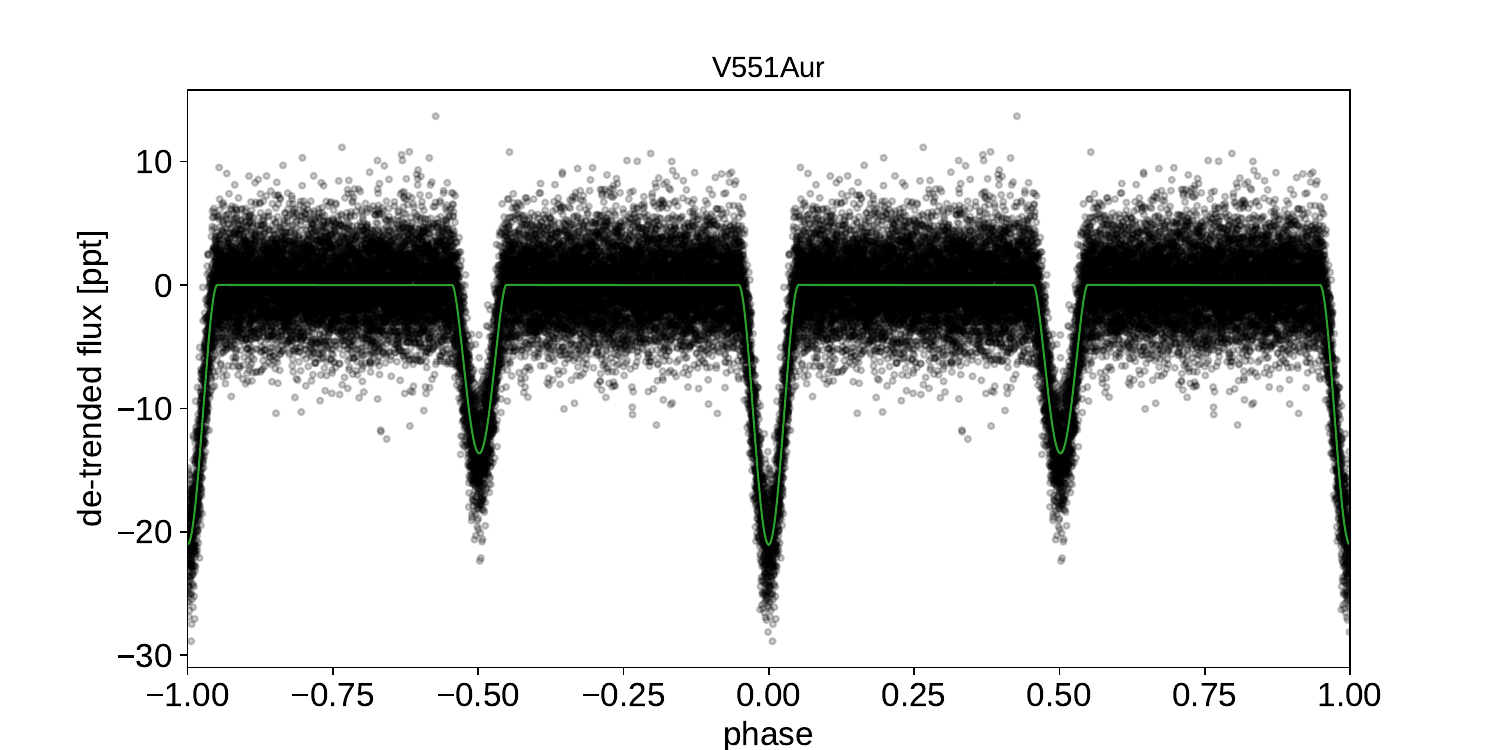}
    \caption{The phased folded LC of V551\,Aur (V6) with the best fitting model.}
    \label{fig:model}
\end{figure}

\begin{figure}
    \includegraphics[width=\columnwidth]{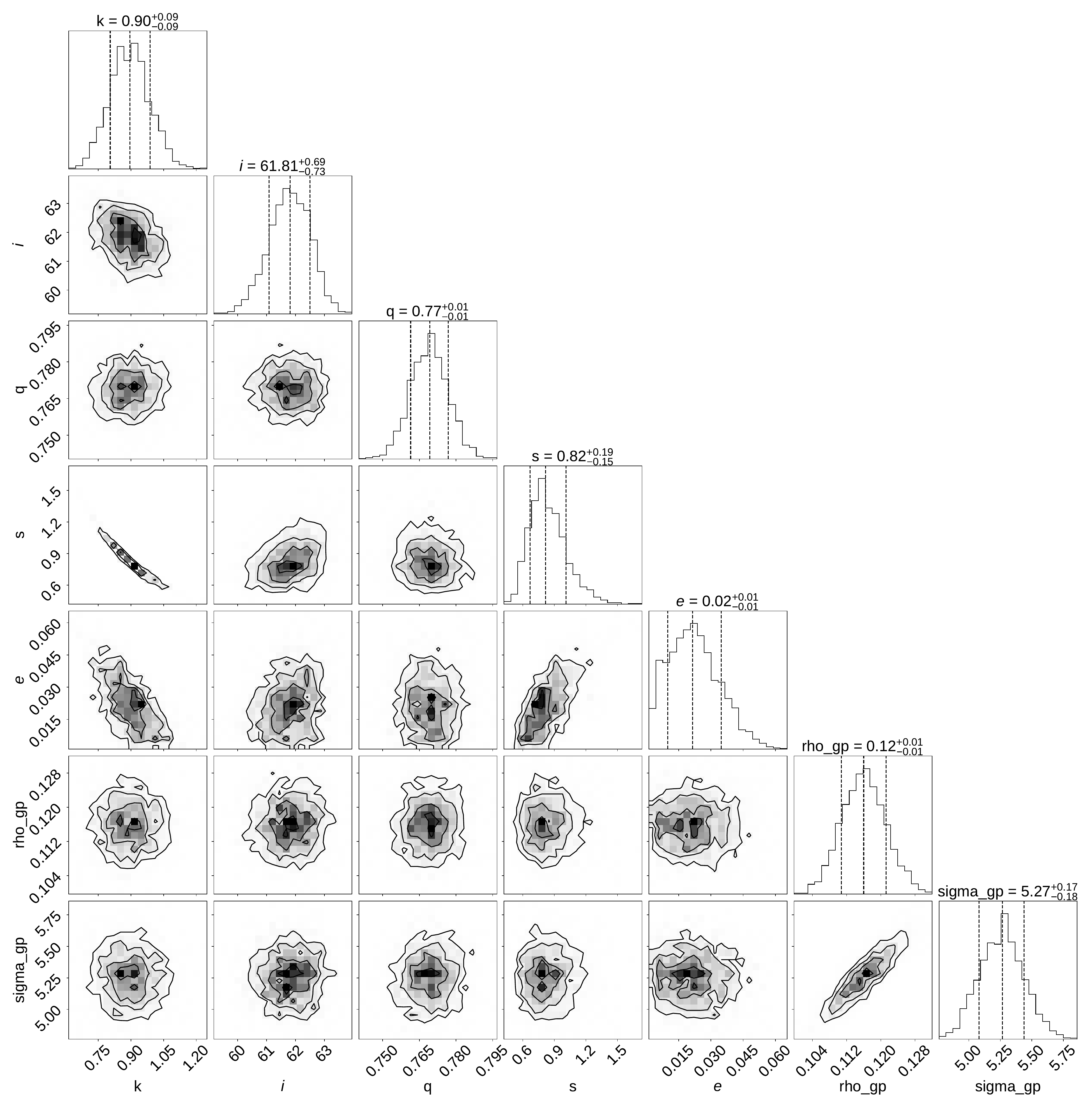}
    \caption{The resultant posterior distribution for the eclipse model and the Gaussian process model for the best fit from \texttt{exoplanet} code for V551 Aur.}
    \label{fig:phoebev551_corner}
\end{figure}

The Appendix Table.~\ref{priors} contains the prior distributions that were chosen for modeling in the \texttt{exoplanet} code. Table.~\ref{Tab:binarypars} provide the best fit parameters for binary modeling for all stars, which were compared to previous estimates by \citet{2018chehlah}. Fig.~\ref{fig:model3}, \ref{fig:model1}, \ref{fig:model} depict the best-fit LCs. Fig.~\ref{fig:eclips1}, \ref{fig:eclips3}, \ref{fig:eclips}, and Fig.~\ref{fig:phoebev551_corner}, \ref{fig:phoebecornerv549}, \ref{fig:phoebecorner174} show the Gaussian process fits and corner plots, respectively.

\begin{table}
\centering
\begin{minipage}{\linewidth}
%\begin{minipage}{150mm}
\caption{The parameters computed with \texttt{exoplanet} code for ZV3, V4 and V6 are given in the table. $T_{0}$ is the time of primary minima, $q$ is the mass ratio, $i$ is the inclination, $k$ is the radius ratio, $e$ is eccentricity and $s$ is the surface brightness ratio. The last two terms are the Gaussian process parameters, $\rho_{gp}$ represents the undamped period of the oscillator and $\sigma_{gp}$ the standard deviation of the process.}
\label{Tab:binarypars}
\end{minipage}
\bigskip
\resizebox{\columnwidth}{!}{ 
\begin{tabular}{cccc}
\hline
\hline
\\
\textbf{Star}& \textbf{Parameter}  &\textbf{This work} & \textbf{Chehlaeh et al.}\\
&  & & \textbf{(2018)}\\
\hline
\\

&Period (days)   & 3.224040 (fixed) &- \\ 
&$T_{0}$ (BJD)   & 2458818.0594 (fixed)& - \\ 
%a & 11$\pm$4 \\ 
&$q$ & 0.480(5) & -\\ 
&$i$ (deg) & 82.10(9)& - \\ 
ZV3&k & 1.0(2) & -\\ 
&$e$ & 0.04(4) & -\\ 
&$s$ & 0.20(9)& - \\ 
&$\rho_{gp}$ & 1.0(2)& - \\ \smallskip
&$\sigma_{gp}$ & 1.7(2) & -\\ 
\hline 
\\
 & Period (days) & 3.300284 (fixed) &3.300281(1)\\ 
&$T_{0}$ (BJD) & 2452308.388 (fixed)  & -\\ 
% a & 10$\pm$4 & - \\ 
&$q$ & 0.626(6) &0.626(7)\\ 
&$i$ (deg) & 80.4(2) & 86.61(6)\\ 
V4&$k$ & 0.6(2) & -\\ 
&$e$ & 0.13(5)  & 0 (fixed)\\ 
&$s$ & 2(1)  & -\\ 
&$\rho_{gp}$ & 0.23(2) & - \\ \smallskip 
&$\sigma_{gp}$ & 4.9(2) & -\\  
\hline
\\
 &Period (days) & 1.1731767 (fixed) & 1.1731752(8)\\ 
&$T_{0}$ (BJD) & 2452307.410 (fixed) & - \\ 
%$a$ ($R_{earth}$)& 6$\pm$2 & - \\ 
&$q$ & 0.769(8) & 0.769(5)\\ 
&$i$ (deg) & 61.8(7) & 73.01(6)\\ 
V6&$k$ & 0.90(9)& - \\ 
&$e$ & 0.02(1) & 0 (fixed)\\ 
&$s$ & 0.8(2)& - \\
&$\rho_{gp}$ & 0.116(5) &-\\ 
&$\sigma_{gp}$ & 5.3(2) &-\\ 

 \\
\hline
\end{tabular}
}
\end{table}

\subsection{V551 Aur : A pulsating eclipsing binary}

V551 Aur is an EB that exhibits pulsational variability in its LC. Previous authors proposed that the variability could be due to pulsations in the binary system's primary component. However, the nature and origin of this star's pulsation remain poorly understood. The associated frequency, 7.7248(5), is very close to the $9^{th}$ harmonic of the orbital frequency, so the pulsation may be a higher order resonance with the orbital frequency. This could indicate a tidally excited $g$-mode on the primary star. The companions' nature is unknown because no spectroscopic observations have been made.

\begin{figure}
    \includegraphics[width=\columnwidth]{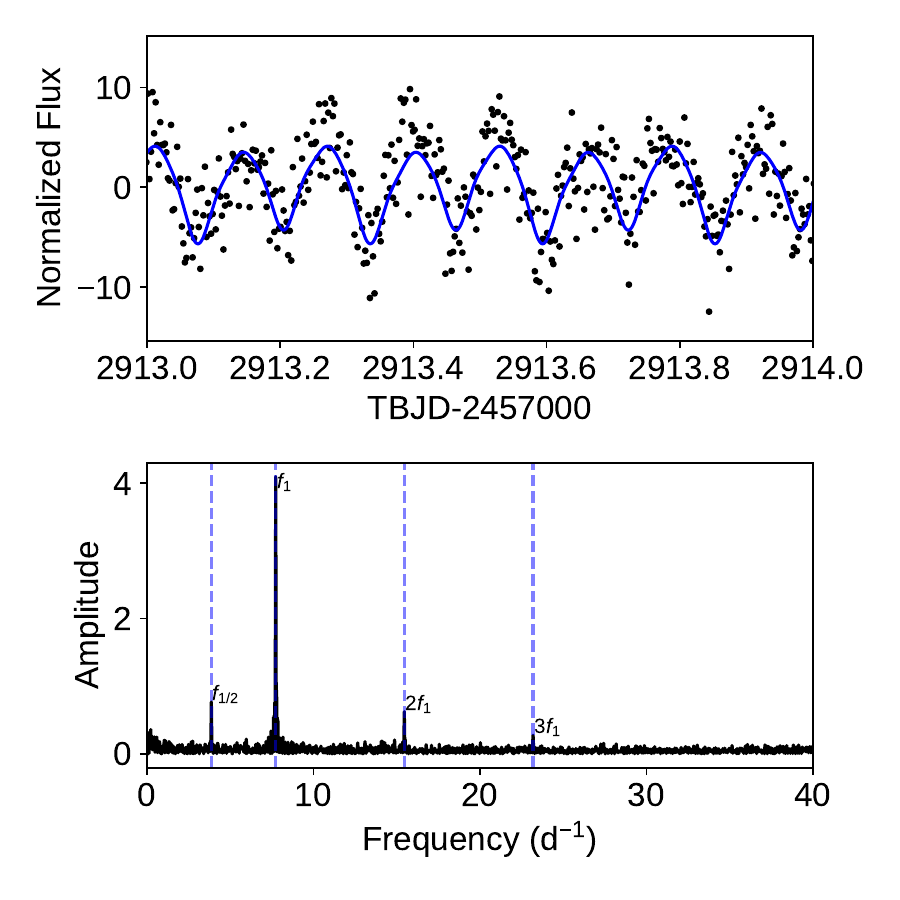}
\caption{The top panel is the residuals after removing the binary model from the LC of V551\,Aur. Here only the pulsation signals is seen. The bottom panel shows the frequency spectrum of the residual. Apart from the dominant peak being the pulsation frequency, one can notice the first two harmonics and a sub-harmonics in the spectrum.}
    \label{V6_sub}
\end{figure}

\begin{figure}
    \includegraphics[width=\columnwidth]{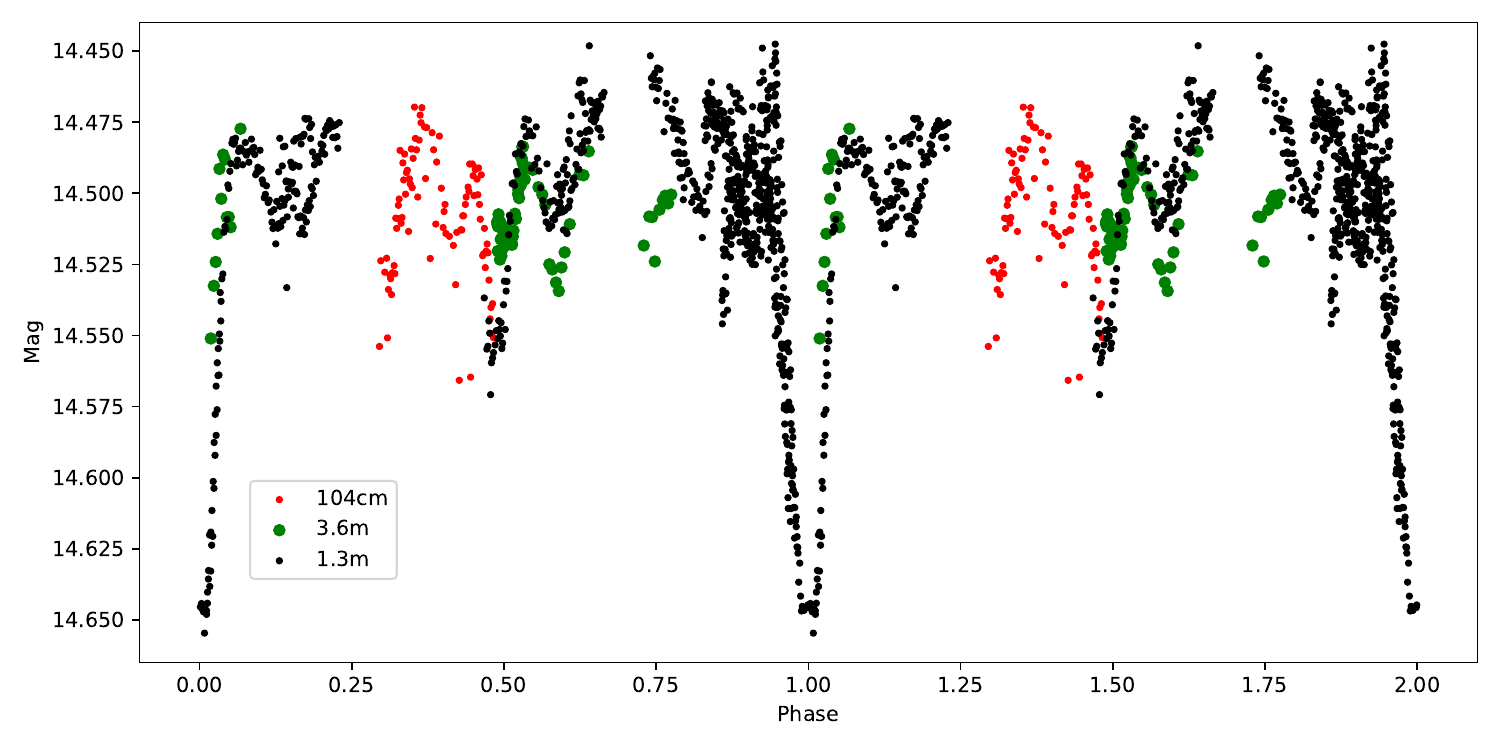}
    \caption{Ground-based phase folded LC of V551\,Aur using the orbital period determined from TESS. The red, green and black data points represents the data taken with 104 cm Sampurnanand, 3.6 m DOT and 1.3 m DFOT telescopes respectively at ARIES, Nainital.}
    \label{fig:V5_comb}
\end{figure}

As described in the previous section, we modelled both eclipses and intrinsic variability simultaneously. The pulsational frequency, first harmonics, second harmonics, and sub-harmonics were identified in the residuals of the eclipse model subtracted LC (refer to Fig.~\ref{V6_sub}). The presence of a subharmonic frequency in the pulsation indicates that the pulsations are non-linear, which is not expected from a pure stable oscillator. This lends credence to the theory that the pulsation in one of the companions is caused by tidal forces. The ground-based phased LC (Fig.~\ref{fig:V5_comb}) shows that during the secondary eclipse, pulsations dominate, indicating that the pulsating member is the primary companion.

\subsection{Spectroscopy}\label{spectro}

\begin{figure}
    \centering
    \includegraphics[scale=0.6]{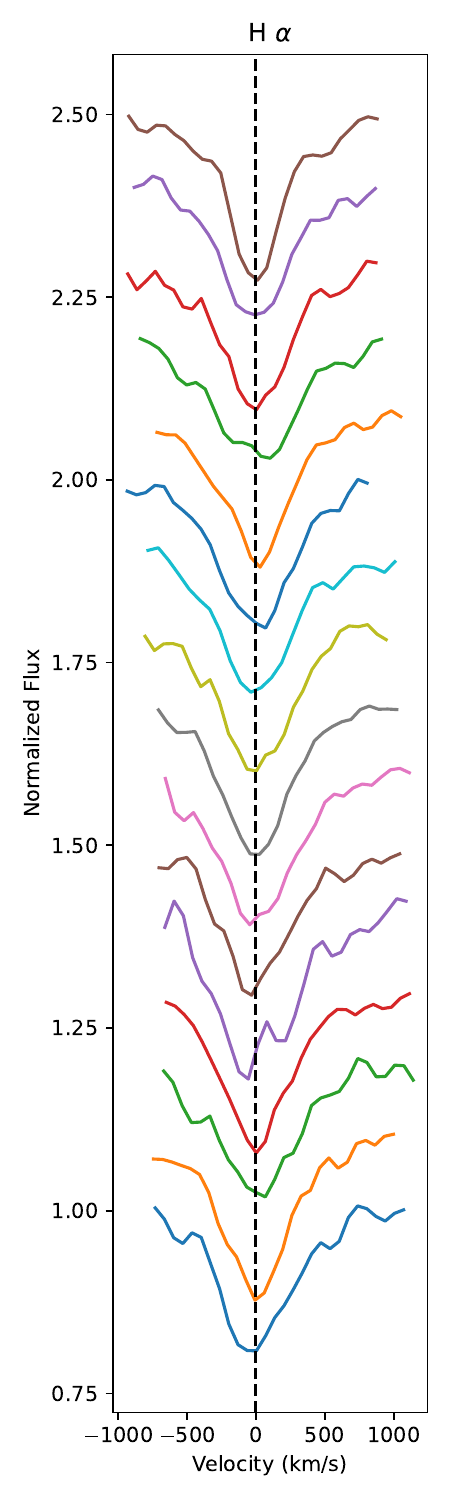}
    \caption{The H$\alpha$ line of low-resolution spectra at different phases obtained with HFOSC equipped on HCT. The normalized flux for each spectra was added by 0.1 for visual clarity. The phase increases from 0 to 1 from bottom to top.}
    \label{Halphalow}
\end{figure}

Low-resolution spectroscopic observations for V551 Aur were carried out with Himalayan Chandra telescope, Hanle Faint Object Spectrograph Camera (HFOSC) \citep{2002hfosc}, during 2021 and 2022. A total of 16 spectra with a resolution of R $\sim$ 1200 were obtained during different phases of the orbital period. The radial velocities proved to be unreliable for our study because of the low-resolution and due to the fact that the components' profiles are always blended and 
furthermore affected by the pulsation. However, close examination of the strong Balmer lines shows significant line profile variations across different phases (see Fig.~\ref{Halphalow} for the H$\alpha$ lines) indicating a spectroscopic binary. A series of high-resolution spectra is required to construct the star's radial velocity curves, which will provide accurate system parameters. Because we do not have direct access to the spectroscopic observational facilities required to observe a $14^{th}$ magnitude star, we propose this as a future project. We used the low-resolution spectra to determine the basic parameters of the star by comparison with synthetic spectra (Table.~\ref{Tab:teff}), using the chi-square minimization technique employed in \texttt{GSSP} code \citep{2015GSSP}. 

\begin{figure}
    \includegraphics[width=\columnwidth]{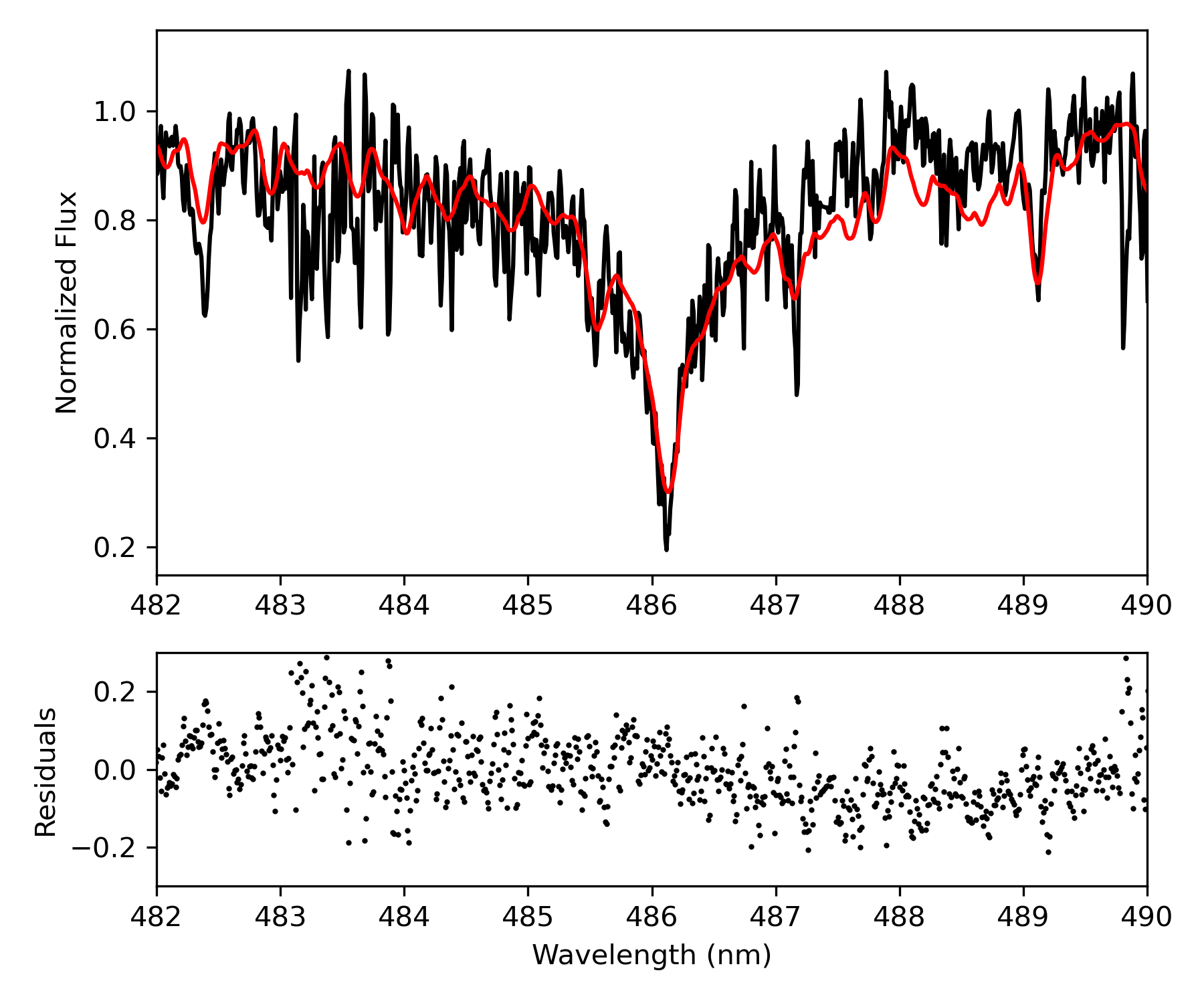}
    \caption{The top panel shows the H$\beta$ line (black) of the medium-resolution spectrum acquired for V551\,Aur at an orbital phase 0.96 and the best fit synthetic spectrum (red). The bottom panel shows the residuals of the fit.}
    \label{fig:V5_hbeta}
\end{figure}

\begin{figure}
    \includegraphics[width=\columnwidth]{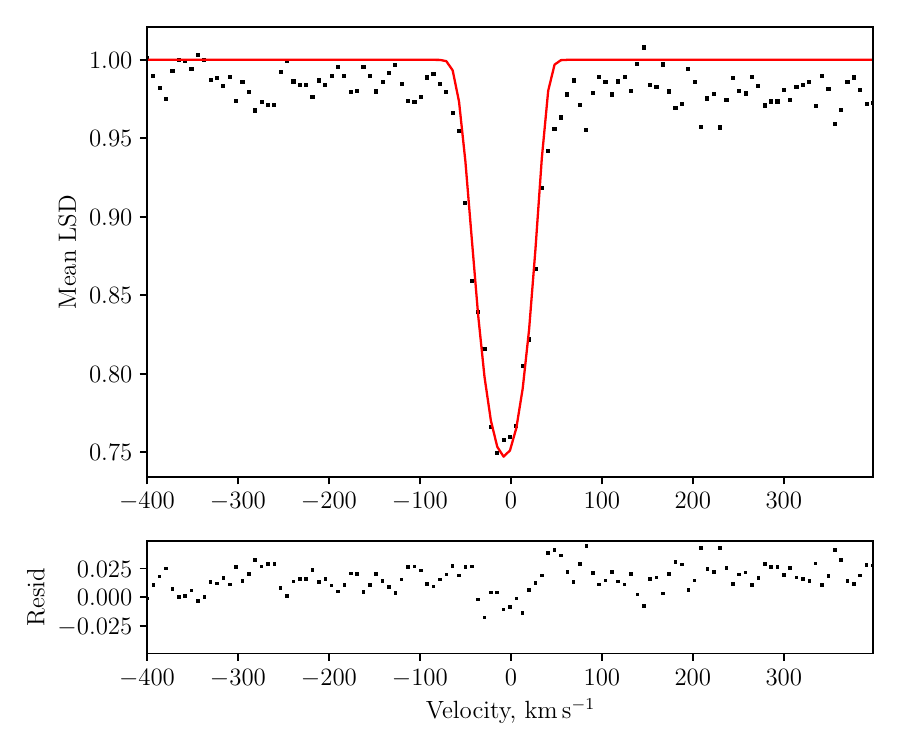}
    \caption{Top panel shows the mean LSD profile of the medium-resolution spectra (black) for V551\,Aur. A rotationally broadened line profile (red) was fit to this data to find vsini and radial velocity. The bottom panel gives the residuals of the fit.}
    \label{fig:lsd}
\end{figure}

A medium-resolution spectrum was taken with the 2.4-m Telescope, Thailand National Observatory (TNO), Thailand, Medium-Resolution Spectrograph with a resolution of R$\sim$18,000 over a range of 380-900 nm. The H$\beta$ line (Fig.~\ref{fig:V5_hbeta}) was used to calculate the \teff\, and\,\logg\, by comparison with synthetic spectra using the \texttt{GSSP} code. Due to binarity and poor resolution, the metallicity estimation is unreliable and cannot be compared to the cluster metallicity. To improve the signal-to-noise ratio of the line profile, we applied the Least Square Deconvolution (LSD) method \citep{1997Donati}. The LSD profile (Fig.~\ref{fig:lsd}) was used to calculate the \vsini\,and radial velocity of the star. The best fit results of our analysis are presented in Table.~\ref{Tab:Medteff}.

\begin{table}
\centering
\begin{minipage}{\linewidth}
%\begin{minipage}{150mm}
\caption{The table shows the spectroscopic parameters that were found for V551 Aur using the medium-resolution spectrum. The parameters were found by fitting synthetic spectra to the H$\beta$ line and line profile fitting to the LSD profile.}
\label{Tab:Medteff}
\end{minipage}
\bigskip
\begin{tabular}{ccc}
\hline
\hline
\\
\textbf{Parameters}  &\textbf{Hbeta} &\textbf{LSD} \\
\hline
 \\
%$[M/H]$ & 0.68 $\pm$ 0.18& -\\ 
\teff & 6458 $\pm$ 361 K&-\\ 
\logg & 3.0 $\pm$ 1.49 &-\\

RV& -& -8.4 $\pm$ 1.2 km/s \\
\vsini & -& 44.0 $\pm$ 2.5 km/s  \\
\\
\hline
\end{tabular}
\end{table}

\section{SUMMARY}\label{results}

Our ground-based observations revealed 25 member variables and 85 field variables within a 9 arcmin radius field of NGC\,2126. Our research was focused on member variables and stars with promising LCs in the TESS data. We discovered 18 new variable cluster members from the ground.

In TESS, we identified counterparts for 11 known variables, including six members and five field stars. Furthermore, we identified a new field star as a rotational variable based on TESS LC. TESS frequency spectra show distinctive pulsation peaks, especially for $\delta$ Scuti stars. These frequencies can be used to conduct asteroseismic studies on member pulsating stars. Aside from that, the cluster has a higher metallicity than the solar metallicity, implying the presence of chemically peculiar stars in the cluster. Thus, a spectroscopic investigation can benefit our understanding of the member variables.

We attempted to solve the TESS contamination problem by using custom apertures for our sources and cross-matching detected frequencies to ground-based frequencies. This method was useful for stars that produced strong signals in TESS.

We used TESS data to model the EBs V4, V6, and ZV3 by combining their TESS LCs. However, this model still lacks the support of spectroscopic radial velocity measurements. Due to a lack of data, ZV3 had not previously been modeled or its orbital period determined. We utilized the continuous LCs from TESS data to model this star, thereby establishing the linear ephemeris and determining the orbital period.

The intrinsic variability and eclipses of the pulsating EB V551 Aur were modeled simultaneously to determine its parameters. This resulted in slightly different parameter estimates than those reported in previous literature. The low-resolution spectroscopy confirms the spectral line profile variation for this system, indicating that it is a spectroscopic binary. The medium-resolution spectra provided better estimates of spectroscopic parameters for V551\,Aur.

\section{CONCLUSIONS}\label{conclusion}

This project was initiated to search and study pulsating variables in the open star cluster NGC 2126, utilizing ground-based observations, Gaia data and TESS FFIs for improved analysis. This study led us to draw the following conclusions :

\begin{itemize}
    \item Our study demonstrates that TESS LCs alone are unreliable in the case of open clusters due to contamination. To address this issue and fully utilize the TESS data, one must integrate Gaia data and ground-based observations. This allows us to identify contaminated sources and make better use of the frequency resolution provided by TESS.

    \item V551\,Aur, which was known to exhibit pulsations in one of its components, was found to be a spectroscopic binary with membership to the cluster. To understand this system and model it accurately, high-resolution spectroscopic follow-up observations covering this star's orbital phase is required. Due to the star's faintness, we intend to use observations with a larger aperture telescope.
    
\end{itemize}

From isochrone fitting, we determined log age $\sim$9.2, a metallicity $\sim$ 0.016, and a distance of $\sim$1.2 kpc for the cluster. Our results are in agreement with previous determinations from \citet{2018chehlah}, but with slight differences (log age $\sim$ 9.1, Z $\sim$ 0.019, d $\sim$ 1.4 kpc in \citet{2018chehlah}), possibly due to the improved magnitude, color, proper motion and parallax from Gaia.

Our observations of NGC\,2126 are limited to a few nights. More observations are demanding for a detailed frequency analysis of the detected targets. Therefore, we plan to conduct further observations for the cluster using the observing facilities at ARIES, Nainital.

\section*{Acknowledgements}

We are grateful to the Indian and Belgian funding agencies DST (DST/INT/Belg/P-09/2017) and BELSPO (BL/33/IN12) for providing financial support to carry out this work. AD acknowledges the financial support received from DST-INSPIRE Fellowship Programme (DST/INSPIRE Fellowship/2020/IF200245). 
%%%%%%%%%%%%%%%%%%%%%%%%%%%%%%%%%%%%%%%%%%%%%%%%%%
\section*{Data Availability}

The data underlying in this article will be shared upon request to the corresponding author. This paper includes data collected by the TESS and SIMBAD databases operated by the NASA Explorer Program and CDS, Strasbourg, France, respectively. This work has made use of data from the European Space Agency (ESA) mission {\it Gaia}(\url{https://www.cosmos.esa.int/gaia}), processed by the {\it Gaia} Data Processing and Analysis Consortium (DPAC, \url{https://www.cosmos.esa.int/web/gaia/dpac/consortium}). Funding for the DPAC has been provided by national institutions, in particular the institutions participating in the {\it Gaia} Multilateral Agreement. This paper includes data collected by the TESS mission, which are publicly available from the Mikulski Archive for Space Telescopes (MAST).

%%%%%%%%%%%%%%%%%%%% REFERENCES %%%%%%%%%%%%%%%%%%

% The best way to enter references is to use BibTeX:

\bibliographystyle{mnras}
\bibliography{example} % if your bibtex file is called example.bib

\begin{thebibliography}{}
\makeatletter
\relax
\def\mn@urlcharsother{\let\do\@makeother \do\$\do\&\do\#\do\^\do\_\do\%\do\~}
\def\mn@doi{\begingroup\mn@urlcharsother \@ifnextchar [ {\mn@doi@}
  {\mn@doi@[]}}
\def\mn@doi@[#1]#2{\def\@tempa{#1}\ifx\@tempa\@empty \href
  {http://dx.doi.org/#2} {doi:#2}\else \href {http://dx.doi.org/#2} {#1}\fi
  \endgroup}
\def\mn@eprint#1#2{\mn@eprint@#1:#2::\@nil}
\def\mn@eprint@arXiv#1{\href {http://arxiv.org/abs/#1} {{\tt arXiv:#1}}}
\def\mn@eprint@dblp#1{\href {http://dblp.uni-trier.de/rec/bibtex/#1.xml}
  {dblp:#1}}
\def\mn@eprint@#1:#2:#3:#4\@nil{\def\@tempa {#1}\def\@tempb {#2}\def\@tempc
  {#3}\ifx \@tempc \@empty \let \@tempc \@tempb \let \@tempb \@tempa \fi \ifx
  \@tempb \@empty \def\@tempb {arXiv}\fi \@ifundefined
  {mn@eprint@\@tempb}{\@tempb:\@tempc}{\expandafter \expandafter \csname
  mn@eprint@\@tempb\endcsname \expandafter{\@tempc}}}

\bibitem[\protect\citeauthoryear{{Aerts}}{{Aerts}}{2021}]{2021Aerts}
{Aerts} C.,  2021, \mn@doi [Reviews of Modern Physics]
  {10.1103/RevModPhys.93.015001}, \href
  {https://ui.adsabs.harvard.edu/abs/2021RvMP...93a5001A} {93, 015001}

\bibitem[\protect\citeauthoryear{{Aerts}, {Christensen-Dalsgaard}  \&
  {Kurtz}}{{Aerts} et~al.}{2010}]{2010astero}
{Aerts} C.,  {Christensen-Dalsgaard} J.,   {Kurtz} D.~W.,  2010,
  {Asteroseismology}.
Berlin: Springer, \mn@doi{10.1007/978-1-4020-5803-5}

\bibitem[\protect\citeauthoryear{{Ashoka} et~al.,}{{Ashoka}
  et~al.}{2000}]{2000ashoka}
{Ashoka} B.~N.,  et~al., 2000, Bulletin of the Astronomical Society of India,
  \href {https://ui.adsabs.harvard.edu/abs/2000BASI...28..251A} {28, 251}

\bibitem[\protect\citeauthoryear{{Balona}}{{Balona}}{2013}]{2013balona}
{Balona} L.~A.,  2013, \mn@doi [\mnras] {10.1093/mnras/stt322}, \href
  {https://ui.adsabs.harvard.edu/abs/2013MNRAS.431.2240B} {431, 2240}

\bibitem[\protect\citeauthoryear{{Bedding} et~al.,}{{Bedding}
  et~al.}{2023}]{2023ApJ...946L..10B}
{Bedding} T.~R.,  et~al., 2023, \mn@doi [\apjl] {10.3847/2041-8213/acc17a},
  \href {https://ui.adsabs.harvard.edu/abs/2023ApJ...946L..10B} {946, L10}

\bibitem[\protect\citeauthoryear{{Brasseur}, {Phillip}, {Fleming}, {Mullally}
  \& {White}}{{Brasseur} et~al.}{2019}]{2019brasseur}
{Brasseur} C.~E.,  {Phillip} C.,  {Fleming} S.~W.,  {Mullally} S.~E.,   {White}
  R.~L.,  2019, {Astrocut: Tools for creating cutouts of TESS images},
  Astrophysics Source Code Library, record ascl:1905.007

\bibitem[\protect\citeauthoryear{{Bressan}, {Marigo}, {Girardi}, {Salasnich},
  {Dal Cero}, {Rubele}  \& {Nanni}}{{Bressan} et~al.}{2012}]{2012Bressan}
{Bressan} A.,  {Marigo} P.,  {Girardi} L.,  {Salasnich} B.,  {Dal Cero} C.,
  {Rubele} S.,   {Nanni} A.,  2012, \mn@doi [\mnras]
  {10.1111/j.1365-2966.2012.21948.x}, \href
  {https://ui.adsabs.harvard.edu/abs/2012MNRAS.427..127B} {427, 127}

\bibitem[\protect\citeauthoryear{{Charbonneau}}{{Charbonneau}}{1995}]{1995charbo}
{Charbonneau} P.,  1995, \mn@doi [\apjs] {10.1086/192242}, \href
  {https://ui.adsabs.harvard.edu/abs/1995ApJS..101..309C} {101, 309}

\bibitem[\protect\citeauthoryear{{Chehlaeh}, {Mkrtichian}, {Lampens},
  {Komonjinda}, {Kim}, {Van Cauteren}, {Kusakin}  \& {Glazunova}}{{Chehlaeh}
  et~al.}{2018}]{2018chehlah}
{Chehlaeh} N.,  {Mkrtichian} D.,  {Lampens} P.,  {Komonjinda} S.,  {Kim} S.~L.,
   {Van Cauteren} P.,  {Kusakin} A.~V.,   {Glazunova} L.,  2018, \mn@doi
  [\mnras] {10.1093/mnras/sty1906}, \href
  {https://ui.adsabs.harvard.edu/abs/2018MNRAS.480.1850C} {480, 1850}

\bibitem[\protect\citeauthoryear{{Chen}, {Girardi}, {Bressan}, {Marigo},
  {Barbieri}  \& {Kong}}{{Chen} et~al.}{2014}]{2014chen}
{Chen} Y.,  {Girardi} L.,  {Bressan} A.,  {Marigo} P.,  {Barbieri} M.,   {Kong}
  X.,  2014, \mn@doi [\mnras] {10.1093/mnras/stu1605}, \href
  {https://ui.adsabs.harvard.edu/abs/2014MNRAS.444.2525C} {444, 2525}

\bibitem[\protect\citeauthoryear{{Chen}, {Bressan}, {Girardi}, {Marigo}, {Kong}
   \& {Lanza}}{{Chen} et~al.}{2015}]{2015chen}
{Chen} Y.,  {Bressan} A.,  {Girardi} L.,  {Marigo} P.,  {Kong} X.,   {Lanza}
  A.,  2015, \mn@doi [\mnras] {10.1093/mnras/stv1281}, \href
  {https://ui.adsabs.harvard.edu/abs/2015MNRAS.452.1068C} {452, 1068}

\bibitem[\protect\citeauthoryear{{Cowsik}, {Srinivasan}  \& {Prabhu}}{{Cowsik}
  et~al.}{2002}]{2002hfosc}
{Cowsik} R.,  {Srinivasan} R.,   {Prabhu} T.,  2002, in {Vernin} J.,
  {Benkhaldoun} Z.,   {Mu{\~n}oz-Tu{\~n}{\'o}n} C.,  eds,  Astronomical Society
  of the Pacific Conference Series Vol. 266, Astronomical Site Evaluation in
  the Visible and Radio Range. p.~424

\bibitem[\protect\citeauthoryear{{Danielski}, {Babusiaux}, {Ruiz-Dern},
  {Sartoretti}  \& {Arenou}}{{Danielski} et~al.}{2018}]{2018gaiaav}
{Danielski} C.,  {Babusiaux} C.,  {Ruiz-Dern} L.,  {Sartoretti} P.,   {Arenou}
  F.,  2018, \mn@doi [\aap] {10.1051/0004-6361/201732327}, \href
  {https://ui.adsabs.harvard.edu/abs/2018A&A...614A..19D} {614, A19}

\bibitem[\protect\citeauthoryear{{Deb} \& {Singh}}{{Deb} \&
  {Singh}}{2009}]{2009A&A...507.1729D}
{Deb} S.,  {Singh} H.~P.,  2009, \mn@doi [\aap] {10.1051/0004-6361/200912851},
  \href {https://ui.adsabs.harvard.edu/abs/2009A&A...507.1729D} {507, 1729}

\bibitem[\protect\citeauthoryear{{Dileep}, {Joshi}  \& {Kurtz}}{{Dileep}
  et~al.}{2024}]{2024dileep}
{Dileep} A.,  {Joshi} S.,   {Kurtz} D.~W.,  2024, \mn@doi [Bulletin de la
  Societe Royale des Sciences de Liege] {10.25518/0037-9565.11653}, \href
  {https://ui.adsabs.harvard.edu/abs/2024BSRSL..93..227D} {93, 227}

\bibitem[\protect\citeauthoryear{{Donati}, {Semel}, {Carter}, {Rees}  \&
  {Collier Cameron}}{{Donati} et~al.}{1997}]{1997Donati}
{Donati} J.~F.,  {Semel} M.,  {Carter} B.~D.,  {Rees} D.~E.,   {Collier
  Cameron} A.,  1997, \mn@doi [\mnras] {10.1093/mnras/291.4.658}, \href
  {https://ui.adsabs.harvard.edu/abs/1997MNRAS.291..658D} {291, 658}

\bibitem[\protect\citeauthoryear{{Dupret}, {Grigahc{\`e}ne}, {Garrido},
  {Gabriel}  \& {Scuflaire}}{{Dupret} et~al.}{2005}]{2005dupret}
{Dupret} M.~A.,  {Grigahc{\`e}ne} A.,  {Garrido} R.,  {Gabriel} M.,
  {Scuflaire} R.,  2005, \mn@doi [\aap] {10.1051/0004-6361:20041817}, \href
  {https://ui.adsabs.harvard.edu/abs/2005A&A...435..927D} {435, 927}

\bibitem[\protect\citeauthoryear{{Feinstein} et~al.,}{{Feinstein}
  et~al.}{2019}]{2019adina}
{Feinstein} A.~D.,  et~al., 2019, \mn@doi [\pasp] {10.1088/1538-3873/ab291c},
  \href {https://ui.adsabs.harvard.edu/abs/2019PASP..131i4502F} {131, 094502}

\bibitem[\protect\citeauthoryear{{Fonnesbeck}, {Patil}, {Huard}  \&
  {Salvatier}}{{Fonnesbeck} et~al.}{2015}]{2015Pymc}
{Fonnesbeck} C.,  {Patil} A.,  {Huard} D.,   {Salvatier} J.,  2015, {PyMC:
  Bayesian Stochastic Modelling in Python}, Astrophysics Source Code Library,
  record ascl:1506.005

\bibitem[\protect\citeauthoryear{{Foreman-Mackey}}{{Foreman-Mackey}}{2018}]{celerite2}
{Foreman-Mackey} D.,  2018, \mn@doi [Research Notes of the American
  Astronomical Society] {10.3847/2515-5172/aaaf6c}, \href
  {http://adsabs.harvard.edu/abs/2018RNAAS...2a..31F} {2, 31}

\bibitem[\protect\citeauthoryear{{Foreman-Mackey}, {Agol}, {Ambikasaran}  \&
  {Angus}}{{Foreman-Mackey} et~al.}{2017}]{celerite1}
{Foreman-Mackey} D.,  {Agol} E.,  {Ambikasaran} S.,   {Angus} R.,  2017,
  \mn@doi [\aj] {10.3847/1538-3881/aa9332}, \href
  {http://adsabs.harvard.edu/abs/2017AJ....154..220F} {154, 220}

\bibitem[\protect\citeauthoryear{{Foreman-Mackey} et~al.,}{{Foreman-Mackey}
  et~al.}{2021}]{exoplanet:joss}
{Foreman-Mackey} D.,  et~al., 2021, arXiv e-prints, \href
  {https://ui.adsabs.harvard.edu/abs/2021arXiv210501994F} {p. arXiv:2105.01994}

\bibitem[\protect\citeauthoryear{{Gaia Collaboration} et~al.,}{{Gaia
  Collaboration} et~al.}{2016}]{gaia2016}
{Gaia Collaboration} et~al., 2016, \mn@doi [A\&A]
  {10.1051/0004-6361/201629272}, 595, A1

\bibitem[\protect\citeauthoryear{{Gaia Collaboration} et~al.,}{{Gaia
  Collaboration} et~al.}{2023}]{gaiadr3}
{Gaia Collaboration} et~al., 2023, \mn@doi [\aap]
  {10.1051/0004-6361/202243940}, \href
  {https://ui.adsabs.harvard.edu/abs/2023A&A...674A...1G} {674, A1}

\bibitem[\protect\citeauthoryear{{G{\'a}sp{\'a}r} et~al.,}{{G{\'a}sp{\'a}r}
  et~al.}{2003}]{gasper2003}
{G{\'a}sp{\'a}r} A.,  et~al., 2003, \mn@doi [\aap]
  {10.1051/0004-6361:20031311}, \href
  {https://ui.adsabs.harvard.edu/abs/2003A&A...410..879G} {410, 879}

\bibitem[\protect\citeauthoryear{Han \& Brandt}{Han \& Brandt}{2023}]{Han_2023}
Han T.,  Brandt T.~D.,  2023, \mn@doi [The Astronomical Journal]
  {10.3847/1538-3881/acaaa7}, 165, 71

\bibitem[\protect\citeauthoryear{{Handler} et~al.,}{{Handler}
  et~al.}{2009}]{2009handler}
{Handler} G.,  et~al., 2009, \mn@doi [\apjl] {10.1088/0004-637X/698/1/L56},
  \href {https://ui.adsabs.harvard.edu/abs/2009ApJ...698L..56H} {698, L56}

\bibitem[\protect\citeauthoryear{{Joshi} et~al.,}{{Joshi}
  et~al.}{2003}]{joshi2003}
{Joshi} S.,  et~al., 2003, \mn@doi [Monthly Notices of the Royal Astronomical
  Society] {10.1046/j.1365-8711.2003.06823.x}, 344, 431

\bibitem[\protect\citeauthoryear{{Joshi}, {Mary}, {Martinez}, {Kurtz},
  {Girish}, {Seetha}, {Sagar}  \& {Ashoka}}{{Joshi} et~al.}{2006}]{2006joshi}
{Joshi} S.,  {Mary} D.~L.,  {Martinez} P.,  {Kurtz} D.~W.,  {Girish} V.,
  {Seetha} S.,  {Sagar} R.,   {Ashoka} B.~N.,  2006, \mn@doi [Astronomy and
  Astrophysics] {10.1051/0004-6361:20064970}, \href
  {https://ui.adsabs.harvard.edu/abs/2006A&A...455..303J} {455, 303}

\bibitem[\protect\citeauthoryear{{Joshi}, {Mary}, {Chakradhari}, {Tiwari}  \&
  {Billaud}}{{Joshi} et~al.}{2009}]{joshi2009}
{Joshi} S.,  {Mary} D.~L.,  {Chakradhari} N.~K.,  {Tiwari} S.~K.,   {Billaud}
  C.,  2009, \mn@doi [Astronomy and Astrophysics]
  {10.1051/0004-6361/200912382}, \href
  {https://ui.adsabs.harvard.edu/abs/2009A&A...507.1763J} {507, 1763}

\bibitem[\protect\citeauthoryear{{Joshi}, Ryabchikova, Kochukhov, Sachkov,
  Tiwari, Chakradhari  \& Piskunov}{{Joshi} et~al.}{2010}]{joshi2010}
{Joshi} S.,  Ryabchikova T.,  Kochukhov O.,  Sachkov M.,  Tiwari S.~K.,
  Chakradhari N.~K.,   Piskunov N.,  2010, \mn@doi [Monthly Notices of the
  Royal Astronomical Society] {10.1111/j.1365-2966.2009.15725.x}, 401, 1299

\bibitem[\protect\citeauthoryear{{Joshi} et~al.,}{{Joshi}
  et~al.}{2012}]{joshi2012}
{Joshi} S.,  et~al., 2012, \mn@doi [Monthly Notices of the Royal Astronomical
  Society] {10.1111/j.1365-2966.2012.21340.x}, 424, 2002

\bibitem[\protect\citeauthoryear{{Joshi}, {Jagirdar}  \& {Joshi}}{{Joshi}
  et~al.}{2016a}]{2016joshi}
{Joshi} Y.~C.,  {Jagirdar} R.,   {Joshi} S.,  2016a, \mn@doi [Research in
  Astronomy and Astrophysics] {10.1088/1674-4527/16/4/063}, \href
  {https://ui.adsabs.harvard.edu/abs/2016RAA....16...63J} {16, 63}

\bibitem[\protect\citeauthoryear{{Joshi} et~al.,}{{Joshi}
  et~al.}{2016b}]{joshi2016}
{Joshi} S.,  et~al., 2016b, \mn@doi [Astronomy and Astrophysics]
  {10.1051/0004-6361/201527242}, \href
  {https://ui.adsabs.harvard.edu/abs/2016A&A...590A.116J} {590, A116}

\bibitem[\protect\citeauthoryear{{Joshi}, Semenko, Moiseeva, Sharma, Joshi,
  Sachkov, Singh  \& Yerra}{{Joshi} et~al.}{2017}]{joshi2017}
{Joshi} S.,  Semenko E.,  Moiseeva A.,  Sharma K.,  Joshi Y.~C.,  Sachkov M.,
  Singh H.~P.,   Yerra B.~K.,  2017, \mn@doi [Monthly Notices of the Royal
  Astronomical Society] {10.1093/mnras/stx087}, 467, 633

\bibitem[\protect\citeauthoryear{{Joshi}, {Maurya}, {John}, {Panchal}, {Joshi}
  \& {Kumar}}{{Joshi} et~al.}{2020a}]{2020YCyoung}
{Joshi} Y.~C.,  {Maurya} J.,  {John} A.~A.,  {Panchal} A.,  {Joshi} S.,
  {Kumar} B.,  2020a, \mn@doi [\mnras] {10.1093/mnras/staa029}, \href
  {https://ui.adsabs.harvard.edu/abs/2020MNRAS.492.3602J} {492, 3602}

\bibitem[\protect\citeauthoryear{{Joshi}, {John}, {Maurya}, {Panchal}, {Kumar}
  \& {Joshi}}{{Joshi} et~al.}{2020b}]{2020YC}
{Joshi} Y.~C.,  {John} A.~A.,  {Maurya} J.,  {Panchal} A.,  {Kumar} B.,
  {Joshi} S.,  2020b, \mn@doi [\mnras] {10.1093/mnras/staa2881}, \href
  {https://ui.adsabs.harvard.edu/abs/2020MNRAS.499..618J} {499, 618}

\bibitem[\protect\citeauthoryear{{Joshi}, {Bangia}, {Jaiswar}, {Pant}, {Reddy}
  \& {Yadav}}{{Joshi} et~al.}{2022a}]{2022JAI....1140004J}
{Joshi} Y.~C.,  {Bangia} T.,  {Jaiswar} M.~K.,  {Pant} J.,  {Reddy} K.,
  {Yadav} S.,  2022a, \mn@doi [Journal of Astronomical Instrumentation]
  {10.1142/S2251171722400049}, \href
  {https://ui.adsabs.harvard.edu/abs/2022JAI....1140004J} {11, 2240004}

\bibitem[\protect\citeauthoryear{{Joshi} et~al.,}{{Joshi}
  et~al.}{2022b}]{joshi2022}
{Joshi} S.,  et~al., 2022b, \mn@doi [Monthly Notices of the Royal Astronomical
  Society] {10.1093/mnras/stab3158}, 510, 5854

\bibitem[\protect\citeauthoryear{King}{King}{1962}]{king1962structure}
King I.,  1962, The Astronomical Journal, 67, 471

\bibitem[\protect\citeauthoryear{{Kumar} et~al.,}{{Kumar}
  et~al.}{2018}]{2018DOT2}
{Kumar} B.,  et~al., 2018, Bulletin de la Societe Royale des Sciences de Liege,
  \href {https://ui.adsabs.harvard.edu/abs/2018BSRSL..87...29K} {87, 29}

\bibitem[\protect\citeauthoryear{{Lampens}}{{Lampens}}{2021}]{2021Galax...9...28L}
{Lampens} P.,  2021, \mn@doi [Galaxies] {10.3390/galaxies9020028}, \href
  {https://ui.adsabs.harvard.edu/abs/2021Galax...9...28L} {9, 28}

\bibitem[\protect\citeauthoryear{{Landolt}}{{Landolt}}{1992}]{1992landolt}
{Landolt} A.~U.,  1992, \mn@doi [\aj] {10.1086/116242}, \href
  {https://ui.adsabs.harvard.edu/abs/1992AJ....104..340L} {104, 340}

\bibitem[\protect\citeauthoryear{{Lata} et~al.,}{{Lata}
  et~al.}{2023}]{2023lata}
{Lata} S.,  et~al., 2023, \mn@doi [\mnras] {10.1093/mnras/stad013}, \href
  {https://ui.adsabs.harvard.edu/abs/2023MNRAS.520.1092L} {520, 1092}

\bibitem[\protect\citeauthoryear{{Lenz} \& {Breger}}{{Lenz} \&
  {Breger}}{2004}]{2004perio04}
{Lenz} P.,  {Breger} M.,  2004, in {Zverko} J.,  {Ziznovsky} J.,  {Adelman}
  S.~J.,   {Weiss} W.~W.,  eds,  IAU Symposium Vol. 224, The A-Star Puzzle. pp
  786--790, \mn@doi{10.1017/S1743921305009750}

\bibitem[\protect\citeauthoryear{{Li} et~al.,}{{Li}
  et~al.}{2024}]{2024A&A...686A.142L}
{Li} G.,  et~al., 2024, \mn@doi [\aap] {10.1051/0004-6361/202348901}, \href
  {https://ui.adsabs.harvard.edu/abs/2024A&A...686A.142L} {686, A142}

\bibitem[\protect\citeauthoryear{{Lightkurve Collaboration}
  et~al.,}{{Lightkurve Collaboration} et~al.}{2018}]{2018Lk}
{Lightkurve Collaboration} et~al., 2018, {Lightkurve: Kepler and TESS time
  series analysis in Python}, Astrophysics Source Code Library (\mn@eprint
  {ascl} {1812.013})

\bibitem[\protect\citeauthoryear{{Liu}, {Zhang}, {Ren}, {Deng}  \& {Luo}}{{Liu}
  et~al.}{2012}]{liu2012}
{Liu} N.,  {Zhang} X.-B.,  {Ren} A.-B.,  {Deng} L.-C.,   {Luo} Z.-Q.,  2012,
  \mn@doi [Research in Astronomy and Astrophysics]
  {10.1088/1674-4527/12/6/007}, \href
  {https://ui.adsabs.harvard.edu/abs/2012RAA....12..671L} {12, 671}

\bibitem[\protect\citeauthoryear{{Luger}, {Agol}, {Foreman-Mackey}, {Fleming},
  {Lustig-Yaeger}  \& {Deitrick}}{{Luger} et~al.}{2019}]{2019Starry}
{Luger} R.,  {Agol} E.,  {Foreman-Mackey} D.,  {Fleming} D.~P.,
  {Lustig-Yaeger} J.,   {Deitrick} R.,  2019, \mn@doi [\aj]
  {10.3847/1538-3881/aae8e5}, \href
  {https://ui.adsabs.harvard.edu/abs/2019AJ....157...64L} {157, 64}

\bibitem[\protect\citeauthoryear{{Mamajek} \& {Hillenbrand}}{{Mamajek} \&
  {Hillenbrand}}{2008}]{2008mamajek}
{Mamajek} E.~E.,  {Hillenbrand} L.~A.,  2008, \mn@doi [\apj] {10.1086/591785},
  \href {https://ui.adsabs.harvard.edu/abs/2008ApJ...687.1264M} {687, 1264}

\bibitem[\protect\citeauthoryear{{Martinez} et~al.,}{{Martinez}
  et~al.}{2001}]{2001martinez}
{Martinez} P.,  et~al., 2001, \mn@doi [Astronomy and Astrophysics]
  {10.1051/0004-6361:20010432}, \href
  {https://ui.adsabs.harvard.edu/abs/2001A&A...371.1048M} {371, 1048}

\bibitem[\protect\citeauthoryear{{Maurya}, {Joshi}, {Panchal}  \&
  {Gour}}{{Maurya} et~al.}{2023}]{2023maurya}
{Maurya} J.,  {Joshi} Y.~C.,  {Panchal} A.,   {Gour} A.~S.,  2023, \mn@doi
  [\aj] {10.3847/1538-3881/acad7e}, \href
  {https://ui.adsabs.harvard.edu/abs/2023AJ....165...90M} {165, 90}

\bibitem[\protect\citeauthoryear{{Montgomery} \& {O'Donoghue}}{{Montgomery} \&
  {O'Donoghue}}{1999}]{1999mont}
{Montgomery} M.~H.,  {O'Donoghue} D.,  1999, Delta Scuti Star Newsletter, \href
  {https://ui.adsabs.harvard.edu/abs/1999DSSN...13...28M} {13, 28}

\bibitem[\protect\citeauthoryear{Murphy}{Murphy}{2019}]{murphy_2019_2533474}
Murphy S.~J.,  2019, Pulsating stars in binary systems: a review,
  \mn@doi{10.5281/zenodo.2533474}, \url
  {https://doi.org/10.5281/zenodo.2533474}

\bibitem[\protect\citeauthoryear{{Pamos Ortega}, {Mirouh}, {Garc{\'\i}a
  Hern{\'a}ndez}, {Su{\'a}rez Yanes}  \& {Barcel{\'o} Forteza}}{{Pamos Ortega}
  et~al.}{2023}]{2023A&A...675A.167P}
{Pamos Ortega} D.,  {Mirouh} G.~M.,  {Garc{\'\i}a Hern{\'a}ndez} A.,
  {Su{\'a}rez Yanes} J.~C.,   {Barcel{\'o} Forteza} S.,  2023, \mn@doi [\aap]
  {10.1051/0004-6361/202346323}, \href
  {https://ui.adsabs.harvard.edu/abs/2023A&A...675A.167P} {675, A167}

\bibitem[\protect\citeauthoryear{{Pamyatnykh}}{{Pamyatnykh}}{2000}]{2000pamya}
{Pamyatnykh} A.~A.,  2000, in {Breger} M.,  {Montgomery} M.,  eds,
  Astronomical Society of the Pacific Conference Series Vol. 210, Delta Scuti
  and Related Stars. p.~215 (\mn@eprint {arXiv} {astro-ph/0005276}),
  \mn@doi{10.48550/arXiv.astro-ph/0005276}

\bibitem[\protect\citeauthoryear{{Panchal}, {Joshi}, {De Cat}, {Lampens},
  {Goswami}  \& {Tiwari}}{{Panchal} et~al.}{2023}]{2023panchal}
{Panchal} A.,  {Joshi} Y.~C.,  {De Cat} P.,  {Lampens} P.,  {Goswami} A.,
  {Tiwari} S.~N.,  2023, \mn@doi [\mnras] {10.1093/mnras/stad533}, \href
  {https://ui.adsabs.harvard.edu/abs/2023MNRAS.521..677P} {521, 677}

\bibitem[\protect\citeauthoryear{{Pandey}, {Yadav}, {Nanjappa}, {Yadav},
  {Reddy}, {Sahu}  \& {Srinivasan}}{{Pandey} et~al.}{2018}]{2018SBP}
{Pandey} S.~B.,  {Yadav} R. K.~S.,  {Nanjappa} N.,  {Yadav} S.,  {Reddy} B.~K.,
   {Sahu} S.,   {Srinivasan} R.,  2018, Bulletin de la Societe Royale des
  Sciences de Liege, \href
  {https://ui.adsabs.harvard.edu/abs/2018BSRSL..87...42P} {87, 42}

\bibitem[\protect\citeauthoryear{{Perren}, {V{\'a}zquez}  \& {Piatti}}{{Perren}
  et~al.}{2015}]{asteca2015}
{Perren} G.~I.,  {V{\'a}zquez} R.~A.,   {Piatti} A.~E.,  2015, \mn@doi [\aap]
  {10.1051/0004-6361/201424946}, \href
  {https://ui.adsabs.harvard.edu/abs/2015A&A...576A...6P} {576, A6}

\bibitem[\protect\citeauthoryear{{Pigulski}}{{Pigulski}}{2006}]{pigulski2006}
{Pigulski} A.,  2006, in {Aerts} C.,  {Sterken} C.,  eds,  Astronomical Society
  of the Pacific Conference Series Vol. 349, Astrophysics of Variable Stars.
  p.~137

\bibitem[\protect\citeauthoryear{{Poznanski}, {Prochaska}  \&
  {Bloom}}{{Poznanski} et~al.}{2012}]{2012ebv}
{Poznanski} D.,  {Prochaska} J.~X.,   {Bloom} J.~S.,  2012, \mn@doi [\mnras]
  {10.1111/j.1365-2966.2012.21796.x}, \href
  {https://ui.adsabs.harvard.edu/abs/2012MNRAS.426.1465P} {426, 1465}

\bibitem[\protect\citeauthoryear{{Pr{\v{s}}a} \& {Zwitter}}{{Pr{\v{s}}a} \&
  {Zwitter}}{2005}]{2005phoebe}
{Pr{\v{s}}a} A.,  {Zwitter} T.,  2005, \mn@doi [\apj] {10.1086/430591}, \href
  {https://ui.adsabs.harvard.edu/abs/2005ApJ...628..426P} {628, 426}

\bibitem[\protect\citeauthoryear{{Ricker}}{{Ricker}}{2014}]{2014JAVSO..42..234R}
{Ricker} G.~R.,  2014, The Journal of the American Association of Variable Star
  Observers, \href {https://ui.adsabs.harvard.edu/abs/2014JAVSO..42..234R} {42,
  234}

\bibitem[\protect\citeauthoryear{Sarkar et~al.,}{Sarkar
  et~al.}{2024}]{2024mrinmoy}
Sarkar M.,  et~al., 2024, \mn@doi [Monthly Notices of the Royal Astronomical
  Society] {10.1093/mnras/stae2258}, p. stae2258

\bibitem[\protect\citeauthoryear{{Stetson}}{{Stetson}}{1987}]{1987PASP...99..191S}
{Stetson} P.~B.,  1987, \mn@doi [\pasp] {10.1086/131977}, \href
  {https://ui.adsabs.harvard.edu/abs/1987PASP...99..191S} {99, 191}

\bibitem[\protect\citeauthoryear{{Stetson}}{{Stetson}}{1992}]{Stetson1992}
{Stetson} P.~B.,  1992, in {Worrall} D.~M.,  {Biemesderfer} C.,   {Barnes} J.,
  eds,  Astronomical Society of the Pacific Conference Series Vol. 25,
  Astronomical Data Analysis Software and Systems I. p.~297

\bibitem[\protect\citeauthoryear{{Tang}, {Bressan}, {Rosenfield}, {Slemer},
  {Marigo}, {Girardi}  \& {Bianchi}}{{Tang} et~al.}{2014}]{Tang2014}
{Tang} J.,  {Bressan} A.,  {Rosenfield} P.,  {Slemer} A.,  {Marigo} P.,
  {Girardi} L.,   {Bianchi} L.,  2014, \mn@doi [\mnras]
  {10.1093/mnras/stu2029}, \href
  {https://ui.adsabs.harvard.edu/abs/2014MNRAS.445.4287T} {445, 4287}

\bibitem[\protect\citeauthoryear{{Tkachenko}}{{Tkachenko}}{2015}]{2015GSSP}
{Tkachenko} A.,  2015, \mn@doi [\aap] {10.1051/0004-6361/201526513}, \href
  {https://ui.adsabs.harvard.edu/abs/2015A&A...581A.129T} {581, A129}

\bibitem[\protect\citeauthoryear{{Torres}}{{Torres}}{2010}]{2010torres}
{Torres} G.,  2010, \mn@doi [\aj] {10.1088/0004-6256/140/5/1158}, \href
  {https://ui.adsabs.harvard.edu/abs/2010AJ....140.1158T} {140, 1158}

\bibitem[\protect\citeauthoryear{{Trust}, {Jurua}, {De Cat}, {Joshi}  \&
  {Lampens}}{{Trust} et~al.}{2021}]{2021otto}
{Trust} O.,  {Jurua} E.,  {De Cat} P.,  {Joshi} S.,   {Lampens} P.,  2021,
  \mn@doi [\mnras] {10.1093/mnras/stab1149}, \href
  {https://ui.adsabs.harvard.edu/abs/2021MNRAS.504.5528T} {504, 5528}

\bibitem[\protect\citeauthoryear{{Trust}, {Mashonkina}, {Jurua}, {De Cat},
  {Tsymbal}  \& {Joshi}}{{Trust} et~al.}{2023}]{2023otto}
{Trust} O.,  {Mashonkina} L.,  {Jurua} E.,  {De Cat} P.,  {Tsymbal} V.,
  {Joshi} S.,  2023, \mn@doi [\mnras] {10.1093/mnras/stad1936}, \href
  {https://ui.adsabs.harvard.edu/abs/2023MNRAS.524.1044T} {524, 1044}

\bibitem[\protect\citeauthoryear{{Wilson}}{{Wilson}}{1994}]{1994PASP..106..921W}
{Wilson} R.~E.,  1994, \mn@doi [\pasp] {10.1086/133464}, \href
  {https://ui.adsabs.harvard.edu/abs/1994PASP..106..921W} {106, 921}

\bibitem[\protect\citeauthoryear{{Wu}, {Tian}, {Balaguer-N{\'u}{\~n}ez},
  {Jordi}, {Zhao}  \& {Guibert}}{{Wu} et~al.}{2002}]{2002Wu}
{Wu} Z.~Y.,  {Tian} K.~P.,  {Balaguer-N{\'u}{\~n}ez} L.,  {Jordi} C.,  {Zhao}
  L.,   {Guibert} J.,  2002, \mn@doi [\aap] {10.1051/0004-6361:20011474}, \href
  {https://ui.adsabs.harvard.edu/abs/2002A&A...381..464W} {381, 464}

\bibitem[\protect\citeauthoryear{{Yadav} et~al.,}{{Yadav}
  et~al.}{2022}]{2022Sampu}
{Yadav} R.~K.~S.,  et~al., 2022, \mn@doi [Journal of Astronomical
  Instrumentation] {10.1142/S2251171722400062}, \href
  {https://ui.adsabs.harvard.edu/abs/2022JAI....1140006Y} {11, 2240006}

\bibitem[\protect\citeauthoryear{{Zhang}, {Deng}  \& {Luo}}{{Zhang}
  et~al.}{2012}]{2012zhang}
{Zhang} X.~B.,  {Deng} L.~C.,   {Luo} C.~Q.,  2012, \mn@doi [\aj]
  {10.1088/0004-6256/144/5/141}, \href
  {https://ui.adsabs.harvard.edu/abs/2012AJ....144..141Z} {144, 141}

\bibitem[\protect\citeauthoryear{{Zong}, {Charpinet}, {Vauclair}, {Giammichele}
   \& {Van Grootel}}{{Zong} et~al.}{2016}]{2016Zong}
{Zong} W.,  {Charpinet} S.,  {Vauclair} G.,  {Giammichele} N.,   {Van Grootel}
  V.,  2016, \mn@doi [A\&A] {10.1051/0004-6361/201526300}, \href
  {https://ui.adsabs.harvard.edu/abs/2016A&A...585A..22Z} {585, A22}

\makeatother
\end{thebibliography}

% Alternatively you could enter them by hand, like this:
% This method is tedious and prone to error if you have lots of references
%\begin{thebibliography}{99}
%\bibitem[\protect\citeauthoryear{Author}{2012}]{Author2012}
%Author A.~N., 2013, Journal of Improbable Astronomy, 1, 1
%\bibitem[\protect\citeauthoryear{Others}{2013}]{Others2013}
%Others S., 2012, Journal of Interesting Stuff, 17, 198
%\end{thebibliography}

%%%%%%%%%%%%%%%%%%%%%%%%%%%%%%%%%%%%%%%%%%%%%%%%%%

%%%%%%%%%%%%%%%%% APPENDICES %%%%%%%%%%%%%%%%%%%%%

\appendix

\section{Field Stars}

\begin{figure}
    \includegraphics[page=1, width=\columnwidth]{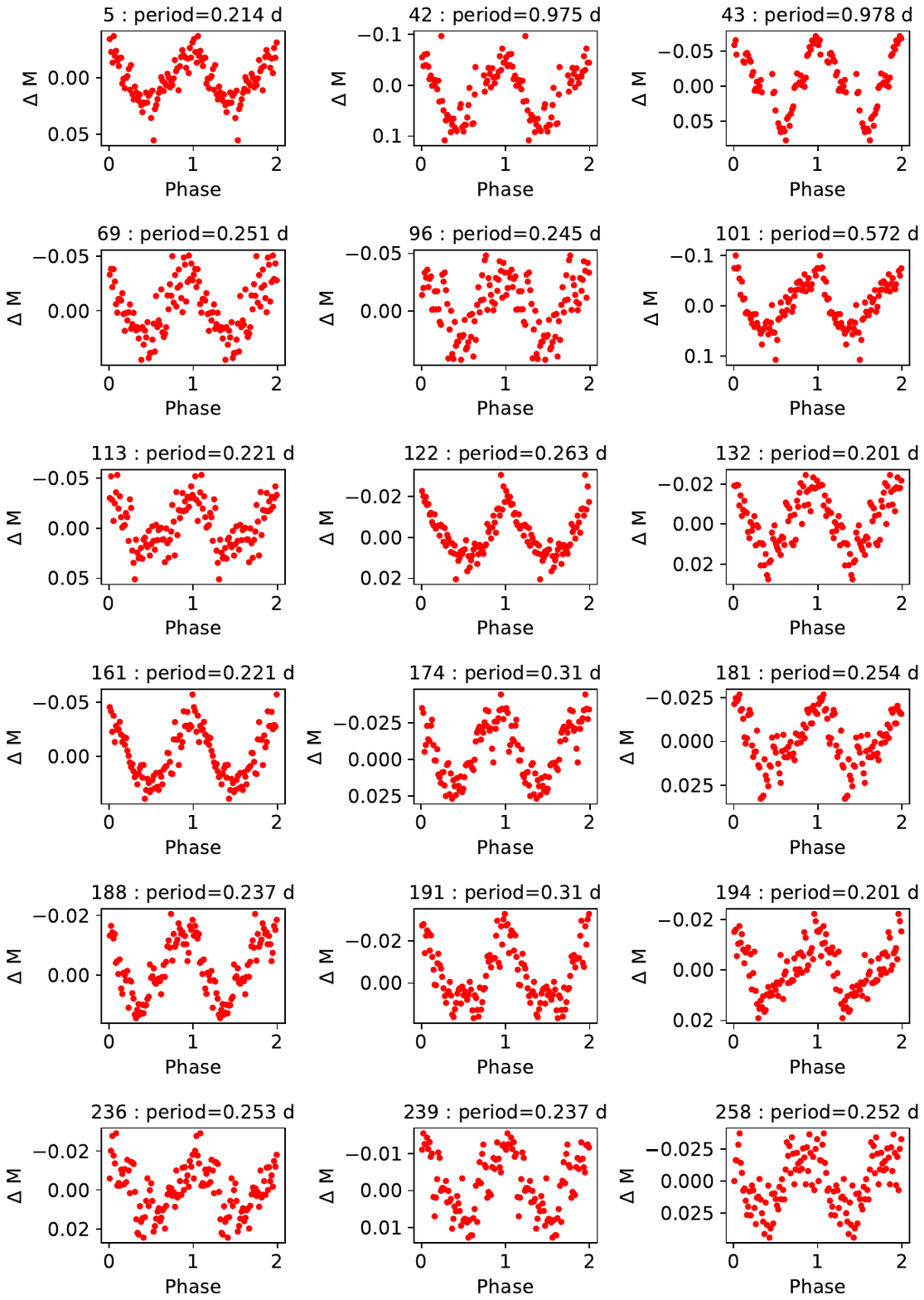}
    \caption{The phased V-band LCs of newly detected field variables in NGC\,2126. Their respective IDs and periods are given as the title of each figure.}
    \label{fig:newvars1}
\end{figure}

\begin{figure}
    \includegraphics[page=2, width=\columnwidth]{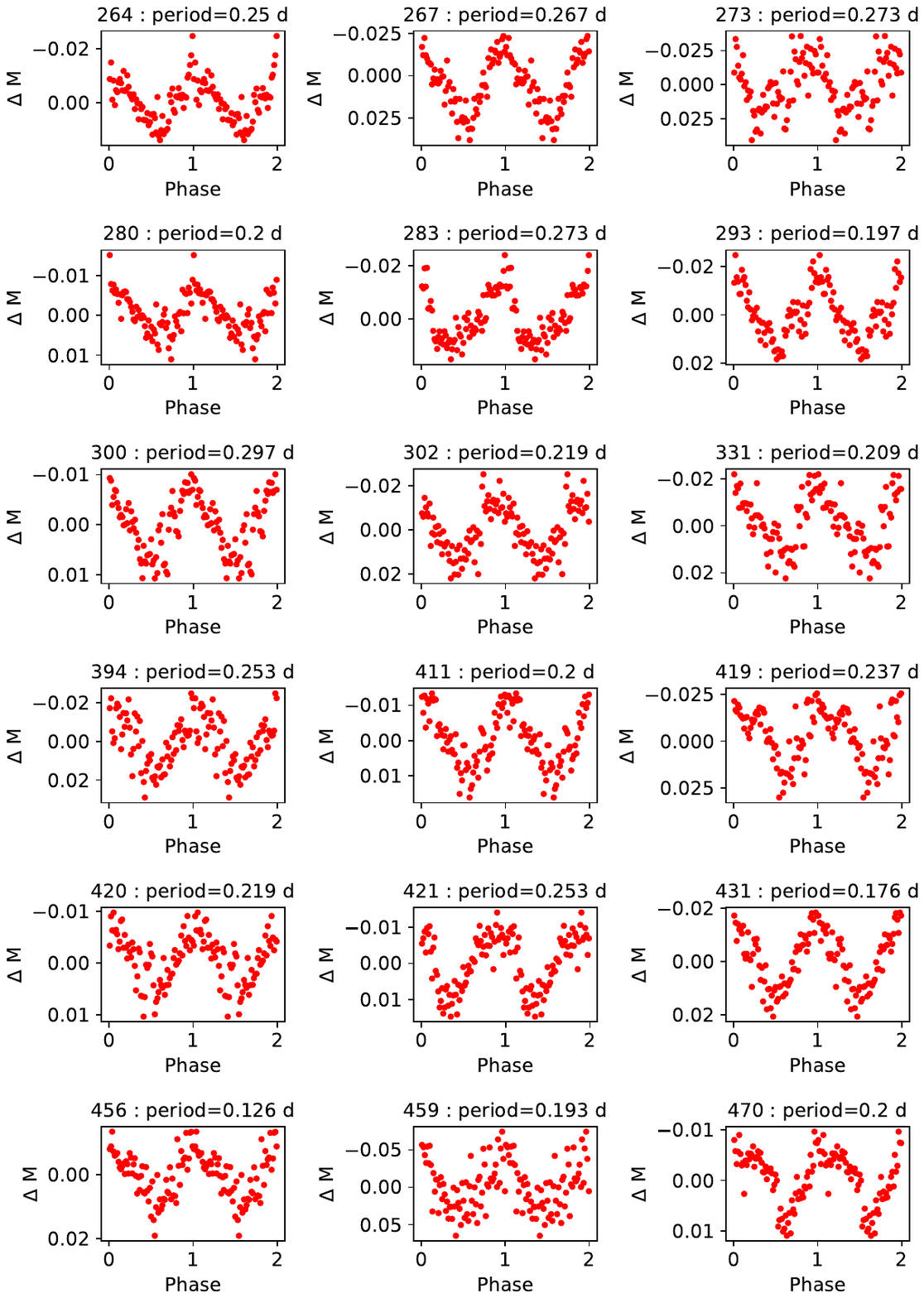}
    \caption{New variables - continued.}
    \label{fig:newvars2}
\end{figure}

\begin{figure}
    \includegraphics[page=3, width=\columnwidth]{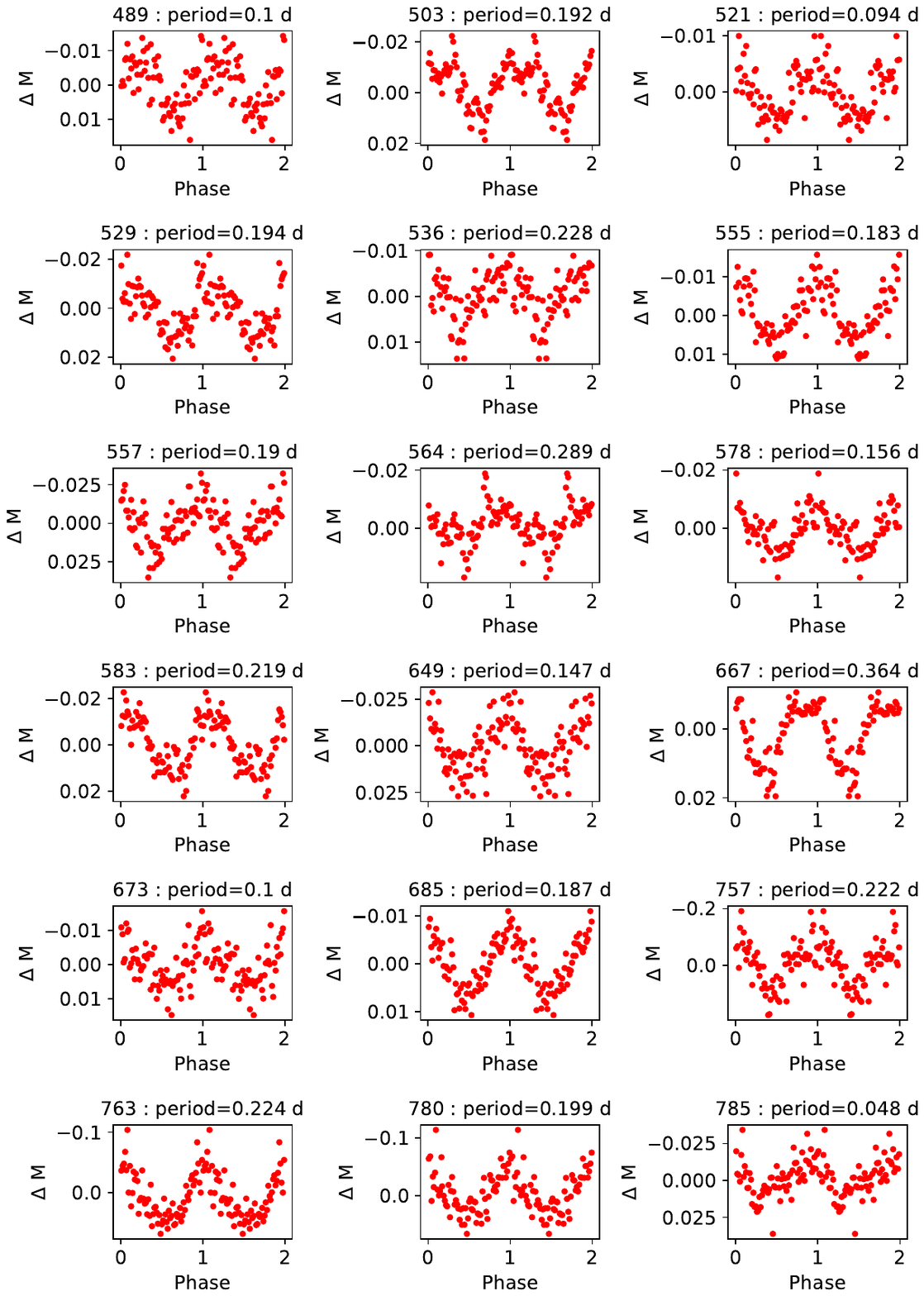}
    \caption{New variables - continued.}
    \label{fig:newvars3}
\end{figure}

\begin{figure}
    \includegraphics[page=4, width=\columnwidth]{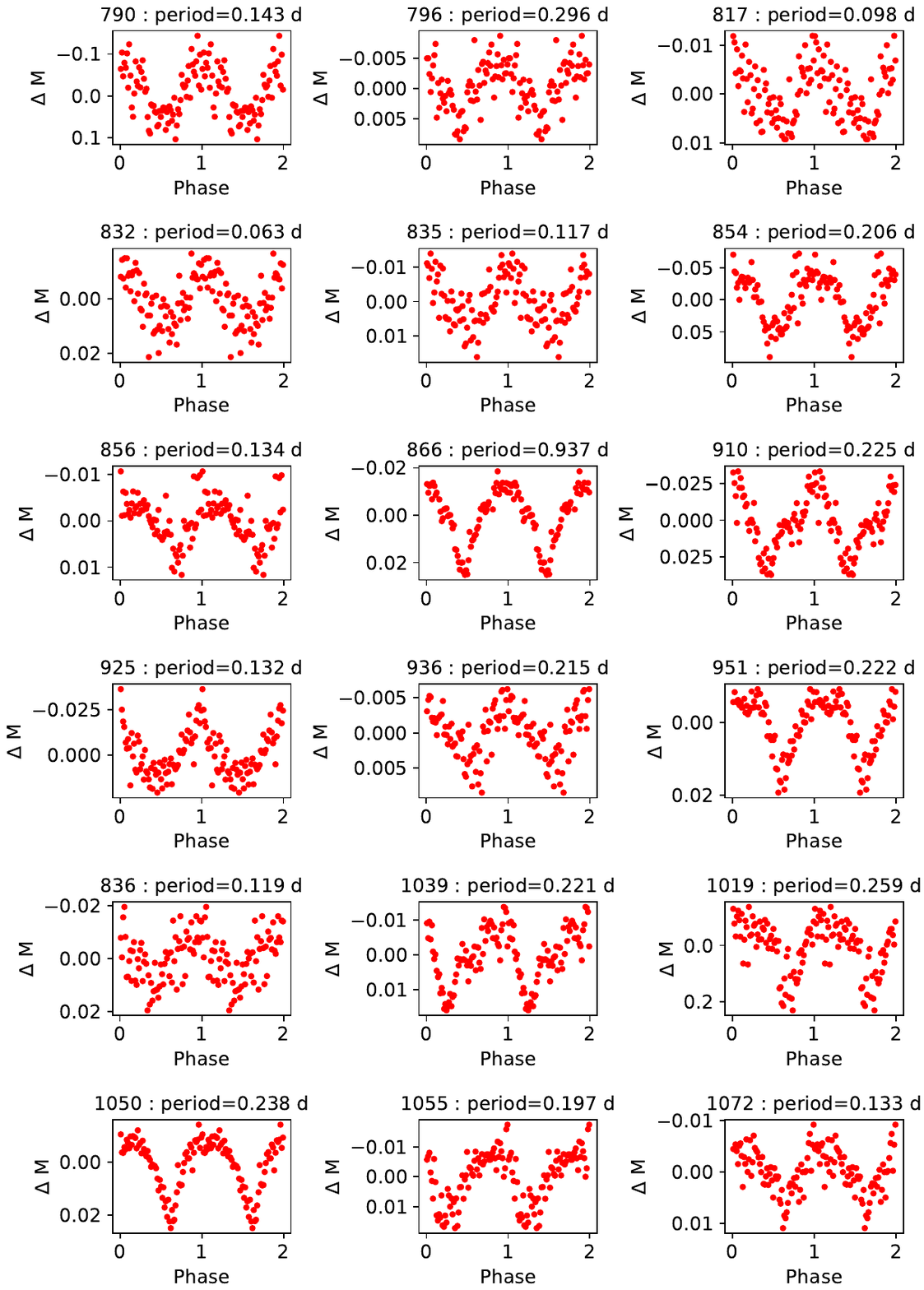}
    \caption{New variables - continued.}
    \label{fig:newvars4}
\end{figure}

\begin{figure}
    \includegraphics[page=5, width=\columnwidth]{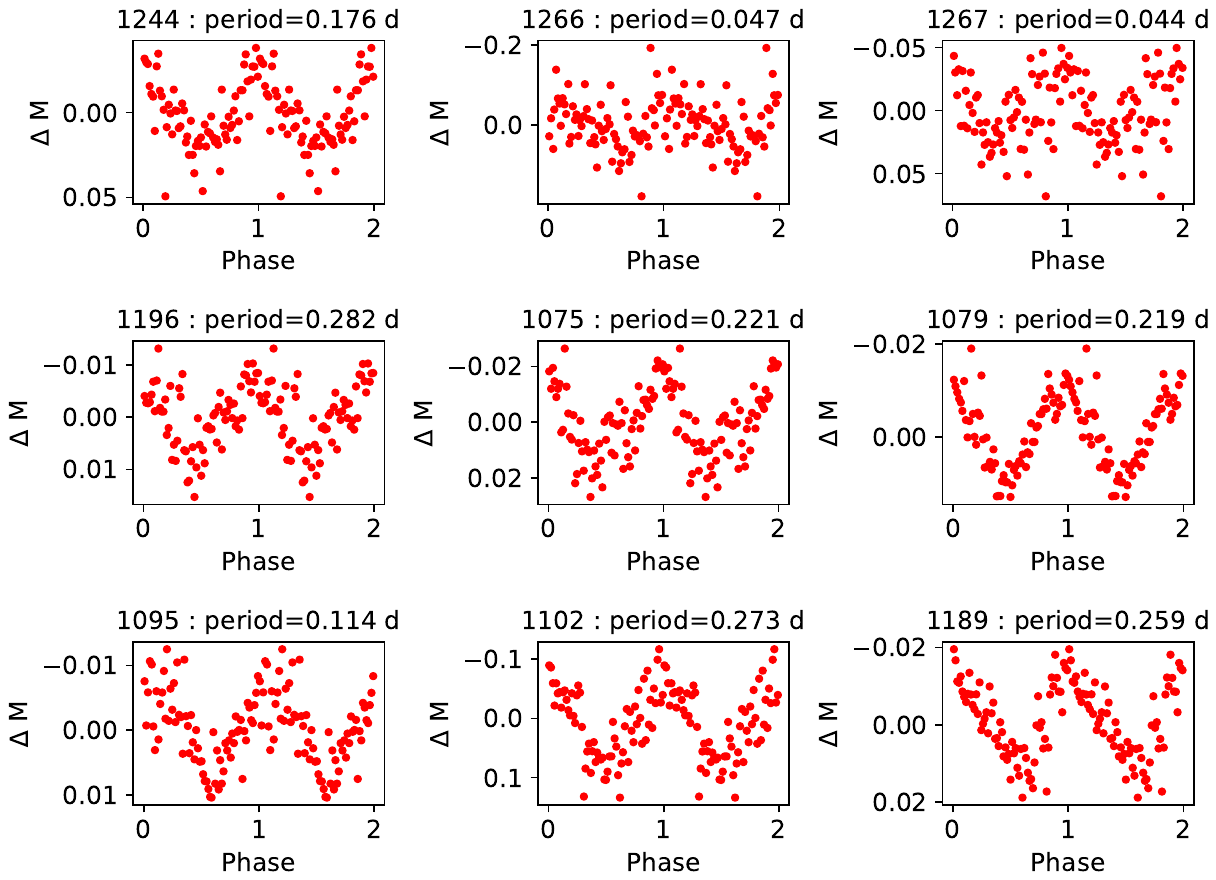}
    \caption{New variables - continued.}
    \label{fig:newvars5}
\end{figure}

\begin{table*}
\centering
\caption{The newly detected field variables from ground observations are listed with their ID, corrected magnitude, dominant frequency detected and their corresponding amplitudes.}
\begin{tabular}{c c c c c c c c c c c c}
\hline
\hline
\\
ID & RA&DEC& Mag &Frequency & Amp   &ID &RA &DEC&  Mag &Frequency & Amp   \\
 &&&  (mag) & (\cd) & (mag)  &  &  &&(mag) & (\cd) & (mag)  \\
\hline
\\
5 &90.693528  &49.732833& 17.272 & 2.340 & 0.027 &  536 & 90.857778 & 49.880417&16.201 & 2.197 & 0.010     \\
42 & 90.818583 &49.742389&18.303 & 0.513 & 0.073  & 555 & 90.452556 & 49.881361& 15.840 & 2.734 & 0.011  \\
43 & 90.720028 & 49.742278&17.485 & 0.511 & 0.063 & 557 & 90.442917 & 49.881528& 17.627 & 2.634 & 0.020  \\
69 & 90.807361 & 49.751944&18.122 & 1.992 & 0.062  & 564 & 90.689056 & 49.884472&13.588 & 1.729 & 0.011  \\ 
96 & 90.649444 & 49.760167&18.465 & 2.042 & 0.036  & 578 & 90.490917 & 49.884139& 16.378 & 3.197 & 0.010  \\
101 & 90.792972 & 49.763167&18.083 & 0.874 & 0.075  & 583 & 90.712944 & 49.886389 & 15.941 & 0.219 & 0.013 \\
113 & 90.806667 & 49.766194&18.301 & 2.265 & 0.033  & 649 & 90.758222 & 49.895194&17.827 & 3.391 & 0.030  \\
122 & 90.430194 & 49.766194&14.958 & 1.898 & 0.022  & 667 & 90.747583 & 49.896778&15.418 & 1.374 & 0.013  \\

132 & 90.841667 & 49.771611&16.458 & 2.487 & 0.018  &  673 & 90.831528 & 49.897778&16.188 & 4.981 & 0.012  \\
161 & 90.663667 & 49.781611&17.450 & 2.258 & 0.038  & 685 & 90.529556 & 49.897194&15.457 & 2.677 & 0.008  \\
174 & 90.683972 & 49.786639&16.978 & 1.615 & 0.034  & 763 & 90.857111 & 49.911806&18.298 & 2.228 & 0.120  \\
181 & 90.815611 & 49.789611&16.520 & 1.972 & 0.024  & 780 & 90.870333 & 49.913639&18.924 & 2.508 & 0.048  \\
188 & 90.691639 & 49.793500&14.827 & 2.112 & 0.018  & 785 & 90.445639 & 49.911389&17.480 & 10.318 & 0.020  \\
191 & 90.770583 & 49.794667&16.312 & 1.615 & 0.031  & 790 & 90.626972 & 49.913194&19.299 & 3.500 & 0.073  \\
194 & 90.868000 & 49.795333&15.880 & 2.487 & 0.018  & 796 & 90.487861 & 49.912889&15.630 & 1.690 & 0.006  \\
236 & 90.814278 & 49.806944&16.793 & 1.978 & 0.016  & 817 & 90.450556 & 49.915833&16.551 & 5.111 & 0.008  \\
239 & 90.580361 & 49.805833&15.727 & 2.112 & 0.011  &  832 & 90.689083 & 49.920194 &16.610 & 7.970 & 0.015  \\
258 & 90.855111 & 49.812944&18.185 & 1.985 & 0.050  & 835 & 90.456444 & 49.918833& 16.667 & 4.260 & 0.012  \\
264 & 90.769750 & 49.813583&15.541 & 1.996 & 0.011  &854 & 90.704528 & 49.925222&17.714 & 2.424 & 0.052  \\
267 & 90.624278 & 49.813194&16.557 & 1.875 & 0.030  & 856 & 90.469889 & 49.923417&15.634 & 3.722 & 0.010 \\
273 & 90.665056 & 49.814778&18.306 & 1.832 & 0.030  & 866 & 90.495306 & 49.926917& 14.498 & 0.533 & 0.019  \\
280 & 90.606861 & 49.816028&14.400 & 2.496 & 0.007  & 910 & 90.795611 & 49.939806& 16.296 & 2.225 & 0.029  \\
283 & 90.738278 & 49.816972&15.930 & 1.830 & 0.017  & 925 & 90.532194 & 49.941667& 17.325 & 3.780 & 0.020  \\
293 & 90.554944 & 49.819472&16.688 & 2.533 & 0.017  & 936 & 90.445556 & 49.944222 & 14.031 & 0.215 & 0.005 \\
300 & 90.638722 & 49.821000&14.776 & 1.685 & 0.009  & 951 & 90.835222 & 49.951389& 14.348 & 2.251 & 0.012  \\
302 & 90.700750 & 49.822222&16.736 & 2.287 & 0.020  & 836 & 90.466111 & 49.919111&17.283 & 4.217 & 0.015  \\ 
331 & 90.569389 & 49.829278&16.639 & 2.390 & 0.016  & 1039 & 90.521861 & 49.974750 &15.712 & 2.263 & 0.018  \\ 
394 & 90.775917 & 49.849889&17.145 & 1.973 & 0.017  & 1019 & 90.484417 & 49.969278&19.470 & 1.932 & 0.126  \\ 
411 & 90.622750 & 49.853278&15.926 & 2.502 & 0.013  & 1050 & 90.847750 & 49.978333&13.020 & 2.104 & 0.014  \\
419 & 90.515722 & 49.854000&16.923 & 2.112 & 0.025  &  1055 & 90.858417 & 49.980889&16.489 & 2.538 & 0.019  \\
420 & 90.546167 & 49.854917&14.862 & 2.281 & 0.007  & 1072 & 90.555861 & 49.983944&15.467 & 3.764 & 0.007  \\ 
421 & 90.842778 & 49.857056&15.732 & 1.973 & 0.011  & 1244 & 90.781139 & 50.027389&17.733 & 2.843 & 0.033  \\
431 & 90.828194 & 49.858472&15.247 & 2.840 & 0.018  &  1266 & 90.614139 & 50.031250&19.602 & 10.637 & 0.054  \\
456 & 90.637583 & 49.862139&16.486 & 3.967 & 0.011  & 1267 & 90.747306 & 50.032417&18.748 & 11.270 & 0.035  \\
459 & 90.705750 & 49.863694&18.543 & 2.588 & 0.040  & 1196 & 90.541778 & 50.012444&14.887 & 1.775 & 0.017  \\
470 & 90.518778 & 49.864222&14.798 & 2.500 & 0.007  & 1075 & 90.873500& 49.987528&17.240 & 2.263 & 0.019  \\
489 & 90.530833 & 49.868667&16.649 & 5.008 & 0.008  & 1079 & 90.830778 & 49.988194& 13.016 & 2.287 & 0.015  \\ 
503 & 90.469333 & 49.871417&15.547 & 2.608 & 0.016  & 1095 & 90.619694 & 49.991056& 16.197 & 4.403 & 0.010  \\
521 & 90.653472 & 49.876083&15.578 & 5.306 & 0.005  & 1102 & 90.526806 & 49.991806& 18.745 & 1.829 & 0.083  \\
529 & 90.475611 & 49.875667&16.745 & 2.573 & 0.015  &1189 & 90.456472 & 50.010611& 16.340 & 1.931 & 0.015  \\

\hline

\end{tabular}
    \label{tab:newvars}
\end{table*}

\section{Color color diagram}

\begin{figure*}
    \includegraphics[width=\textwidth]{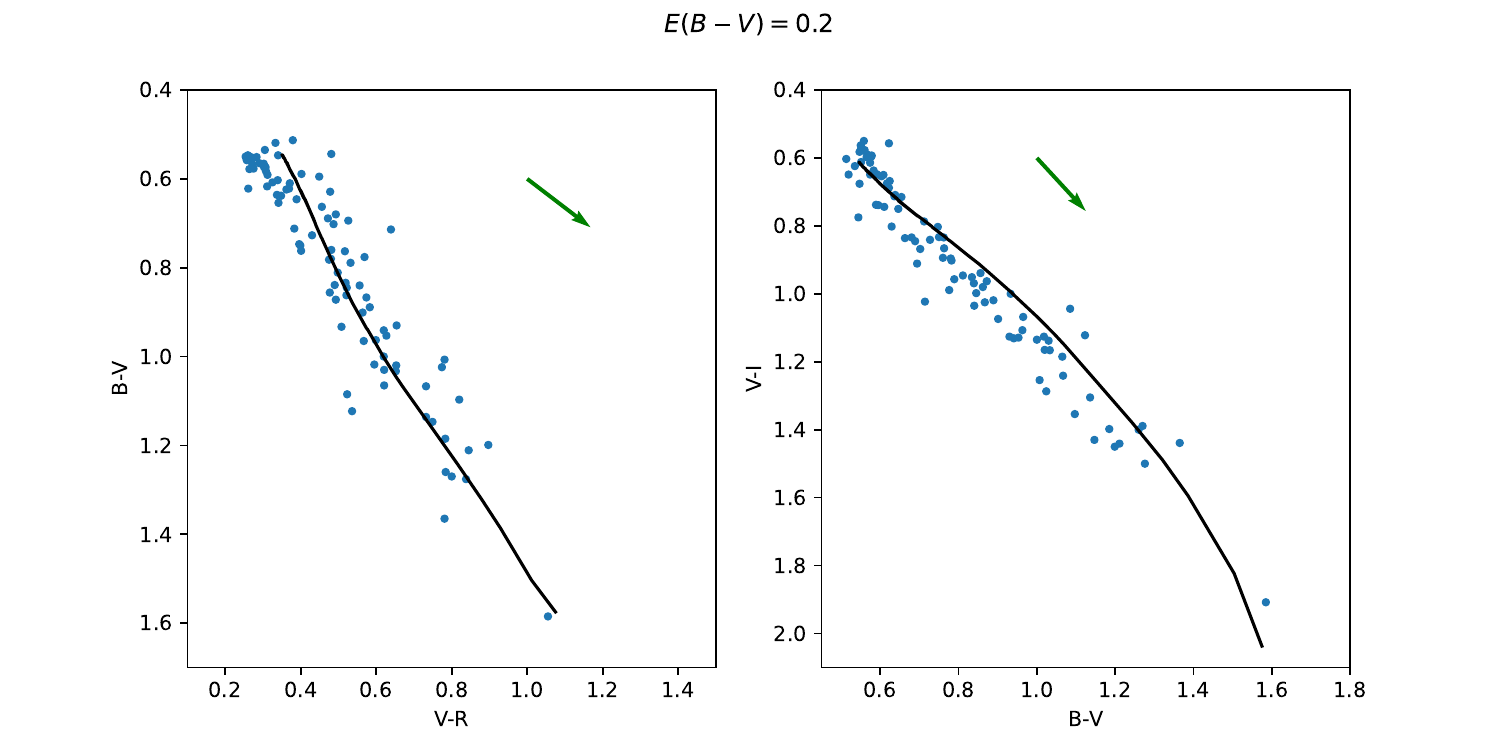}
    \caption{Left panel : The V-R vs B-V diagram for NGC 2126 member stars. Right panel : The B-V vs V-I diagram for NGC 2126 member stars. The black lines in the plot are the theoretical isochrones and the green arrow represents the reddening vector. }
    \label{fig:color_color}
\end{figure*}

\section{TESS Photometry}

\begin{table*}
\centering
\caption{
%List of detected pulsational frequencies for Sectors 19 and 59 for the stars without contamination. 
List of detected pulsational frequencies for Sectors 19, 59, and 73  for the stars without contamination. Their respective frequencies, amplitudes and SNR are listed. The significance criterion was based on the minimum Signal-to-Noise ratio of 5.6 and above.}
\begin{tabular}{ccccccc}
\hline
\hline
\\
ID &Sector& FrequncyID    & Frequency  && Amplitude &  SNR  \\
   &      &               & (\cd)      && (mmag)    &       \\
\hline
\\
V1  & 19 &$f_{V1,19,1} $&$ 1.2475\pm0.0008$&  &$ 7.8 \pm0.3 $&11.1\\
    &    &$f_{V1,19,2} $&$ 1.156 \pm0.001 $&  &$ 4.5 \pm0.3 $& 6.4\\
    & 59 &$f_{V1,59,1} $&$ 1.2355\pm0.0003$&  &$ 5.53\pm0.08$&19.2\\
    &    &$f_{V1,59,2} $&$ 1.1597\pm0.0003$&  &$ 5.37\pm0.08$&17.9\\
    &    &$f_{V1,59,3} $&$ 1.1804\pm0.0004$&  &$ 4.01\pm0.08$&13.6\\
    &    &$f_{V1,59,4} $&$ 1.2893\pm0.0008$&  &$ 2.00\pm0.08$& 7.0\\
    &    &$f_{V1,59,5} $&$ 1.4460\pm0.0009$&  &$ 1.68\pm0.08$& 6.1\\
    & 73 &$f_{V1,73,1} $&$ 1.1485\pm0.0004$&  &$ 5.7 \pm0.1 $& 9.8\\
    &    &$f_{V1,73,2} $&$ 1.2362\pm0.0005$&  &$ 4.7 \pm0.1 $& 8.9\\ \smallskip
    &    &$f_{V1,73,3} $&$ 1.1136\pm0.0008$&  &$ 3.0 \pm0.1 $& 5.1\\
 	          	 		       	    	          	        
V2  & 19 &$f_{V2,19,1} $&$ 1.0771\pm0.0007$&  &$ 8.3 \pm0.3 $&13.9\\
    &    &$f_{V2,19,2} $&$ 1.013 \pm0.001 $&  &$ 5.7 \pm0.3 $& 9.9\\
    &    &$f_{V2,19,3} $&$ 1.176 \pm0.001 $&  &$ 5.1 \pm0.3 $& 9.3\\
    & 59 &$f_{V2,59,1} $&$ 1.0645\pm0.0002$&  &$14.3 \pm0.2 $&25.9\\
    &    &$f_{V2,59,2} $&$ 1.1768\pm0.0007$&  &$ 5.2 \pm0.2 $& 8.9\\
    &    &$f_{V2,59,3} $&$ 1.019 \pm0.001 $&  &$ 3.6 \pm0.2 $& 6.1\\
    & 73 &$f_{V2,73,1} $&$ 1.0652\pm0.0003$&  &$14.0 \pm0.2 $&18.9\\ \smallskip
    &    &$f_{V2,73,2} $&$ 1.1724\pm0.0008$&  &$ 5.1 \pm0.2 $& 7.3\\
 	          	 		       	    	          	        
V3  & 19 &$f_{V3,19,1} $&$12.816 \pm0.003 $&  &$ 4.5 \pm0.5 $& 8.1\\
    &    &$f_{V3,19,2} $&$10.372 \pm0.003 $&  &$ 3.7 \pm0.5 $& 5.5\\
    & 59 &$f_{V3,59,1} $&$12.814 \pm0.001 $&  &$ 4.0 \pm0.2 $&16.2\\
    &    &$f_{V3,59,2} $&$10.374 \pm0.001 $&  &$ 3.4 \pm0.2 $& 8.3\\
    &    &$f_{V3,59,3} $&$13.514 \pm0.002 $&  &$ 1.8 \pm0.2 $& 6.7\\
    & 73 &$f_{V3,73,1} $&$10.374 \pm0.002 $&  &$ 3.6 \pm0.3 $& 7.8\\ \smallskip
    &    &$f_{V3,73,2} $&$12.815 \pm0.002 $&  &$ 3.5 \pm0.3 $& 7.9\\
	          	 		       	    	          	        
V5  & 19 &$f_{V5,19,1} $&$12.091 \pm0.001 $&* &$ 2.2 \pm0.1 $&10.4\\
    &    &$f_{V5,19,2} $&$10.440 \pm0.002 $&  &$ 1.6 \pm0.1 $& 7.5\\
    & 59 &$f_{V5,59,1} $&$11.4365\pm0.0003$&* &$ 2.24\pm0.03$&34.2\\
    &    &$f_{V5,59,2} $&$12.0923\pm0.0004$&* &$ 1.63\pm0.03$&27.3\\
    &    &$f_{V5,59,3} $&$10.4384\pm0.0006$&  &$ 1.06\pm0.03$&23.5\\ 
    &    &$f_{V5,59,4} $&$21.4049\pm0.0007$&  &$ 0.84\pm0.03$&15.1\\
    &    &$f_{V5,59,5} $&$11.8053\pm0.0008$&  &$ 0.76\pm0.03$&12.2\\
    &    &$f_{V5,59,6} $&$10.8224\pm0.0009$&  &$ 0.70\pm0.03$&12.9\\
    &    &$f_{V5,59,7} $&$18.2956\pm0.0009$&  &$ 0.67\pm0.03$&12.5\\
    &    &$f_{V5,59,8} $&$19.0526\pm0.0009$&  &$ 0.64\pm0.03$&13.3\\
    &    &$f_{V5,59,9} $&$11.3121\pm0.0009$&  &$ 0.64\pm0.03$& 9.7\\
    &    &$f_{V5,59,10}$&$10.957 \pm0.001 $&  &$ 0.42\pm0.03$& 7.2\\
    &    &$f_{V5,59,11}$&$23.452 \pm0.001 $&  &$ 0.41\pm0.03$& 9.6\\
    &    &$f_{V5,59,12}$&$21.906 \pm0.002 $&  &$ 0.38\pm0.03$& 6.5\\
    &    &$f_{V5,59,13}$&$12.639 \pm0.002 $&  &$ 0.34\pm0.03$& 6.5\\
    &    &$f_{V5,59,14}$&$21.325 \pm0.002 $&  &$ 0.33\pm0.03$& 6.0\\ 
    & 73 &$f_{V5,73,1} $&$12.0915\pm0.0005$&* &$ 2.54\pm0.06$&24.0\\
    &    &$f_{V5,73,2} $&$11.4449\pm0.0006$&* &$ 1.79\pm0.06$&18.9\\
    &    &$f_{V5,73,3} $&$10.4380\pm0.0007$&  &$ 1.66\pm0.06$&19.1\\
    &    &$f_{V5,73,4} $&$21.4070\pm0.0008$&  &$ 1.47\pm0.06$&12.4\\
    &    &$f_{V5,73,5} $&$11.8051\pm0.0009$&  &$ 1.29\pm0.06$&12.3\\
    &    &$f_{V5,73,6} $&$19.0531\pm0.0009$&  &$ 1.30\pm0.06$&18.3\\
    &    &$f_{V5,73,7} $&$18.298 \pm0.001 $&  &$ 1.16\pm0.06$&13.5\\
    &    &$f_{V5,73,8} $&$11.314 \pm0.001 $&  &$ 1.17\pm0.06$&12.9\\
    &    &$f_{V5,73,9} $&$10.824 \pm0.001 $&  &$ 1.13\pm0.06$&12.7\\
    &    &$f_{V5,73,10}$&$10.953 \pm0.001 $&  &$ 0.92\pm0.06$&10.3\\
    &    &$f_{V5,73,11}$&$23.453 \pm0.001 $&  &$ 0.80\pm0.06$& 9.4\\
    &    &$f_{V5,73,12}$&$11.372 \pm0.002 $&  &$ 0.65\pm0.06$& 7.1\\ \smallskip
    &    &$f_{V5,73,13}$&$23.250 \pm0.002 $&  &$ 0.50\pm0.06$& 6.1\\

\hline

\multicolumn{4}{l}{\small *Frequency values consistent with \cite{2018chehlah}.} \\
%\multicolumn{4}{l}{\small **Frequency values consistent with \cite{2012zhang}.} \\
\end{tabular}
    \label{tab:pulsa1}
\end{table*}

\addtocounter{table}{-1}

\begin{table*}
\centering
\caption{
Continued.}
\begin{tabular}{ccccccc}
\hline
\hline
\\
ID &Sector& FrequncyID    & Frequency  && Amplitude &  SNR  \\
   &      &               & (\cd)      && (mmag)    &       \\
\hline
\\
ZV1 & 19 &$f_{ZV1,19,1} $&$12.253 \pm0.002 $&* &$ 1.2 \pm0.1 $& 9.2\\
    &    &$f_{ZV1,19,2} $&$15.461 \pm0.003 $&* &$ 0.9 \pm0.1 $& 6.8\\
    & 59 &$f_{ZV1,59,1} $&$12.2504\pm0.0006$&* &$ 4.3 \pm0.1 $&21.9\\
    &    &$f_{ZV1,59,2} $&$15.458 \pm0.001$& * &$ 2.0 \pm0.1 $&12.6\\
    &    &$f_{ZV1,59,3} $&$16.120 \pm0.002 $&  &$ 1.4 \pm0.1 $& 9.4\\
    &    &$f_{ZV1,59,4} $&$11.165 \pm0.002 $&  &$ 1.4 \pm0.1 $& 7.9\\
    &    &$f_{ZV1,59,5} $&$15.943 \pm0.002 $&  &$ 1.4 \pm0.1 $& 9.5\\
    &    &$f_{ZV1,59,6} $&$13.512 \pm0.002 $&  &$ 1.3 \pm0.1 $& 6.1\\
    &    &$f_{ZV1,59,7} $&$11.660 \pm0.003 $&  &$ 1.1 \pm0.1 $& 5.8\\
    & 73 &$f_{ZV1,73,1} $&$12.2531\pm0.0008$&* &$ 2.11\pm0.08$&16.0\\
    &    &$f_{ZV1,73,2} $&$15.461 \pm0.002 $&* &$ 0.97\pm0.08$& 6.8\\ \smallskip
    &    &$f_{ZV1,73,3} $&$11.171 \pm0.003 $&  &$ 0.65\pm0.08$& 5.2\\
	          	 		       	    	          	        
N2  & 19 &$f_{N2,19,1} $&$14.552 \pm0.002 $&  &$ 1.4 \pm0.1 $& 7.3\\
    &    &$f_{N2,19,2} $&$17.094 \pm0.003 $&  &$ 1.3 \pm0.1 $& 7.8\\
    &    &$f_{N2,19,3} $&$ 4.608 \pm0.003 $&  &$ 1.1 \pm0.1 $& 6.1\\
    & 59 &$f_{N2,59,1} $&$14.5554\pm0.0007$&* &$ 2.09\pm0.07$&22.7\\
    &    &$f_{N2,59,2} $&$17.0921\pm0.0008$&  &$ 1.89\pm0.07$&21.1\\
    &    &$f_{N2,59,3} $&$16.304 \pm0.001 $&  &$ 1.10\pm0.07$&13.6\\
    &    &$f_{N2,59,4} $&$18.627 \pm0.002 $&  &$ 0.89\pm0.07$&10.0\\
    &    &$f_{N2,59,5} $&$15.134 \pm0.002 $&  &$ 0.82\pm0.07$& 9.5\\
    &    &$f_{N2,59,6} $&$ 5.206 \pm0.002 $&  &$ 0.79\pm0.07$& 5.7\\
    &    &$f_{N2,59,7} $&$11.785 \pm0.003*$&  &$ 0.61\pm0.07$& 6.1\\
    &    &$f_{N2,59,8}$&$15.098 \pm0.003 $&  &$ 0.57\pm0.07$& 6.6\\
    &    &$f_{N2,59,9}$&$15.774 \pm0.003 $&  &$ 0.49\pm0.07$& 5.9\\
    & 73 &$f_{N2,73,1} $&$14.554 \pm0.001 $&* &$ 1.89\pm0.09$&14.5\\
    &    &$f_{N2,73,2} $&$17.088 \pm0.001 $&* &$ 1.55\pm0.09$&17.0\\
    &    &$f_{N2,73,3} $&$16.301 \pm0.002 $&  &$ 1.08\pm0.09$& 9.9\\ \smallskip
    &    &$f_{N2,73,4} $&$15.138 \pm0.002 $&  &$ 0.91\pm0.09$& 6.8\\
    
839  & 19 &$f_{839,19,1} $&$1.3530 \pm 0.0006$& &$21.6 \pm 0.6$&15.0 \\
    &  &$f_{839,19,2} $&$2.715 \pm 0.003$& &$4.5 \pm 0.6$&5.2 \\

\hline

\multicolumn{4}{l}{\small *Frequency values consistent with \cite{2018chehlah}.} \\
%\multicolumn{4}{l}{\small **Frequency values consistent with \cite{2012zhang}.} \\
\end{tabular}
\end{table*}

\begin{table*}
\centering
\caption{List of detected pulsational frequencies for Sectors 19, 59 and 73 for the stars that are contaminated. Their respective frequencies, amplitudes and SNR are listed. The significance criterion was based on the minimum Signal-to-Noise ratio of 5.6 and above.}
\begin{tabular}{ccccccc}
\hline
\hline
\\
ID &Sector& FrequncyID    & Frequency  && Amplitude  & SNR  \\
   &      &               & (\cd)      && (mmag)    &      \\
\hline
\\
V6 & 19 &$ f_{V6,19,1}  $&$ 7.7248\pm0.0005$&* &$3.08\pm0.07$&46.8\\
   & 59 &$ f_{V6,59,1}  $&$ 7.7266\pm0.0004$&* &$4.10\pm0.07$&41.5\\
   &    &$ f_{V6,59,1}/2$&$ 3.862 \pm0.002 $&  &$0.75\pm0.07$& 8.2\\
   &    &$2f_{V6,59,1}  $&$15.450 \pm0.002 $&* &$0.63\pm0.07$& 7.9\\
   & 73 &$ f_{V6,73,1}  $&$ 7.7270\pm0.0002$&* &$4.77\pm0.06$&49.3\\
   &    &$ f_{V6,73,1}/2$&$ 3.864 \pm0.002 $&  &$0.74\pm0.06$& 7.8\\ \smallskip
   &    &$2f_{V6,73,1}  $&$15.454 \pm0.002 $&* &$0.56\pm0.06$& 8.8\\
     	           	                       	      	     	    
ZV2& 19 &$ f_{ZV2,19,1}  $&$14.851 \pm0.002 $&* &$1.6 \pm0.2 $& 7.7\\
   & 59 &$ f_{ZV2,59,1}  $&$14.852 \pm0.001 $&* &$2.4 \pm0.2 $&14.2\\
   &    &$ f_{ZV2,59,2}  $&$14.802 \pm0.003 $&  &$1.3 \pm0.2 $& 7.9\\
   &    &$ f_{ZV2,59,3}  $&$14.409 \pm0.003 $&  &$1.0 \pm0.2 $& 6.8\\
   &    &$ f_{ZV2,59,4}  $&$ 0.318 \pm0.002 $&  &$1.2 \pm0.1 $& 5.8\\
   &    &$ f_{ZV2,59,5}  $&$ 1.815 \pm0.002 $&**&$1.1 \pm0.1 $& 5.7\\
   & 73 &$ f_{ZV2,73,1}  $&$14.8394\pm0.0008$&* &$2.53\pm0.09$&26.1\\
   &    &$ f_{ZV2,73,2}  $&$14.405 \pm0.002 $&  &$1.01\pm0.09$&10.3\\ \smallskip
   &    &$ f_{ZV2,73,3}  $&$14.793 \pm0.002 $&  &$1.07\pm0.09$&11.0\\
     	           	                       	      	     	    
N1 & 19 &$ f_{N1,19,1}  $&$13.5991\pm0.0008$&* &$1.81\pm0.07$&19.4\\
   &    &$ f_{N1,19,2}  $&$17.176 \pm0.003 $&* &$0.76\pm0.07$&10.4\\
   & 59 &$ f_{N1,59,1}  $&$13.5969\pm0.0009$&* &$1.70\pm0.07$&19.7\\
   &    &$ f_{N1,59,2}  $&$17.177 \pm0.002 $&* &$0.74\pm0.07$& 8.4\\ 
   & 73 &$ f_{N1,73,1}  $&$13.5972\pm0.0003$&* &$3.38\pm0.05$&28.4\\
   &    &$ f_{N1,73,2}  $&$17.1767\pm0.0006$&* &$1.67\pm0.05$&24.6\\
   &    &$ f_{N1,73,3}  $&$21.203 \pm0.001 $&* &$0.71\pm0.05$& 9.1\\
   &    &$ f_{N1,73,4}  $&$21.348 \pm0.002 $&  &$0.59\pm0.05$& 7.8\\

\hline

\multicolumn{4}{l}{\small *Frequency values consistent with \cite{2018chehlah}.} \\
\multicolumn{4}{l}{\small **Frequency values consistent with \cite{2012zhang}.} \\
\end{tabular}
    \label{tab:pulsa3}
\end{table*}

\begin{figure}
    \includegraphics[width=\columnwidth]{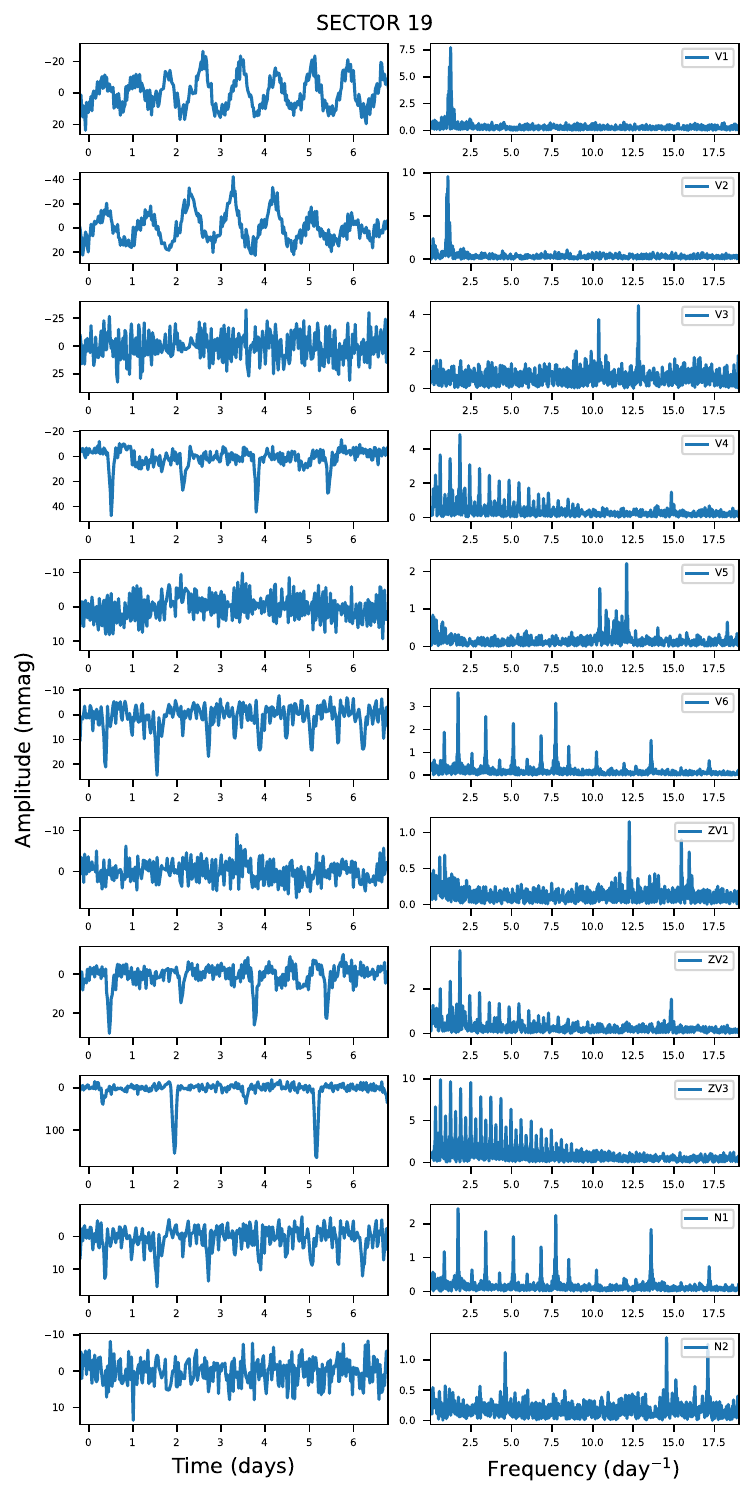}
    \caption{The Sector 19 LCs and the corresponding frequency spectra of known variables in NGC 2126.}
    \label{fig:lcs19}
\end{figure}

\begin{figure}
    \includegraphics[width=\columnwidth]{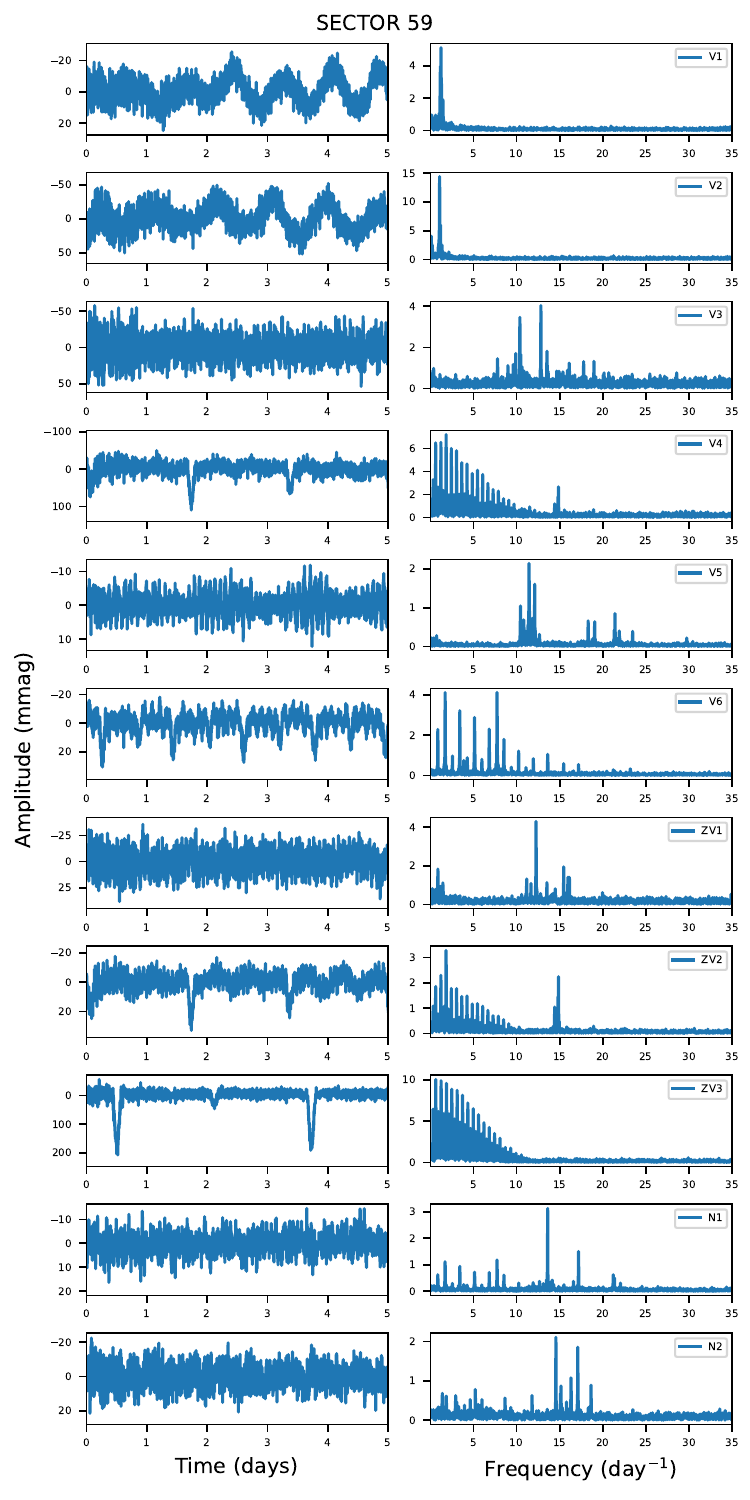}
    \caption{The Sector 59 LCs and the corresponding frequency spectra of known variables in NGC 2126.}
    \label{fig:lcs59}
\end{figure}

\begin{figure}
    \includegraphics[width=\columnwidth]{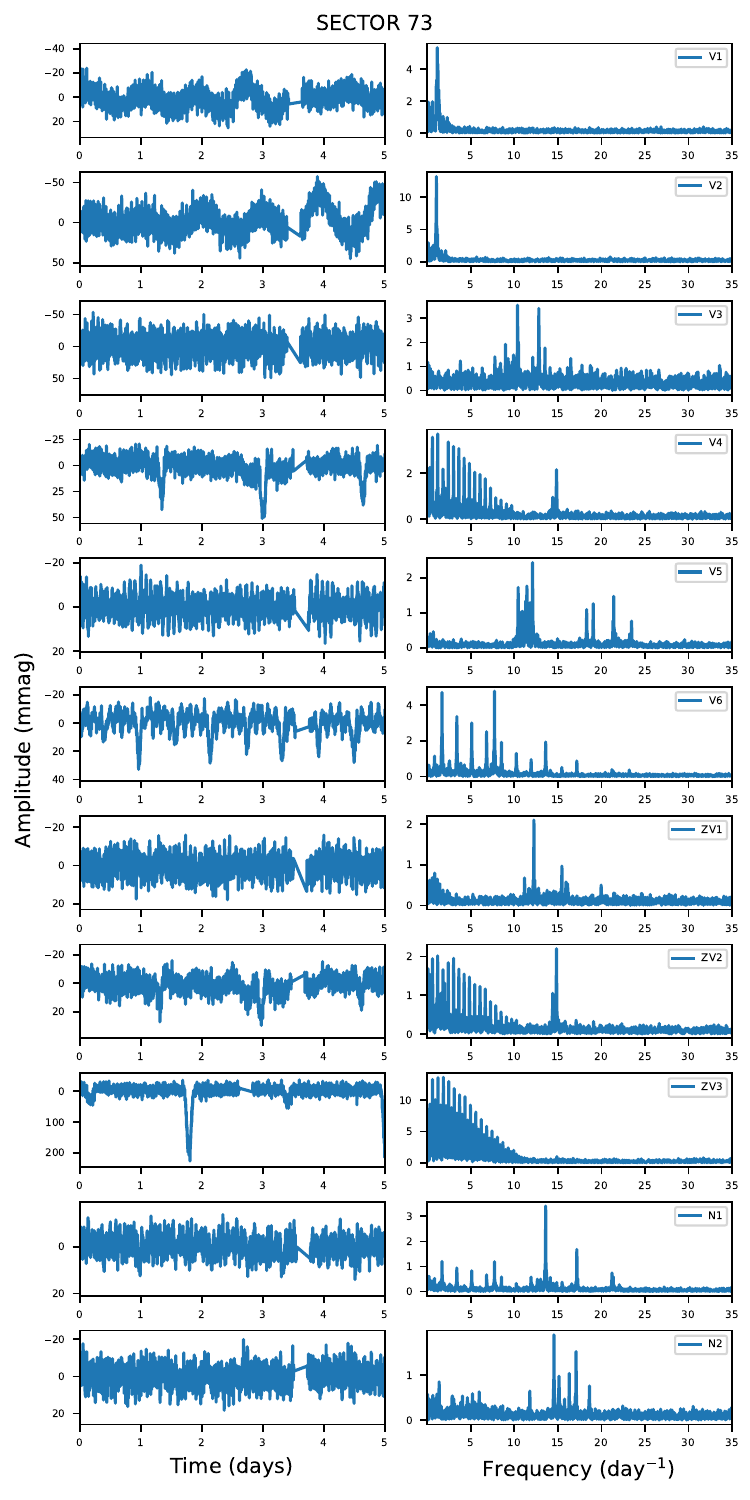}
    \caption{The Sector 73 LCs and the corresponding frequency spectra of known variables in NGC 2126.}
    \label{fig:lcs73}
\end{figure}

\section{Binary Ephermeris}
\begin{figure}
    \includegraphics[width=\columnwidth]{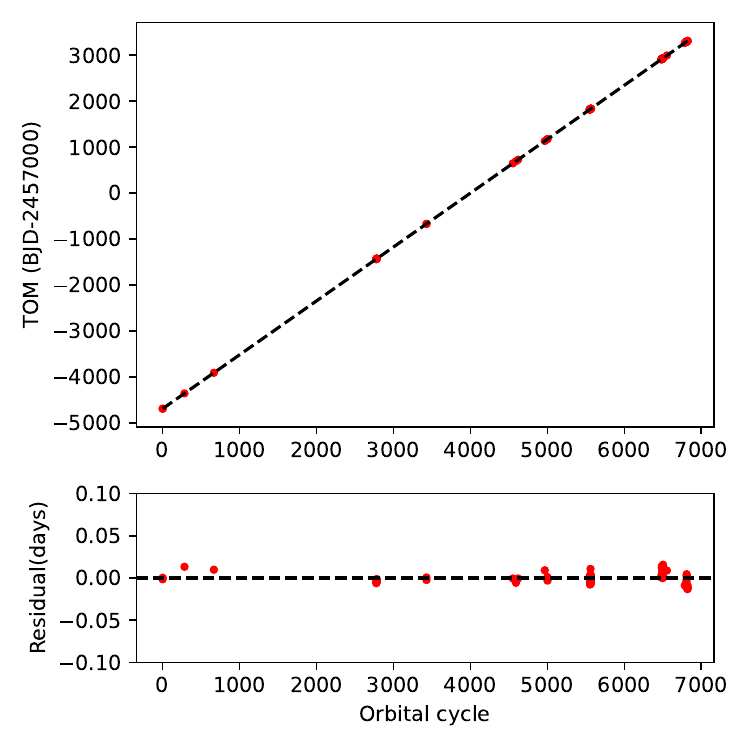}
\caption{The top panel shows the linear ephemeris of V551 Aur by fitting a straight line to the observed times of primary minima (TOM) from all the TESS sectors along with the ground-based minima detected in this study and listed in \citet{2018chehlah}. The bottom panel shows the residuals of the fit.}
    \label{fig:Eph1}
\end{figure}

\begin{figure}
    \includegraphics[width=\columnwidth]{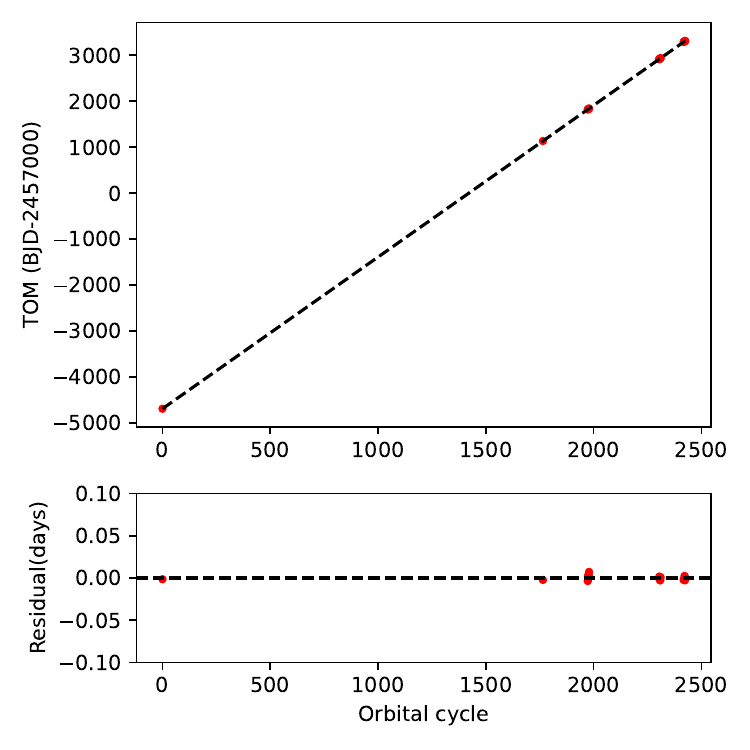}
\caption{The top panel shows the linear ephemeris of V549 Aur by fitting a straight line to the observed time of primary minima (TOM) from all the TESS sectors and two eclipse minima given in \citet{2018chehlah}. The bottom panel shows the residuals of the fit.}
    \label{fig:Ephv549}
\end{figure}

\section{EB modeling}

\begin{table*}
\centering
\begin{minipage}{\linewidth}
\caption{The table gives the adopted prior distributions for the \texttt{PyMC3} runs with 6000 samples. Here, $M1$ is the mass of the primary star, $R1$ is the radius of the primary star, $k$ is the radius ratio, $q$ is the mass ratio, $s$ is the surface brightness ratio and $ecs$ represents the two orthogonal components of eccentricity $e\cos{\omega}$ and $e\sin{\omega}$.} 
\label{priors}
\end{minipage}
\bigskip
\begin{tabular}{cccc}
\hline
\hline
\\
\textbf{Parameters}  &\textbf{V6} &\textbf{V4} &\textbf{ZV3}\\
\hline
\\
$log\_M1$ & Normal(log(1.375), 1.0) & Normal(log(1.05), 1.0) & Normal(log(1.20), 1.0)\\
$log\_R1$ & Normal(log(1.728), 1.0)& Normal(log(0.83), 1.0)& Normal(log(1.039), 1.0)\\
$log\_k$  & Normal(0.0, 10.0)& Normal(0.0, 10.0)& Normal(0.0, 10.0)\\
$log\_q$ & Normal(log(0.769), 0.01)& Normal(log(0.626), 0.01)& Normal(log(0.48), 0.01)\\
$log\_s$ & Normal(log(0.5), 10.0)& Normal(log(0.5), 10.0)& Normal(log(0.5), 10.0)\\
$ecs$ & UnitDisk([1e-5, 0.0])& UnitDisk([1e-5, 0.0])& UnitDisk([1e-5, 0.0])\\
\hline
\end{tabular}
\end{table*}

\begin{figure}
    \includegraphics[width=\columnwidth]{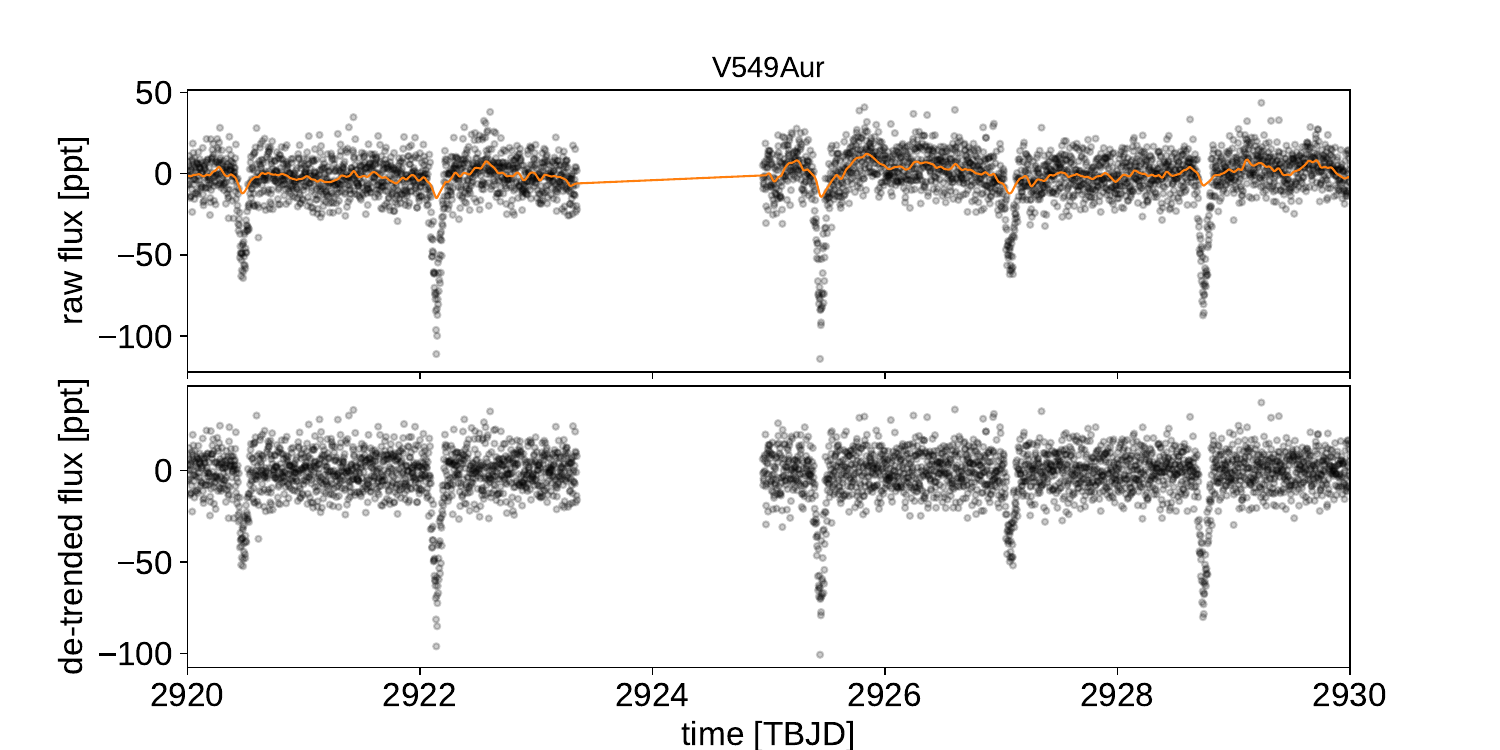}
    \caption{Top panel shows the Gaussian process model of V549 Aur. Bottom panel shows the residuals after subtracting the model}
    \label{fig:eclips1}
\end{figure}

\begin{figure}
    \includegraphics[width=\columnwidth]{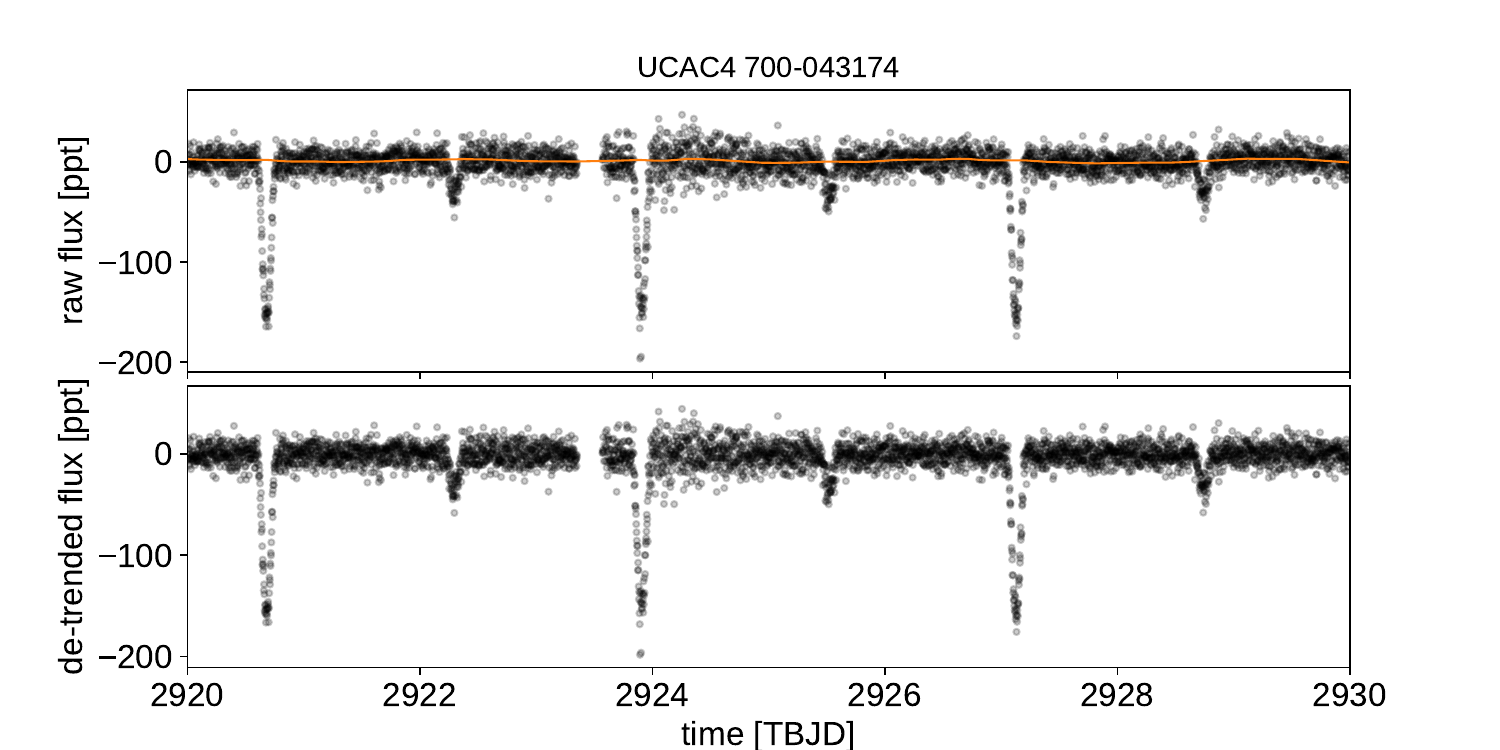}
    \caption{Top panel shows the Gaussian process model of V549 Aur. Bottom panel shows the residuals after subtracting the model}
    \label{fig:eclips3}
\end{figure}

\begin{figure}
    \includegraphics[width=\columnwidth]{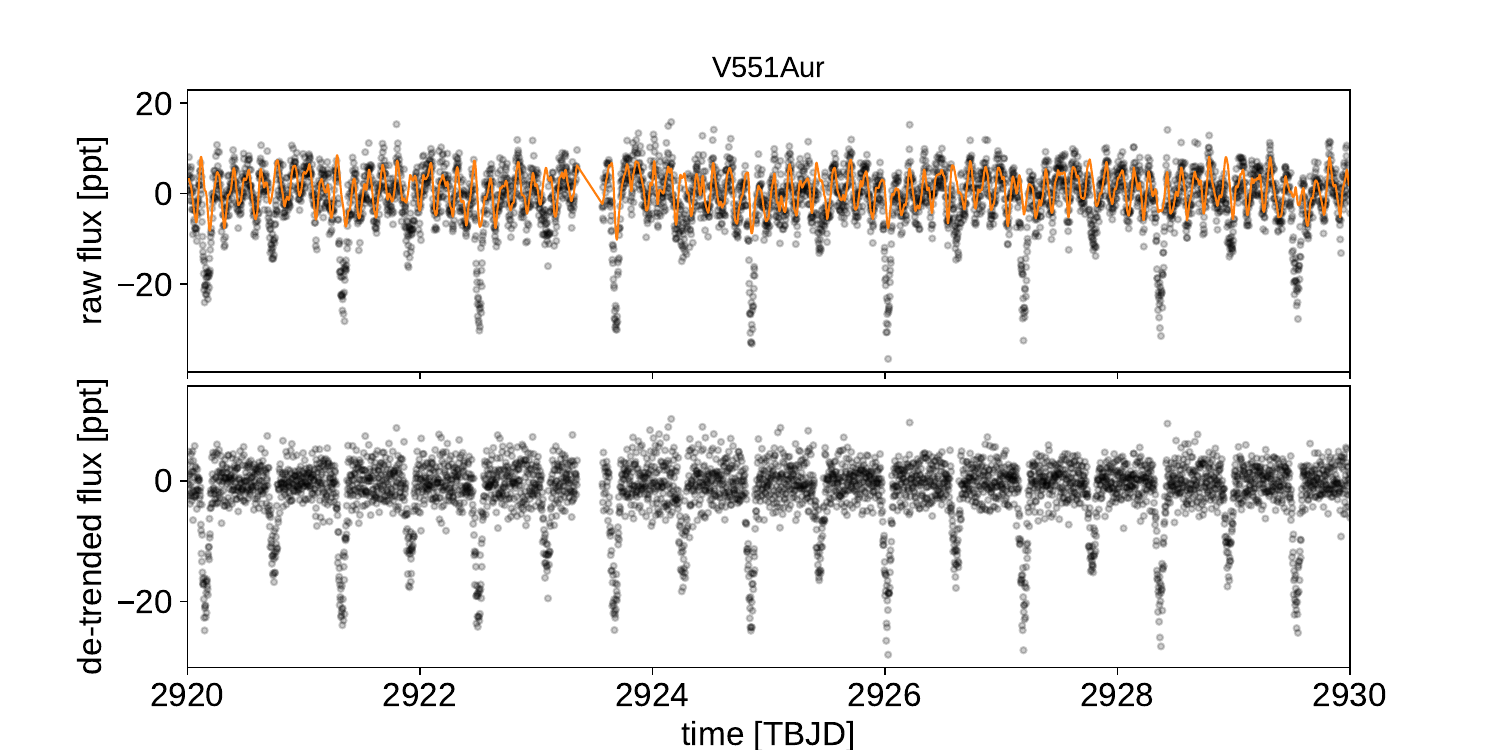}
    \caption{The top panel shows the Gaussian process model of V551 Aur for the pulsational variability. The bottom panel shows the residual LC after subtracting the pulsation model.}
    \label{fig:eclips}
\end{figure}

\begin{figure}
    \includegraphics[width=\columnwidth]{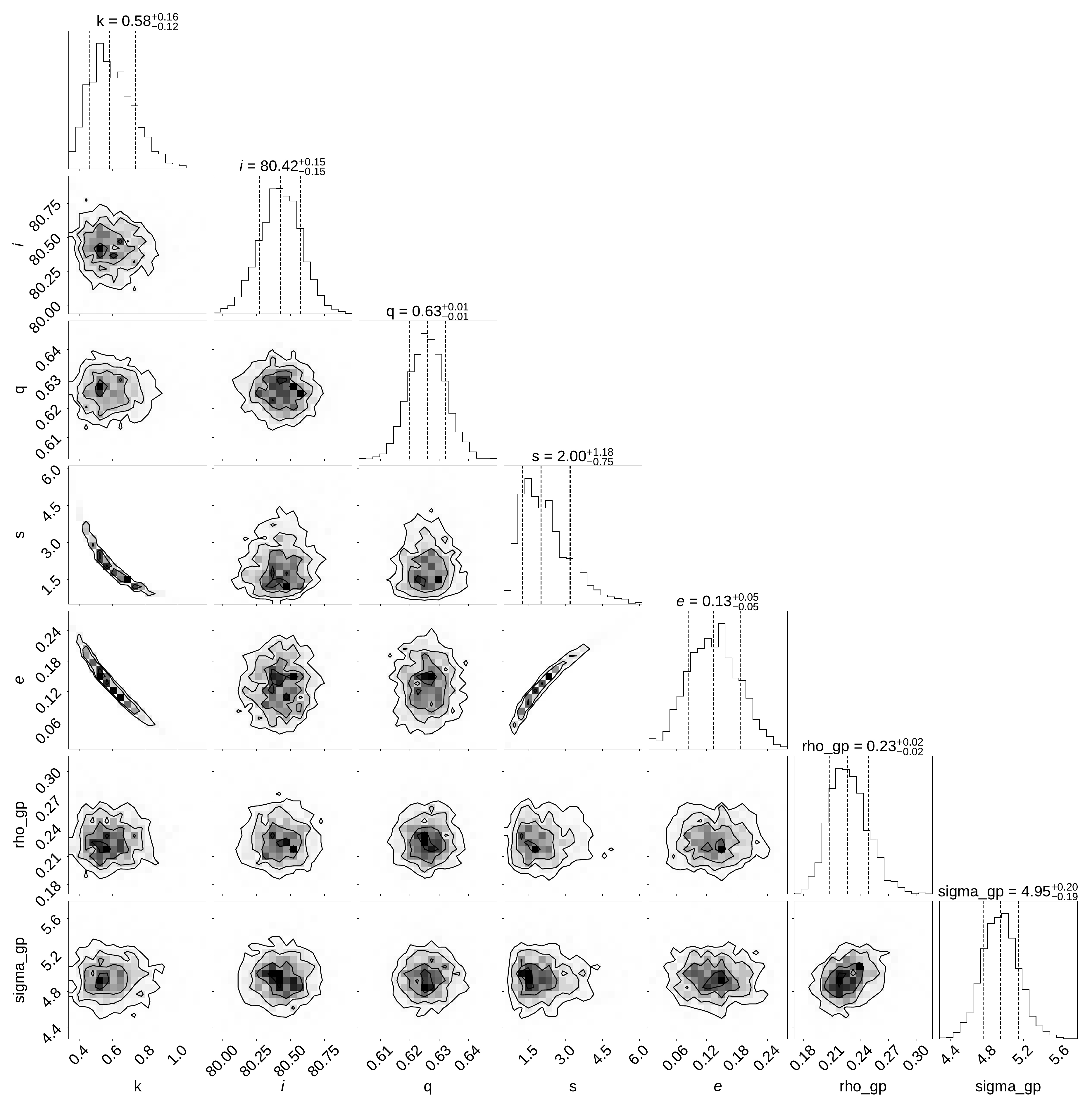}
    \caption{Figure shows the posterior distribution for the eclipse model and Gaussian process model from \texttt{exoplanet} code for V549 Aur.}
    \label{fig:phoebecornerv549}
\end{figure}

\begin{figure}
   \includegraphics[width=\columnwidth]{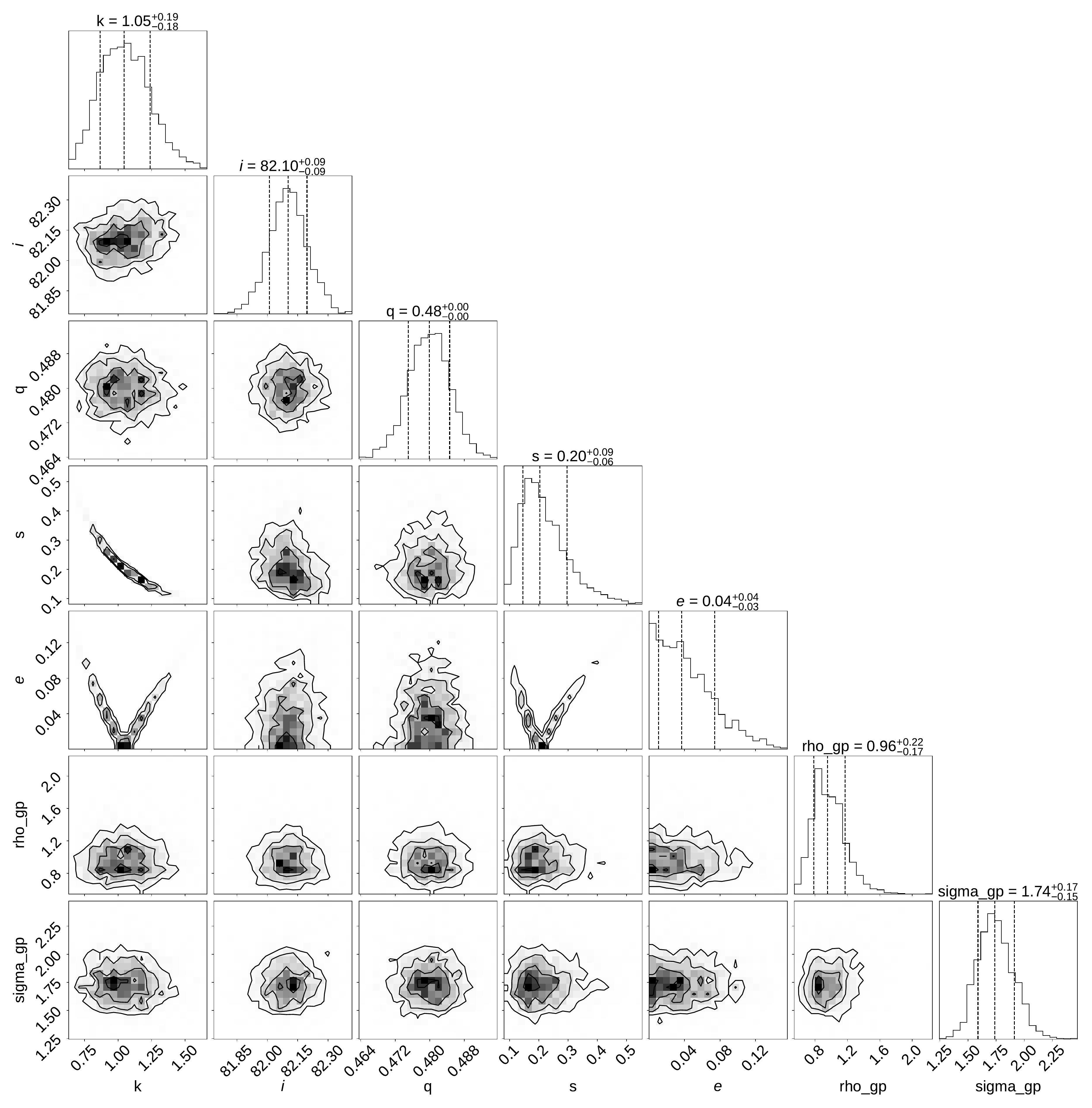}
    \caption{Figure shows the posterior distribution for the eclipse model and the Gaussian process model from \texttt{exoplanet} code for ZV3.}
    \label{fig:phoebecorner174}
\end{figure}

\section{Low resolution spectroscopy}

\begin{table}
\centering
\begin{minipage}{\linewidth}
%\begin{minipage}{150mm}
\caption{
The table lists the effective temperature (\teff) and surface gravity (\logg) using Gaia DR3 gsphot and the series of low-resolution spectra. The values of \teff\ and \logg\ were derived by fitting the observed spectra with synthetic spectra. The last row gives the mean \teff\ and \logg\ values of those derived from the individual spectra.}
%temperature and log $g$ from all the results.

\label{Tab:teff}
\end{minipage}
\bigskip
\begin{tabular}{ccc}
\hline
\hline
\\
\textbf{Source}  &\textbf{\teff} &\textbf{\logg}  \\
\hline
\\
Gaia DR3 (gsphot)   &   5943 & 4.0 \\
 &   \\
Low-resolution   &    &\\
spectrum (this study)   &  &  \\
   &    &\\
2021-08-15 & 5853 $\pm$ 213 & 3 $\pm$ 1\\ 
2021-09-13 & 5898 $\pm$ 178 & 4 $\pm$ 1\\ 
2021-09-18 & 5938 $\pm$ 177 & 3 $\pm$ 1\\ 
2021-09-19 & 5792 $\pm$ 213 & 3 $\pm$ 1\\ 
2021-10-05 & 5868 $\pm$ 299 & 4 $\pm$ 3 \\ 
2021-10-12 & 5792 $\pm$ 272& 4 $\pm$ 2 \\ 
2021-10-19 & 5772 $\pm$ 424 & 3 $\pm$ 3\\ 
2021-10-22 & 5832 $\pm$ 348& 4 $\pm$ 3 \\ 
2021-10-26 & 5746 $\pm$ 221 & 3 $\pm$ 1\\ 
2021-11-15 & 5898 $\pm$ 173 & 3 $\pm$ 1\\ 
2021-11-21 & 5878 $\pm$ 365& 3 $\pm$ 3 \\ 
2021-12-08 & 5853 $\pm$ 247& 4 $\pm$ 2 \\ 
2021-12-20 & 5777 $\pm$ 211& 3 $\pm$ 1 \\ 
2022-01-03 & 5640 $\pm$ 241 & 3 $\pm$ 2\\ 
2022-02-08 & 5827 $\pm$ 252 & 3 $\pm$ 1\\ 
2022-02-17 & 5853 $\pm$ 208& 3 $\pm$ 1  \\ 

mean &  5826 $\pm$ 262& 3$\pm$ 2\\

\hline
\end{tabular}
\end{table}

%%%%%%%%%%%%%%%%%%%%%%%%%%%%%%%%%%%%%%%%%%%%%%%%%%

% Don't change these lines
\bsp	% typesetting comment
\label{lastpage}
\end{document}